\documentclass{jcgt}

\setciteauthor{Jamie Portsmouth, Peter Kutz, and Stephen Hill}
\setcitetitle{EON: A Practical Energy-Preserving Rough Diffuse BRDF}

\accepted{2024-07-12}{2024-11-21}{2025-03-31}{Christophe Hery}{14}{1}{116}{139}{2025}
\seturl{http://jcgt.org/published/0014/01/06/}

\usepackage[normalem]{ulem}
\usepackage{siunitx}
\usepackage{csquotes}
\usepackage{url}
\usepackage{subcaption}
\usepackage{xcolor}
\usepackage{algorithm}
\usepackage{algorithmicx}
\usepackage{algpseudocode}
\usepackage{mathtools}
\usepackage{xfrac}
\usepackage{comment}
\usepackage{natbib}
\usepackage{hyperref}

\includecomment{introduction}
\includecomment{existing-models}
\includecomment{eon}
\includecomment{eon-sampling}
\includecomment{conclusion}

\lstdefinelanguage{GLSL}%
{%
	morekeywords={%
		false,FALSE,NULL,true,TRUE,%
		__LINE__,__FILE__,__VERSION__,GL_core_profile,GL_es_profile,GL_compatibility_profile,%
		precision,highp,mediump,lowp,%
		break,case,continue,default,discard,do,else,for,if,return,switch,while,define%
		void,bool,int,uint,float,double,vec2,vec3,vec4,dvec2,dvec3,dvec4,bvec2,bvec3,bvec4,ivec2,ivec3,ivec4,uvec2,uvec3,uvec4,mat2,mat3,mat4,mat2x2,mat2x3,mat2x4,mat3x2,mat3x3,mat3x4,mat4x2,mat4x3,mat4x4,dmat2,dmat3,dmat4,dmat2x2,dmat2x3,dmat2x4,dmat3x2,dmat3x3,dmat3x4,dmat4x2,dmat4x3,dmat4x4,sampler1D,sampler2D,sampler3D,image1D,image2D,image3D,samplerCube,imageCube,sampler2DRect,image2DRect,sampler1DArray,sampler2DArray,image1DArray,image2DArray,samplerBuffer,imageBuffer,sampler2DMS,image2DMS,sampler2DMSArray,image2DMSArray,samplerCubeArray,imageCubeArray,sampler1DShadow,sampler2DShadow,sampler2DRectShadow,sampler1DArrayShadow,sampler2DArrayShadow,samplerCubeShadow,samplerCubeArrayShadow,isampler1D,isampler2D,isampler3D,iimage1D,iimage2D,iimage3D,isamplerCube,iimageCube,isampler2DRect,iimage2DRect,isampler1DArray,isampler2DArray,iimage1DArray,iimage2DArray,isamplerBuffer,iimageBuffer,isampler2DMS,iimage2DMS,isampler2DMSArray,iimage2DMSArray,isamplerCubeArray,iimageCubeArray,atomic_uint,usampler1D,usampler2D,usampler3D,uimage1D,uimage2D,uimage3D,usamplerCube,uimageCube,usampler2DRect,uimage2DRect,usampler1DArray,usampler2DArray,uimage1DArray,uimage2DArray,usamplerBuffer,uimageBuffer,usampler2DMS,uimage2DMS,usampler2DMSArray,uimage2DMSArray,usamplerCubeArray,uimageCubeArray,struct,%
		gl_BackColor,gl_BackLightModelProduct,gl_BackLightProduct,gl_BackMaterial,gl_BackSecondaryColor,gl_ClipDistance,gl_ClipPlane,gl_ClipVertex,gl_Color,gl_DepthRange,gl_DepthRangeParameters,gl_EyePlaneQ,gl_EyePlaneR,gl_EyePlaneS,gl_EyePlaneT,gl_Fog,gl_FogCoord,gl_FogFragCoord,gl_FogParameters,gl_FragColor,gl_FragCoord,gl_FragData,gl_FragDepth,gl_FrontColor,gl_FrontFacing,gl_FrontLightModelProduct,gl_FrontLightProduct,gl_FrontMaterial,gl_FrontSecondaryColor,gl_InstanceID,gl_Layer,gl_LightModel,gl_LightModelParameters,gl_LightModelProducts,gl_LightProducts,gl_LightSource,gl_LightSourceParameters,gl_MaterialParameters,gl_ModelViewMatrix,gl_ModelViewMatrixInverse,gl_ModelViewMatrixInverseTranspose,gl_ModelViewMatrixTranspose,gl_ModelViewProjectionMatrix,gl_ModelViewProjectionMatrixInverse,gl_ModelViewProjectionMatrixInverseTranspose,gl_ModelViewProjectionMatrixTranspose,gl_MultiTexCoord0,gl_MultiTexCoord1,gl_MultiTexCoord2,gl_MultiTexCoord3,gl_MultiTexCoord4,gl_MultiTexCoord5,gl_MultiTexCoord6,gl_MultiTexCoord7,gl_Normal,gl_NormalMatrix,gl_NormalScale,gl_ObjectPlaneQ,gl_ObjectPlaneR,gl_ObjectPlaneS,gl_ObjectPlaneT,gl_Point,gl_PointCoord,gl_PointParameters,gl_PointSize,gl_Position,gl_PrimitiveIDIn,gl_ProjectionMatrix,gl_ProjectionMatrixInverse,gl_ProjectionMatrixInverseTranspose,gl_ProjectionMatrixTranspose,gl_SecondaryColor,gl_TexCoord,gl_TextureEnvColor,gl_TextureMatrix,gl_TextureMatrixInverse,gl_TextureMatrixInverseTranspose,gl_TextureMatrixTranspose,gl_Vertex,gl_VertexID,%
		gl_MaxClipPlanes,gl_MaxCombinedTextureImageUnits,gl_MaxDrawBuffers,gl_MaxFragmentUniformComponents,gl_MaxLights,gl_MaxTextureCoords,gl_MaxTextureImageUnits,gl_MaxTextureUnits,gl_MaxVaryingFloats,gl_MaxVertexAttribs,gl_MaxVertexTextureImageUnits,gl_MaxVertexUniformComponents,%
		abs,acos,all,any,asin,atan,ceil,clamp,cos,cross,degrees,dFdx,dFdy,distance,dot,equal,exp,exp2,faceforward,floor,fract,ftransform,fwidth,greaterThan,greaterThanEqual,inversesqrt,length,lessThan,lessThanEqual,log,log2,matrixCompMult,max,min,mix,mod,noise1,noise2,noise3,noise4,normalize,not,notEqual,outerProduct,pow,radians,reflect,refract,shadow1D,shadow1DLod,shadow1DProj,shadow1DProjLod,shadow2D,shadow2DLod,shadow2DProj,shadow2DProjLod,sign,sin,smoothstep,sqrt,step,tan,texture1D,texture1DLod,texture1DProj,texture1DProjLod,texture2D,texture2DLod,texture2DProj,texture2DProjLod,texture3D,texture3DLod,texture3DProj,texture3DProjLod,textureCube,textureCubeLod,transpose,%
		rgb
	},
	sensitive=true,%
	morecomment=[s]{/*}{*/},%
	morecomment=[l]//,%
	morestring=[b]",%
	morestring=[b]',%
	moredelim=*[directive]\#,%
	moredirectives={define,defined,elif,else,if,ifdef,endif,line,error,ifndef,include,pragma,undef,warning,extension,version}%
}[keywords,comments,strings,directives]%

\definecolor{backcolour}{rgb}{1, 1, 1}
\definecolor{codegreen}{rgb}{0,0.6,0}
\definecolor{codegray}{rgb}{0.5,0.5,0.5}
\definecolor{codepurple}{rgb}{0.58,0,0.82}
\definecolor{codeblue}{rgb}{0,0.3,0.6}

\DeclareUnicodeCharacter{2212}{-}

\DeclarePairedDelimiter{\norm}{\lVert}{\rVert}

\makeatletter
\newcommand\tinyplus{\@setfontsize\tinyplus{7}{8}}
\makeatother

\begin{document}

\title{EON: A Practical Energy-Preserving \\Rough Diffuse BRDF}

\author{Jamie Portsmouth~\href{https://orcid.org/0000-0003-4261-7730}{\includegraphics[width=8pt]{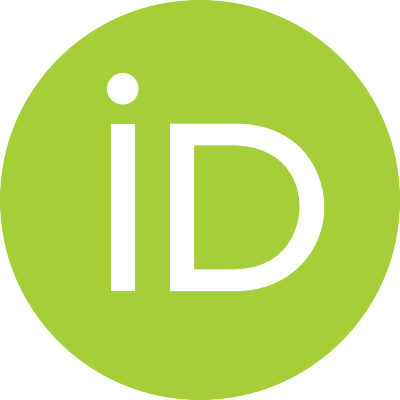}}\\Autodesk
        \and Peter Kutz~\href{https://orcid.org/0009-0003-4097-1223}{\includegraphics[width=8pt]{ORCIDlogo}}\\Adobe
        \and Stephen Hill~\href{https://orcid.org/0009-0004-8278-7771}{\includegraphics[width=8pt]{ORCIDlogo}}\\Lucasfilm
       }

\teaser{
  \vspace*{-0.5cm}
  \centering
    \includegraphics[width=0.32\linewidth]{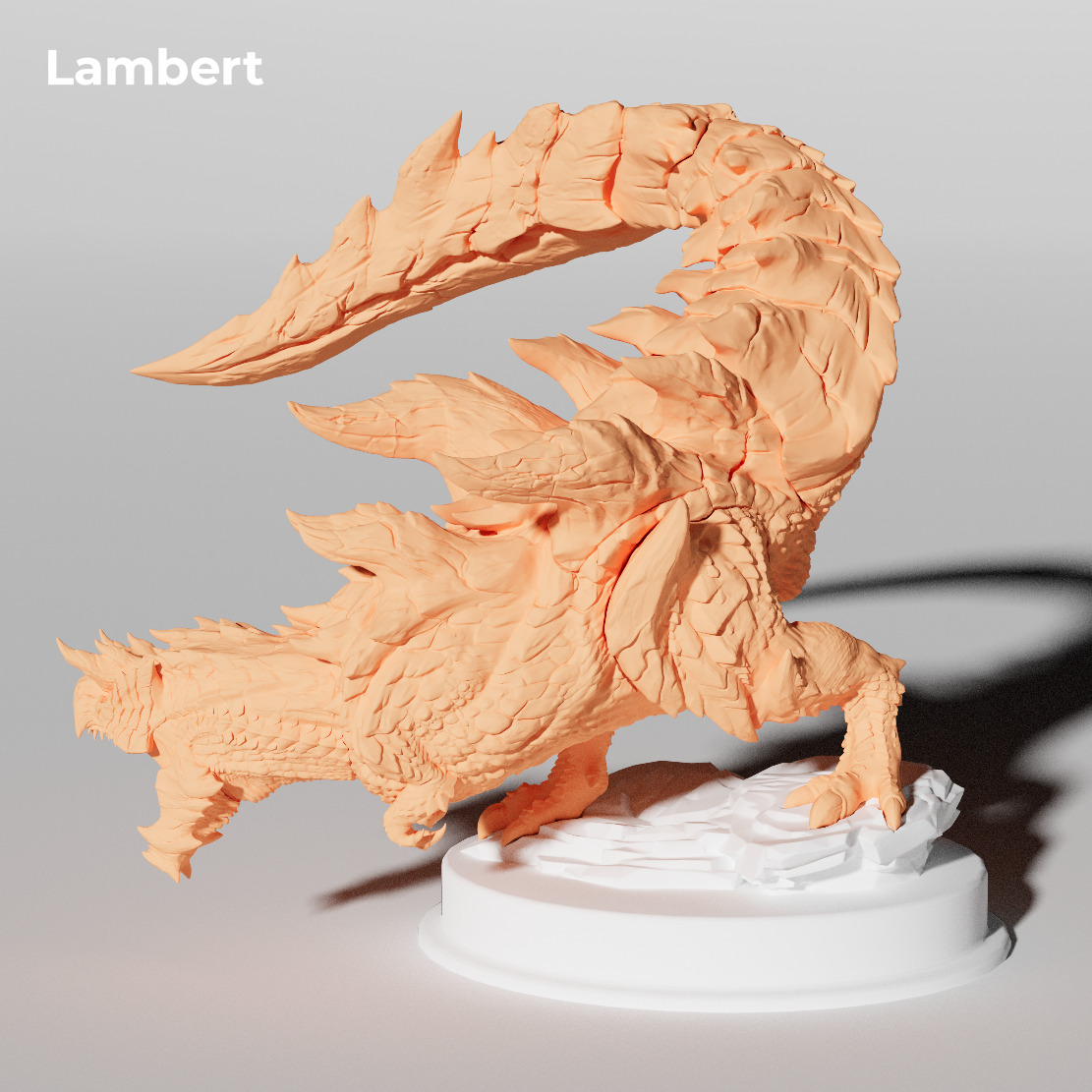}
    \includegraphics[width=0.32\linewidth]{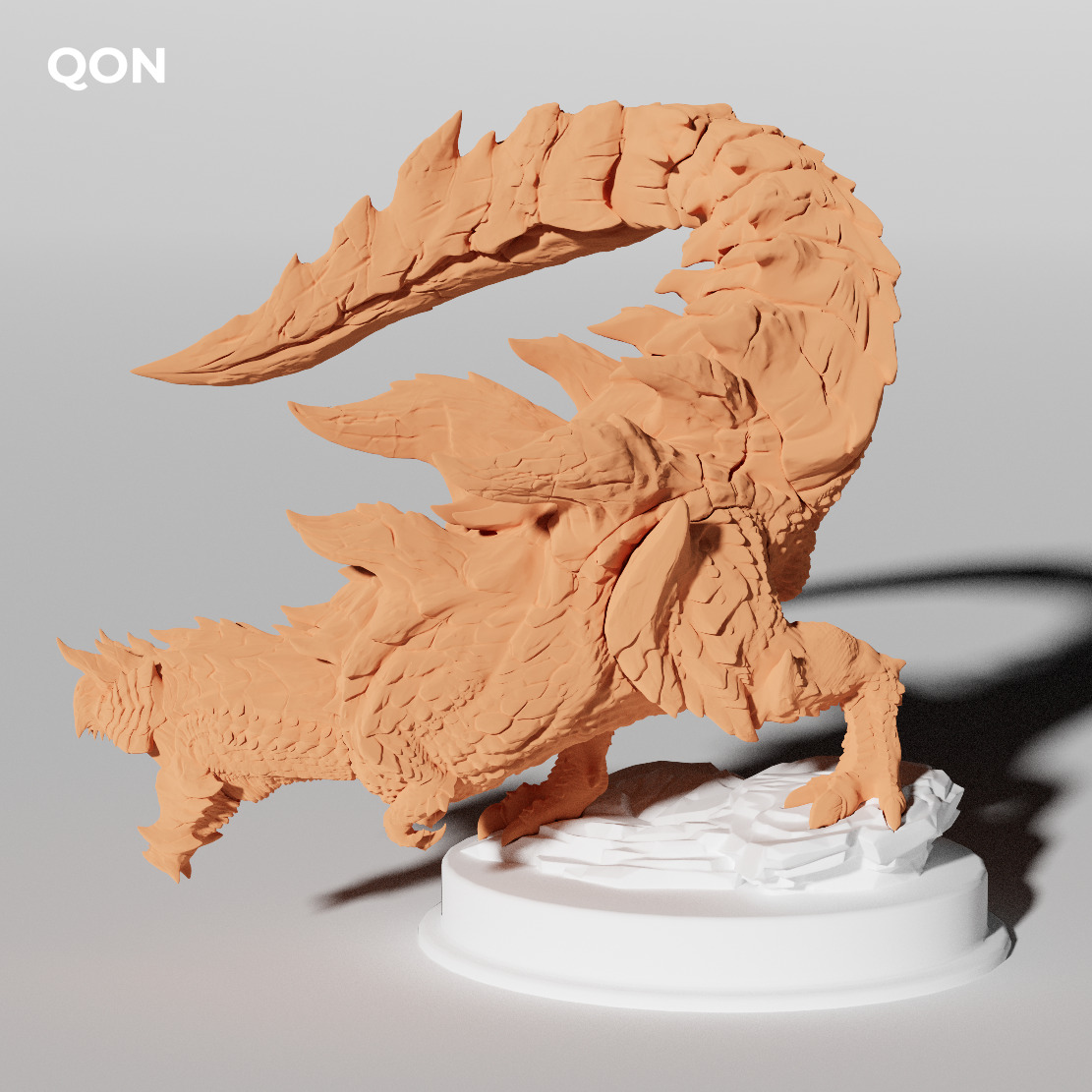}
    \includegraphics[width=0.32\linewidth]{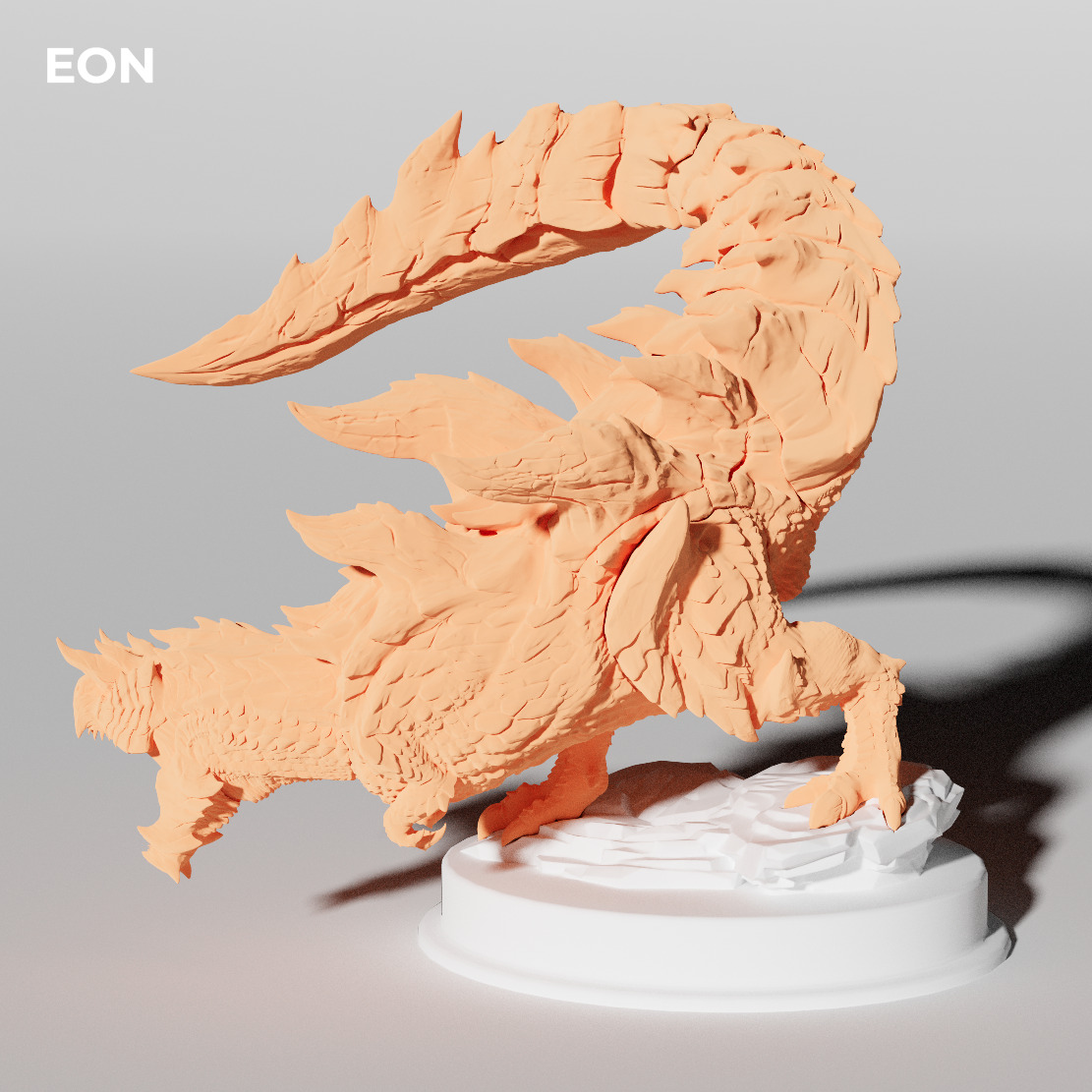} \\
    \vspace{0.07cm}
    \includegraphics[width=0.32\linewidth]{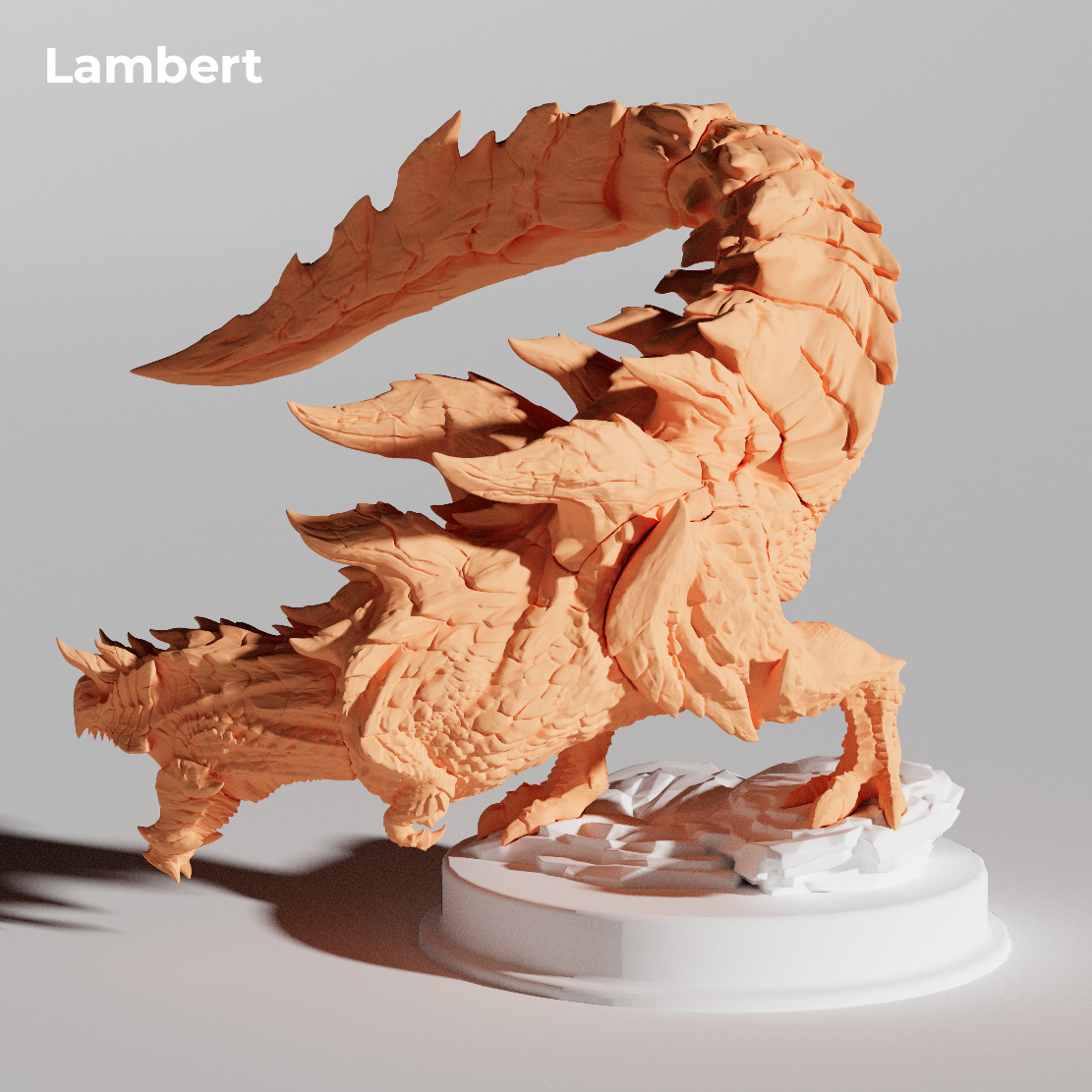}
    \includegraphics[width=0.32\linewidth]{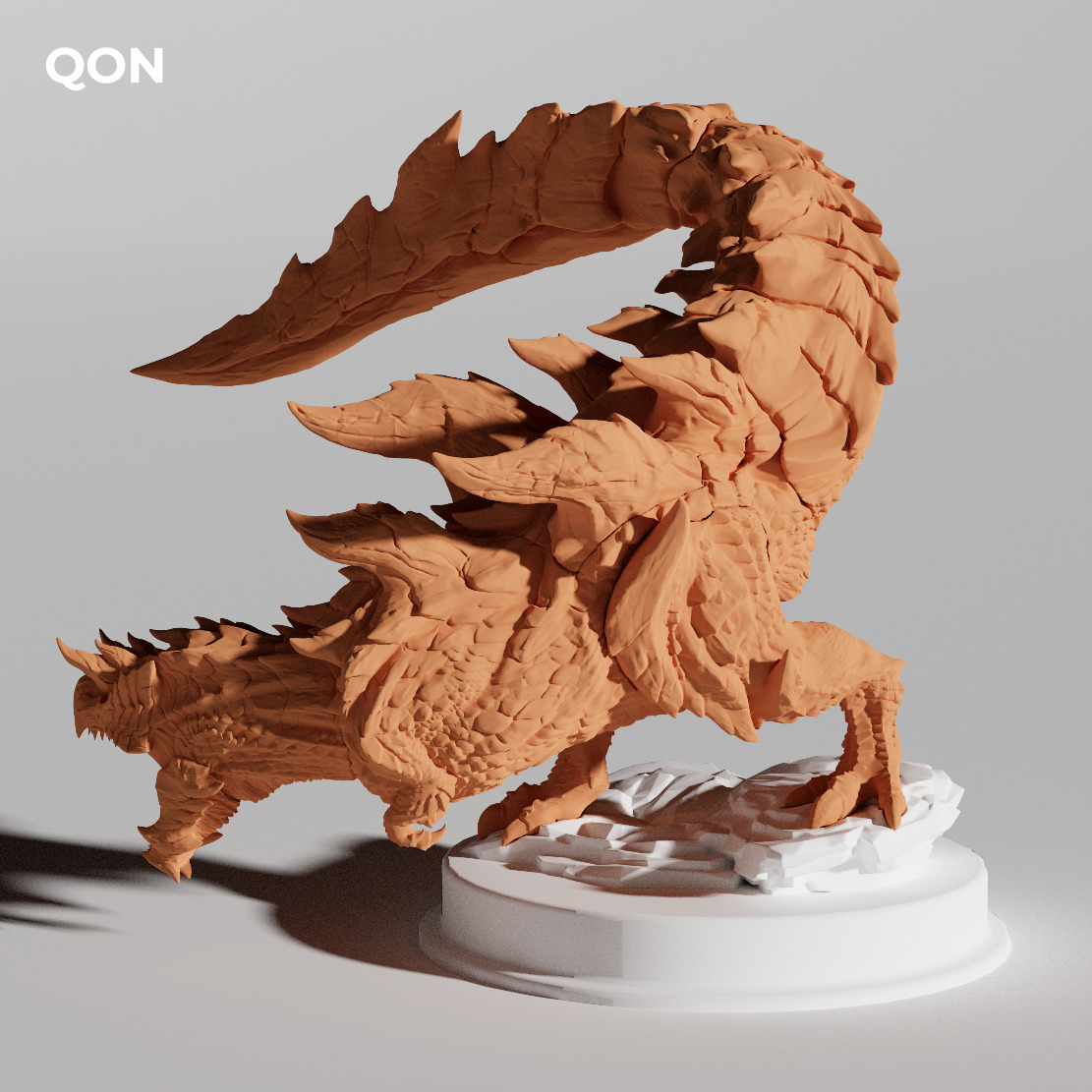}
    \includegraphics[width=0.32\linewidth]{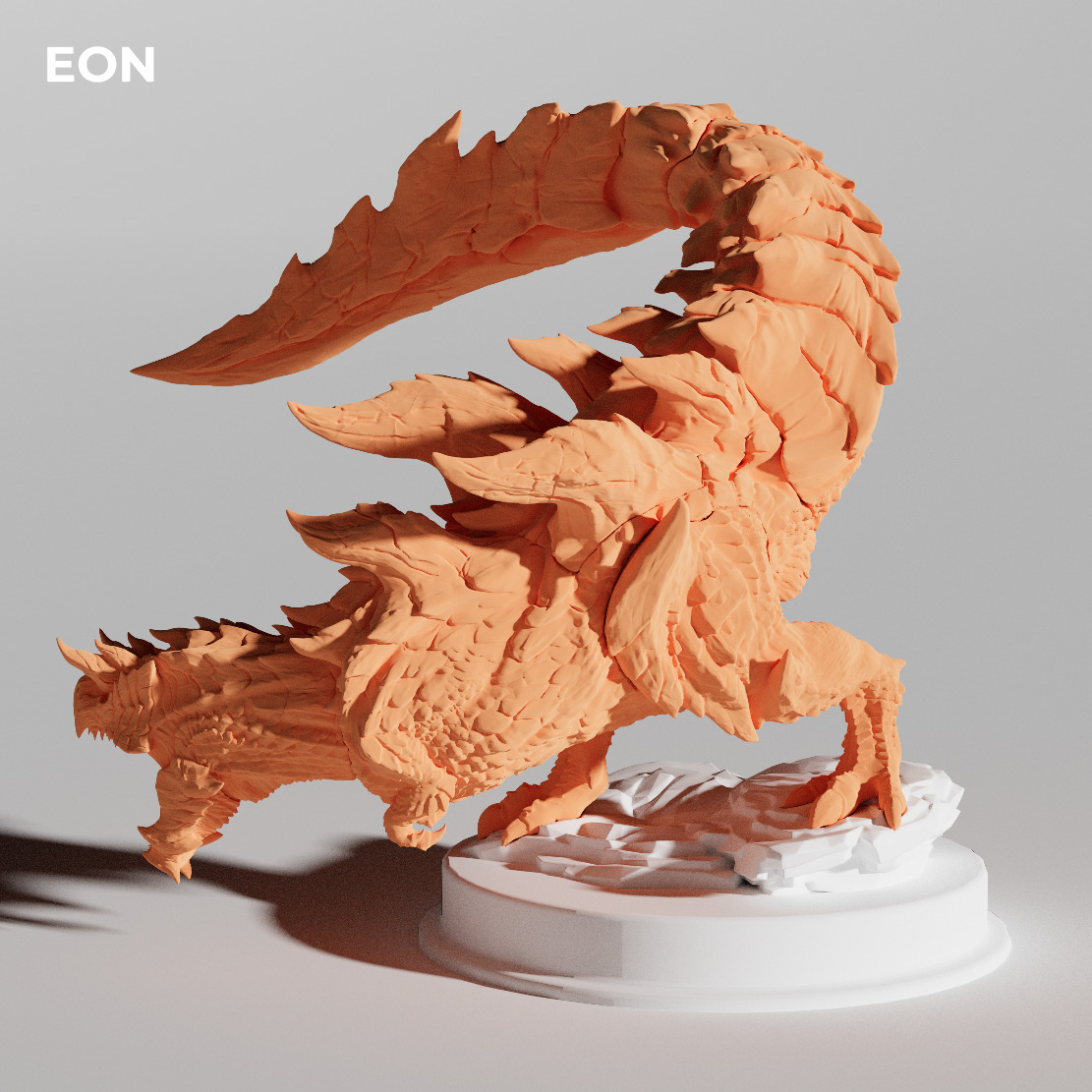}
  \caption{Lambert model (left), the classic ``qualitative'' Oren--Nayar QON model (center), and our new EON model (right), all at maximal roughness and the same single-scattering albedo color. The top and bottom rows show front and side-lit cases respectively. The EON model compensates for the energy loss of the QON model, producing similar brightness to the Lambert model combined with the characteristic ``flattening'' and enhanced backscattering of the Oren--Nayar model. \vspace{-0.2cm}
  \label{fig:teaser}}
}

\maketitle
\thispagestyle{firstpagestyle}

\begin{abstract}
\small
We introduce the \emph{energy-preserving Oren--Nayar} (EON) model for reflection from rough surfaces.
Unlike the popular qualitative Oren--Nayar model (QON) and its variants, our model is energy preserving via analytical energy compensation. We include self-contained GLSL source code for efficient evaluation of the new model and importance sampling based on a novel technique we term \emph{Clipped Linearly Transformed Cosine} (CLTC) sampling.
\end{abstract}


\section{Introduction}
\label{sec:introduction}
\begin{introduction}

\setlength{\headheight}{20pt}

The original rough diffuse surface reflectance model of \citet{OrenNayar94} has been highly influential in computer graphics. They noted that real materials do not necessarily exhibit Lambertian behavior, but instead appear ``flatter'' with less edge darkening and lower contrast.
To explain this, they introduced a V-cavity--based microfacet model (in the geometrical optics limit), parameterized by a scalar roughness parameter $\sigma$ and single-scattering albedo color $\boldsymbol{\rho}$, where Lambertian behavior is recovered in the $\sigma \rightarrow 0$ limit.
This exhibits a peak in backscattering at grazing configurations (i.e., when the incident and outgoing photon directions are both almost parallel to the surface and almost opposite in direction), which aligns with the behavior of real materials.
Figure~\ref{fig:retroreflection} shows examples of the flat and bright appearance of the various Oren--Nayar models compared to the Lambertian model, caused by this backscattering.
\begin{figure}[!b]
  \captionsetup[subfigure]{font=small,labelfont={bf,sf}}
  \centering
  \begin{subfigure}[t]{0.48\linewidth}
    \includegraphics[width=\linewidth]{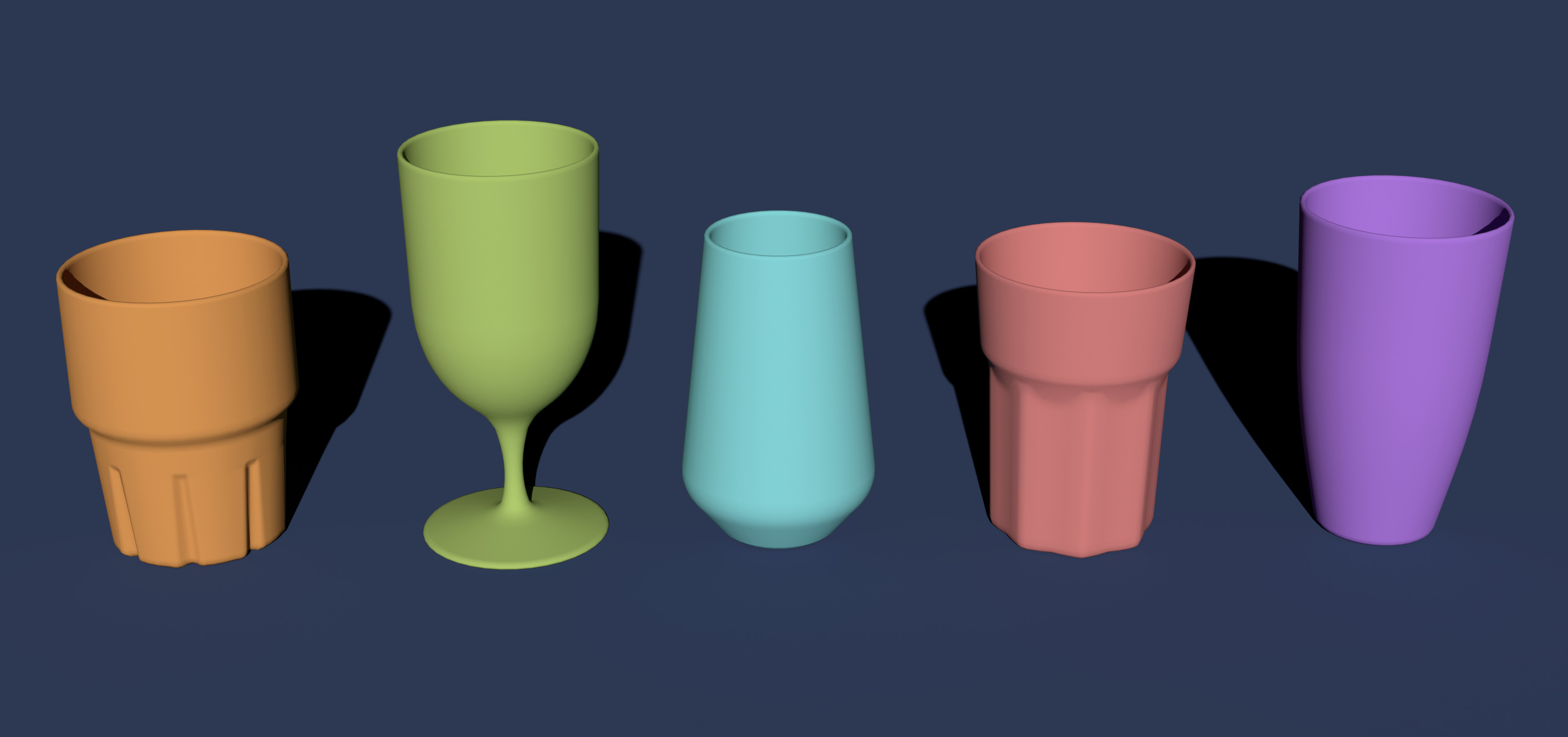}
    \subcaption{Lambert}
  \end{subfigure}%
  \hspace{2mm} 
  \begin{subfigure}[t]{0.48\linewidth}
    \includegraphics[width=\linewidth]{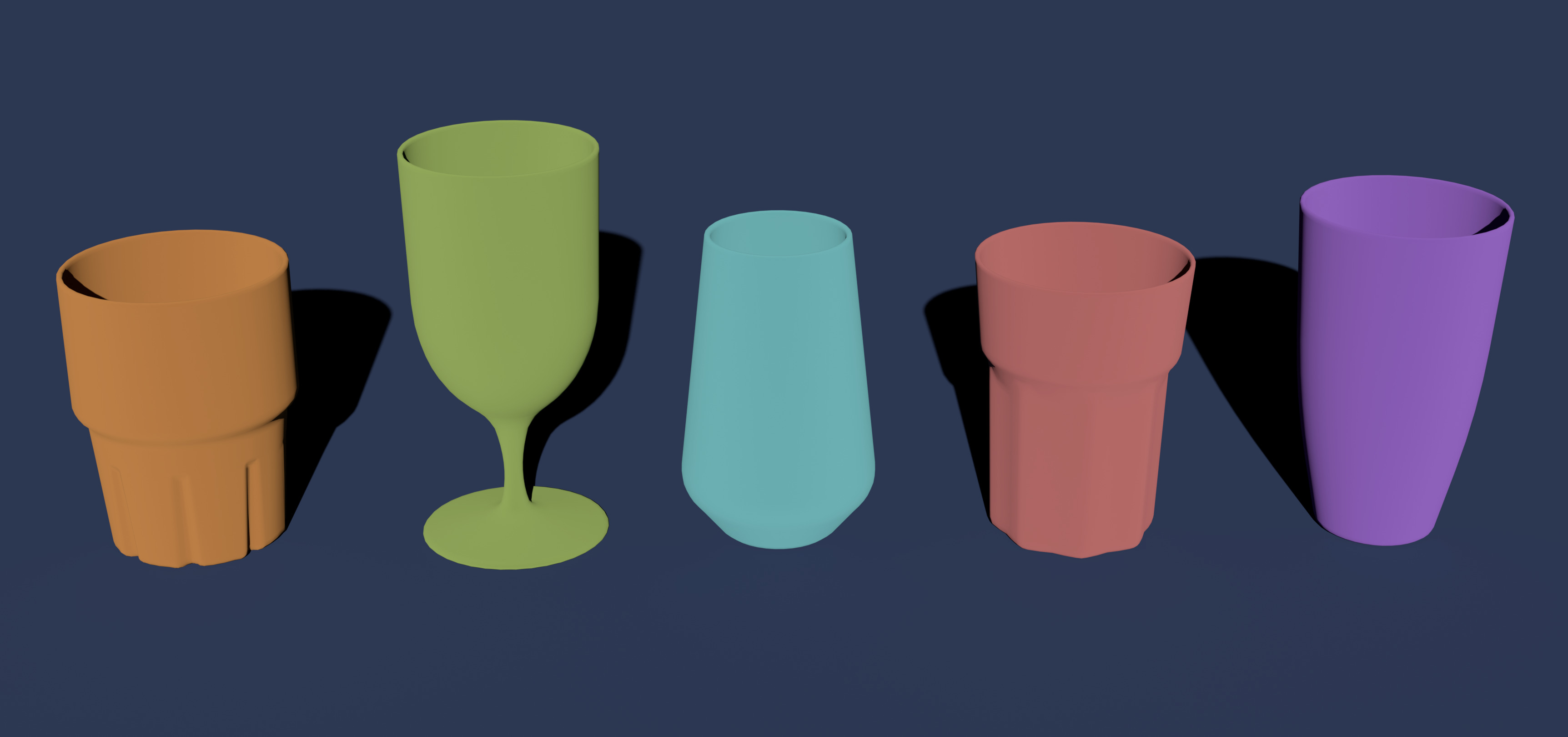}
    \subcaption{fullON}
  \end{subfigure}\\ 
  \par\medskip
  \begin{subfigure}[t]{0.48\linewidth}
    \includegraphics[width=\linewidth]{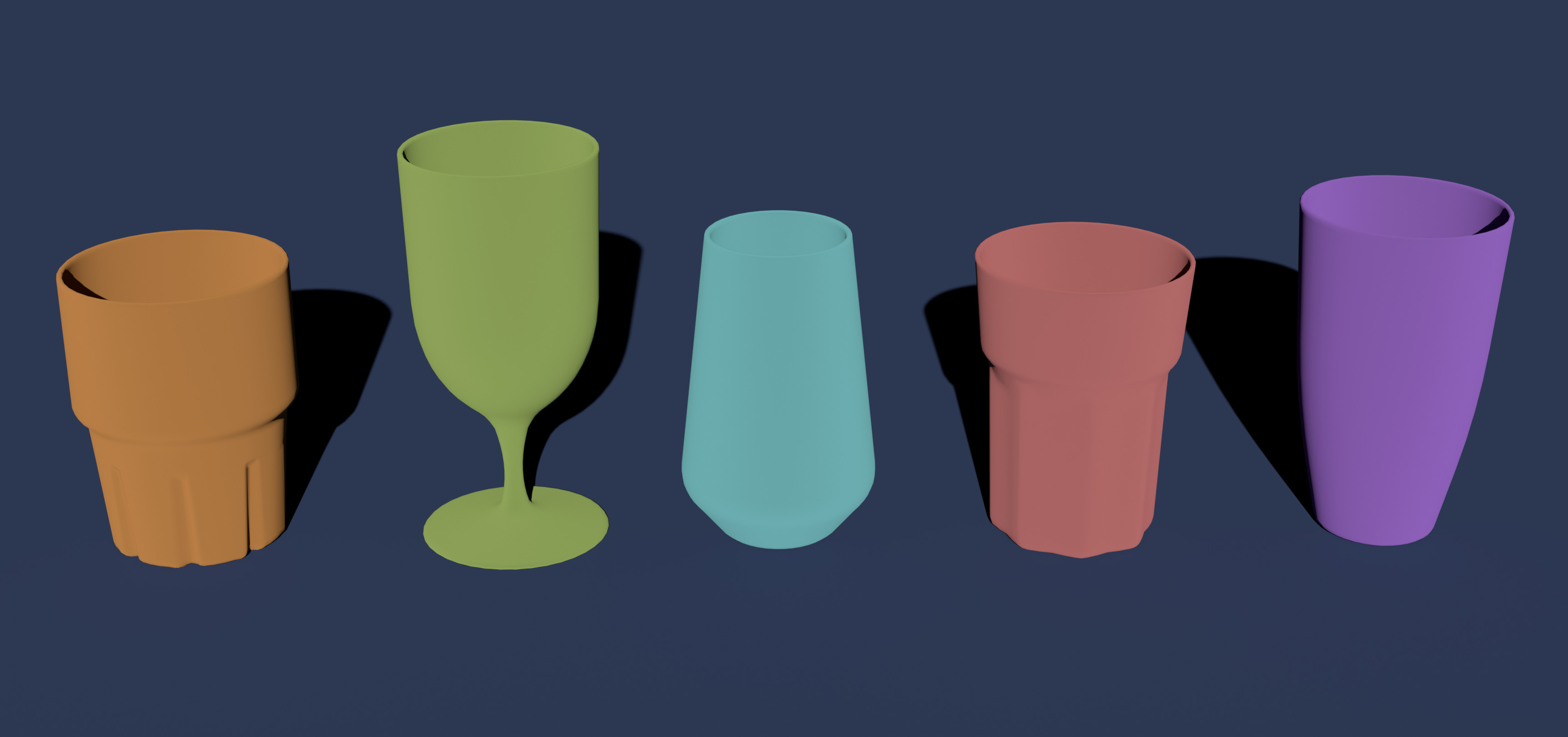}
    \subcaption{QON}
  \end{subfigure}%
  \hspace{2mm} 
  \begin{subfigure}[t]{0.48\linewidth}
    \includegraphics[width=\linewidth]{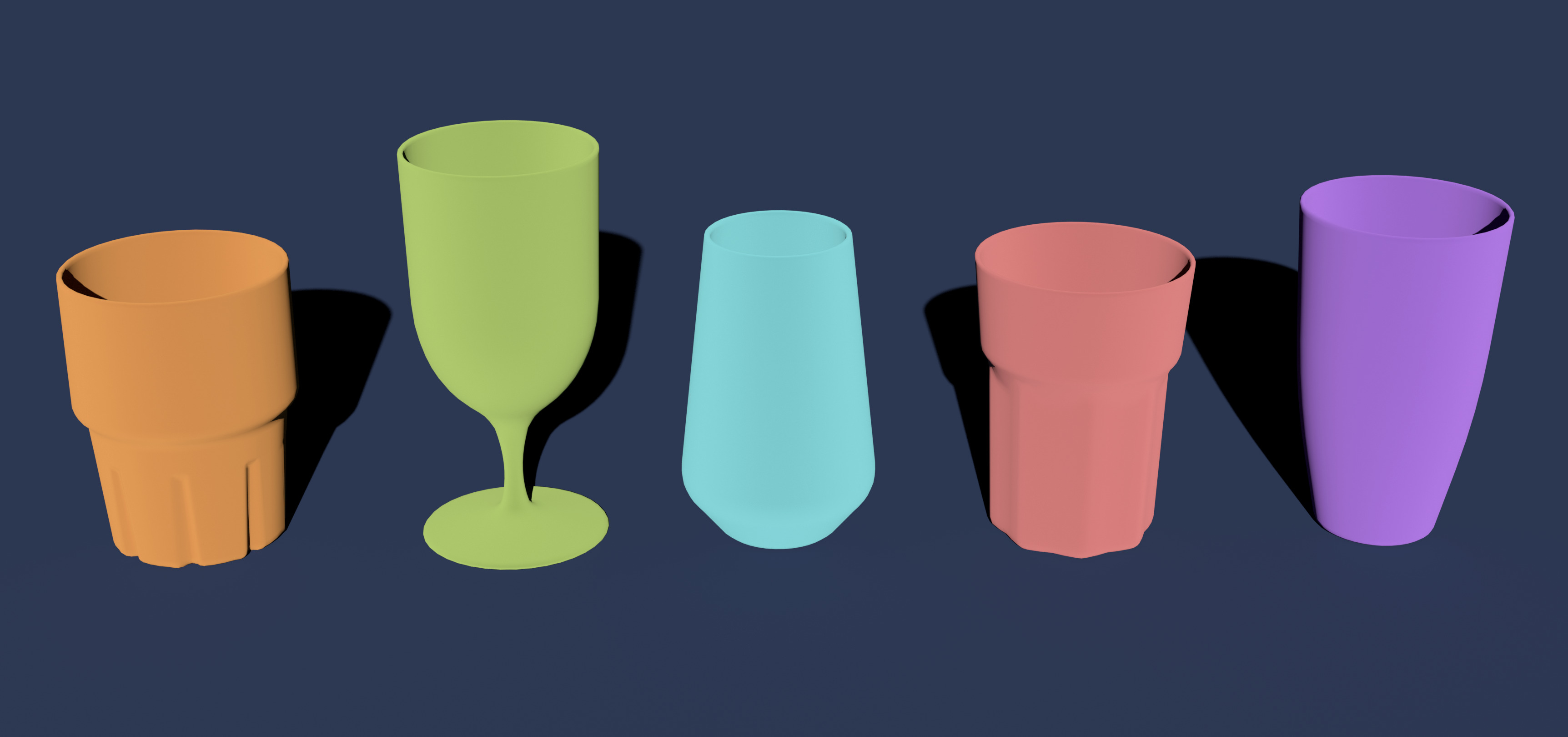}
    \subcaption{EON}
  \end{subfigure}
  \caption{Backscattering from smooth (top left) and maximally rough diffuse BRDFs.}
  \label{fig:retroreflection}
\end{figure}

\citet{OrenNayar94} presented their full model as well as a more approximate ``qualitative'' version. The full version (henceforth the \emph{full Oren--Nayar} model, or fullON) accounts for both single and double scattering (due to interreflections) on the microfacet surface.
Figure~\ref{fig:brdf_plots} shows the fullON BRDF as the blue curve, where the backscattering peak is most prominent in the grazing configuration in the upper right panel.
\begin{figure}[tbh]
  \centering
  \includegraphics[width=0.99\linewidth]{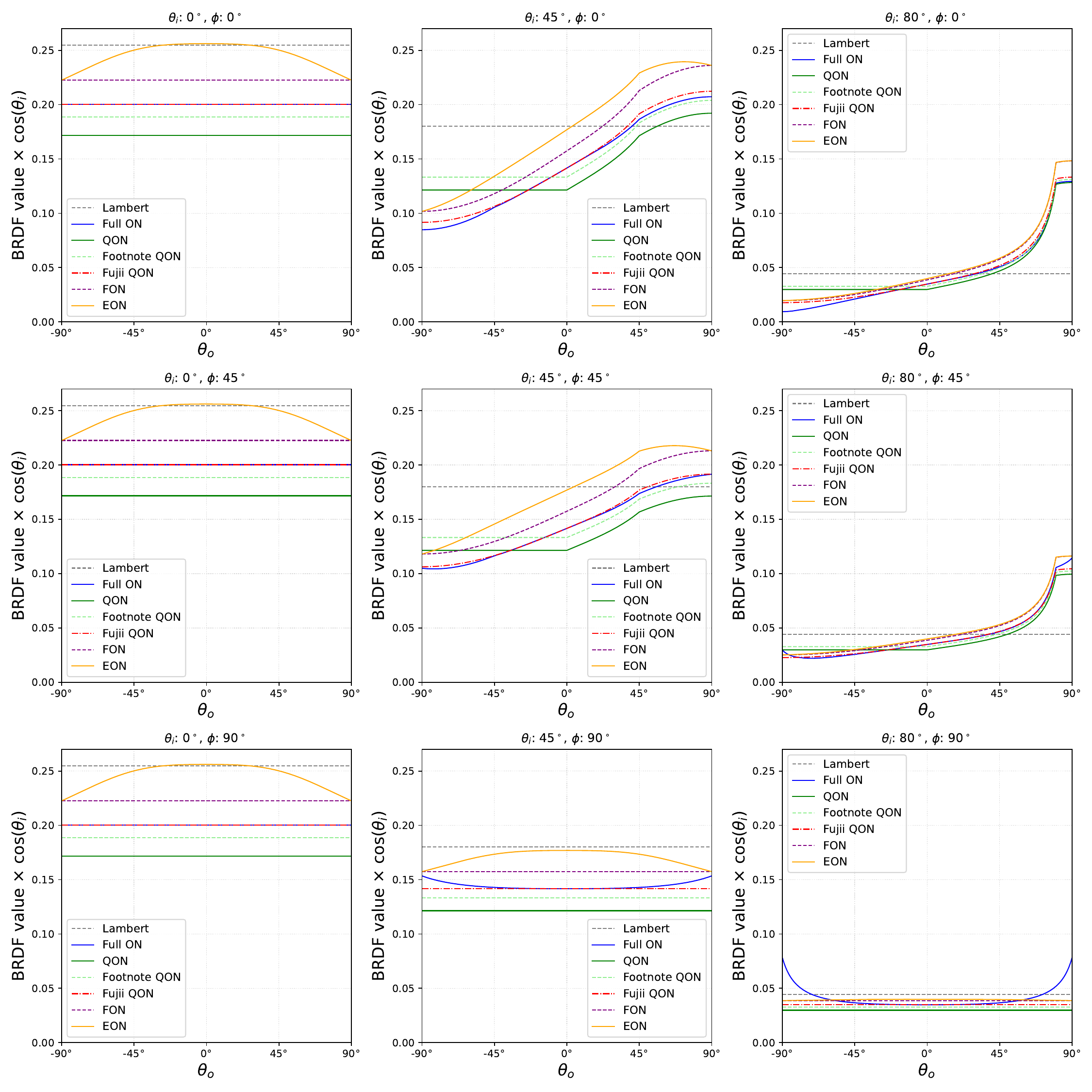}
  \caption{Comparison of the various Oren--Nayar models discussed in Section~\ref{sec:introduction}, plotted for various input and output angles at 80\% albedo and 50\% roughness (corresponding to BRDF input parameters $\sigma=\pi/4$ or $r=1/2$ depending on the model). Note that the BRDF values are multiplied by a foreshortening factor of $\cos\theta_i$, where $\theta_i$ is the zenith angle of the light direction. These plots align with those in the article by \protect\citet{FUJII}, which uses the same conventions as those in the original paper by \protect\citet{OrenNayar94}.}
  \label{fig:brdf_plots}
\end{figure}

The cheaper ``qualitative'' variant of the model---henceforth the \emph{qualitative Oren--Nayar} model or QON for short---ignores the double scattering interreflections. We review the mathematical form of the QON model in Section~\ref{sec:QON}. This QON model has been widely used in the industry as the fullON model is more complex to implement and costly to evaluate (e.g., in MaterialX \cite{Smythe2016}, OSL \cite{Gritz2010}, MDL \cite{Kettner2015}, and Arnold \cite{Georgiev2018}).

Both of the original Oren--Nayar models, fullON and QON, are reciprocal but not energy preserving. Indeed, both models violate energy preservation (i.e., lose energy) and energy conservation (i.e., gain energy) for certain configurations of directions and roughness values. In addition, the QON variant (the dark green curve in Figure~\ref{fig:brdf_plots}) is overly bright when backlit and exhibits a subtle discontinuity artifact in the shading, which appears as a dark ring when rendering a sphere \cite{FUJII}.

A modification to the original models of Oren--Nayar was later proposed by \citet{FUJII}, who noted that the fullON model can be reasonably approximated via simple modification of the coefficients of the QON model (this \emph{Fujii QON} model is the dot-dashed bright red curve in Figure~\ref{fig:brdf_plots}). This model improves upon QON---incorporating a tinted interreflection component, eliminating excessive forward scattering, and eliminating the dark ring artifact---but still suffers from the energy loss issue of the fullON and QON models.
Fujii thus proposed a second, simplified qualitative model that we term the \emph{Fujii Oren--Nayar} model or FON, which we discuss in detail in Section~\ref{sec:FON}. The FON model is the dashed darker red curve in Figure~\ref{fig:brdf_plots}. This model was designed to closely match the shape of the fullON model at lower computational cost while reducing energy loss and fixing the discontinuity artifact. While this model is brighter than fullON in most cases, it is also energy conserving. However, the FON model still does not preserve energy and thus still appears to darken for high roughness values (and thus fails a white-furnace test). \citet{FUJII} also presented a numerical comparison of all these models, which matches our Figure~\ref{fig:brdf_plots}.\footnote
{
  Note, the light green curve \emph{Footnote QON} corresponds to this footnote in \citet{OrenNayar94}:
  \begin{displayquote}\itshape
    Discrepancies caused by the lack of the interreflection component in the qualitative
    model can be partially compensated by replacing the constant 0.33 in coefficient A with~0.57.
  \end{displayquote}
} This plot therefore also serves as a check that our interpretation of the FON model is correct, since the statement of it was somewhat unclear.

In Section~\ref{sec:EON} we present an analytical extension to the FON model that is reciprocal and explicitly energy preserving (thus passing a white-furnace test), which we term \emph{energy-preserving Oren--Nayar}, or EON. EON is the orange curve in Figure~\ref{fig:brdf_plots}. This augments the FON model with a term representing the full multiple-scattering contribution missing from the other models. It is analytical and straightforward to implement, and a high-quality fit can be found to its functional form for improved efficiency. While EON (like FON) is not directly derived from an underlying microphysical theory, it provides a more practical model than any of the preceding Oren--Nayar forms while preserving the qualitative appearance, which is both energy conserving \emph{and} energy preserving (and thus brighter and more visually appealing), while being as efficient and easy to implement as the existing standard QON model. In Section~\ref{sec:EON_sampling} we provide an analytical importance sampling scheme for EON that takes into account the backscattering peak (using a technique we introduce termed \emph{Clipped Linearly Transformed Cosine} sampling, or CLTC), significantly improving sample variance compared to cosine-weighted hemispherical sampling. Figure~\ref{fig:brdf_renders} contains renders of all of the BRDFs that are featured in Figure~\ref{fig:brdf_plots} for comparison.

\begin{figure}[htb]
  \captionsetup[subfigure]{font=small,labelfont={bf,sf}}
  \centering
  \begin{subfigure}[t]{0.25\linewidth}
    \includegraphics[width=\linewidth]{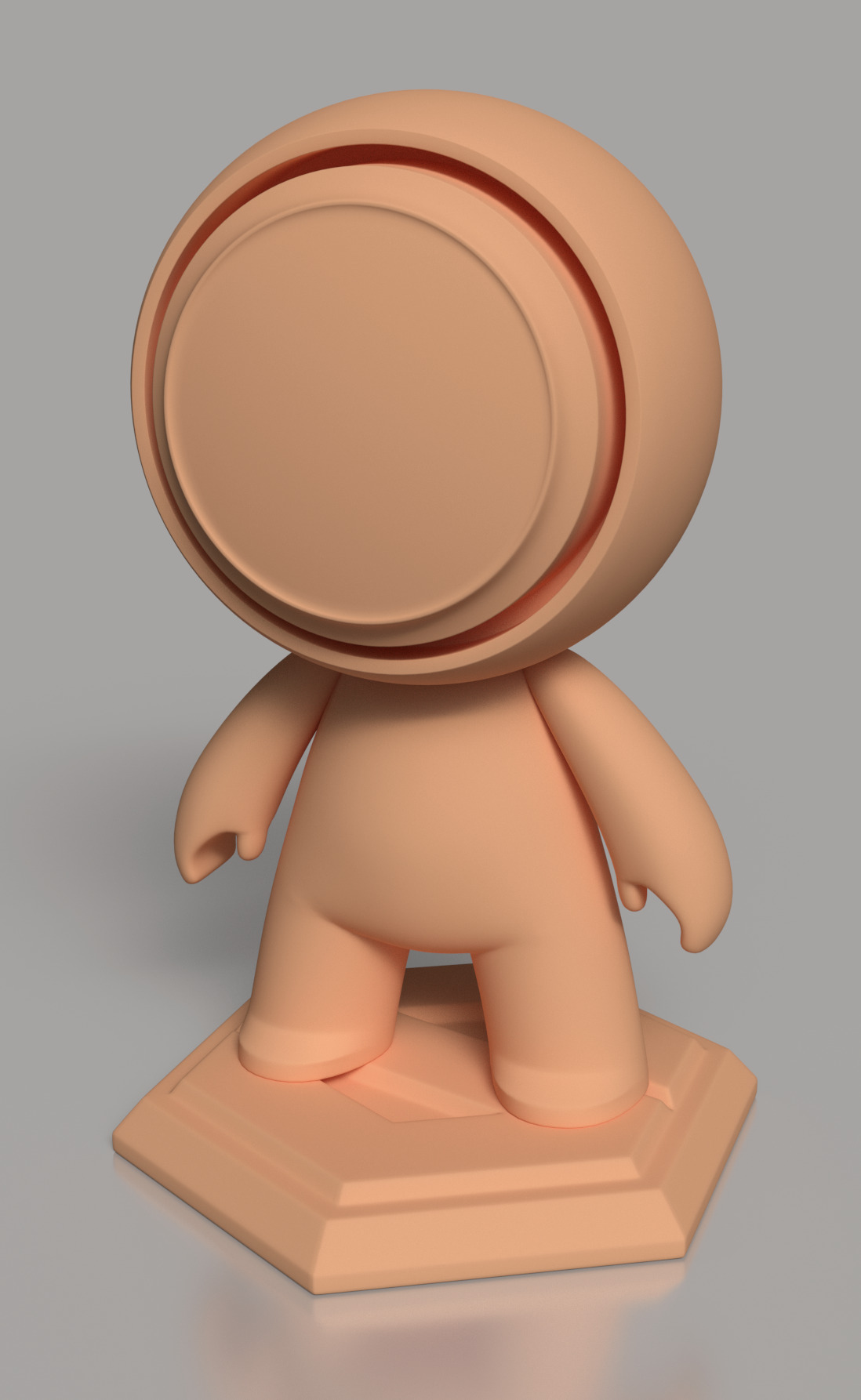}
    \subcaption{Lambert}
  \end{subfigure}%
  \begin{subfigure}[t]{0.25\linewidth}
    \includegraphics[width=\linewidth]{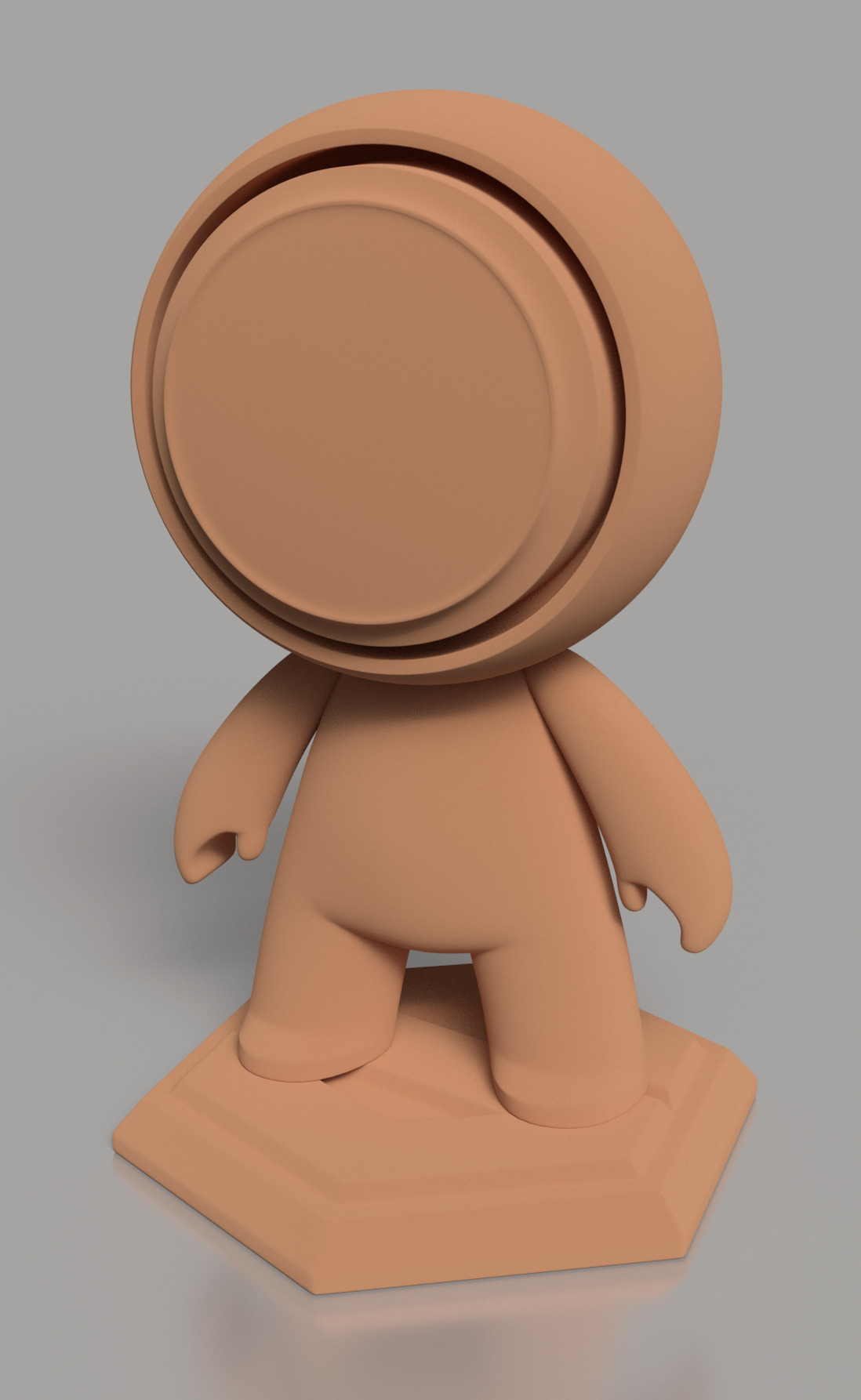}
    \subcaption{fullON}
  \end{subfigure}%
  \begin{subfigure}[t]{0.25\linewidth}
    \includegraphics[width=\linewidth]{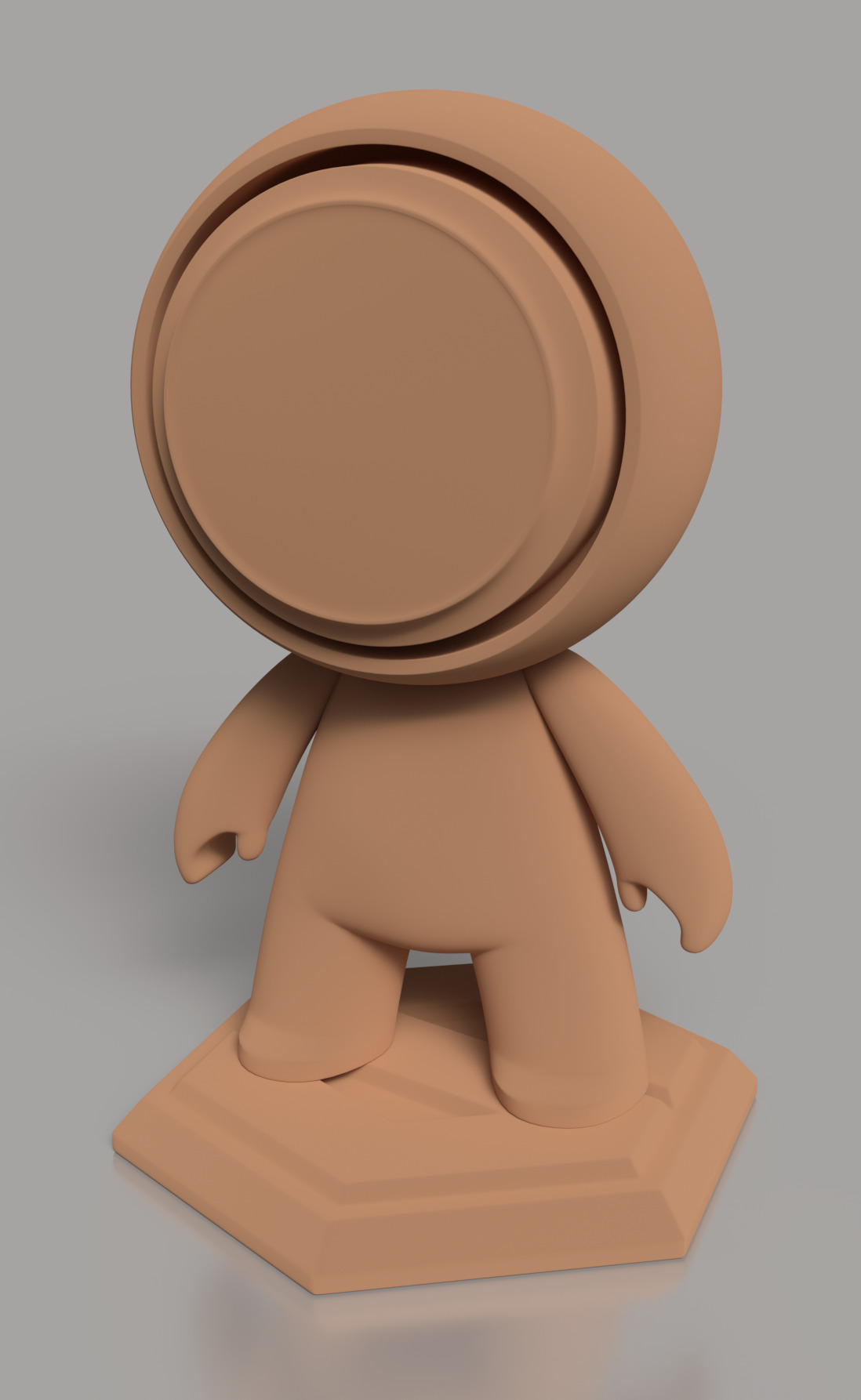}
    \subcaption{QON}
  \end{subfigure}%
  \begin{subfigure}[t]{0.25\linewidth}
    \includegraphics[width=\linewidth]{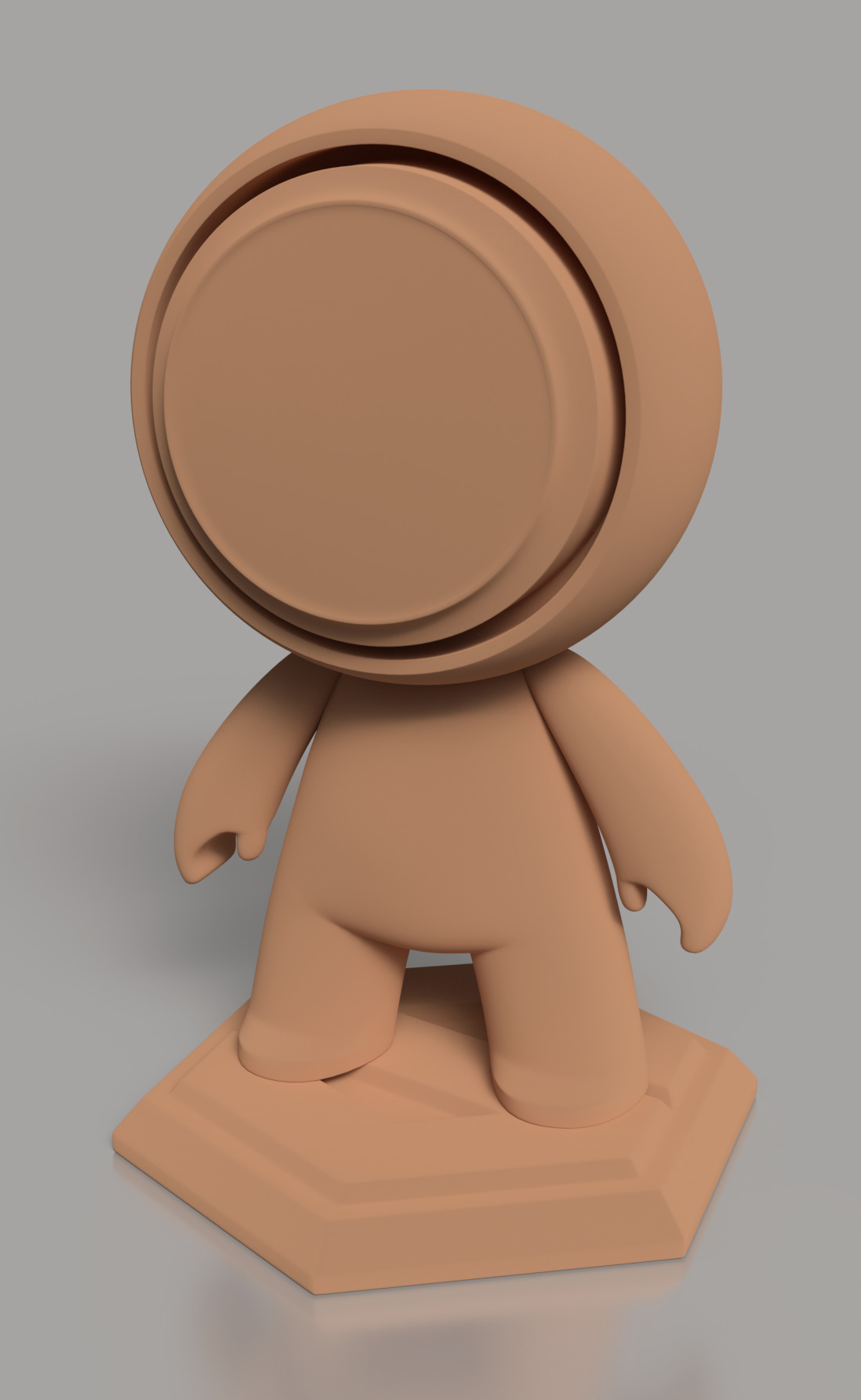}
    \subcaption{Footnote QON}
  \end{subfigure}\\ 
  \par\bigskip
  \begin{subfigure}[t]{0.25\linewidth}
    \includegraphics[width=\linewidth]{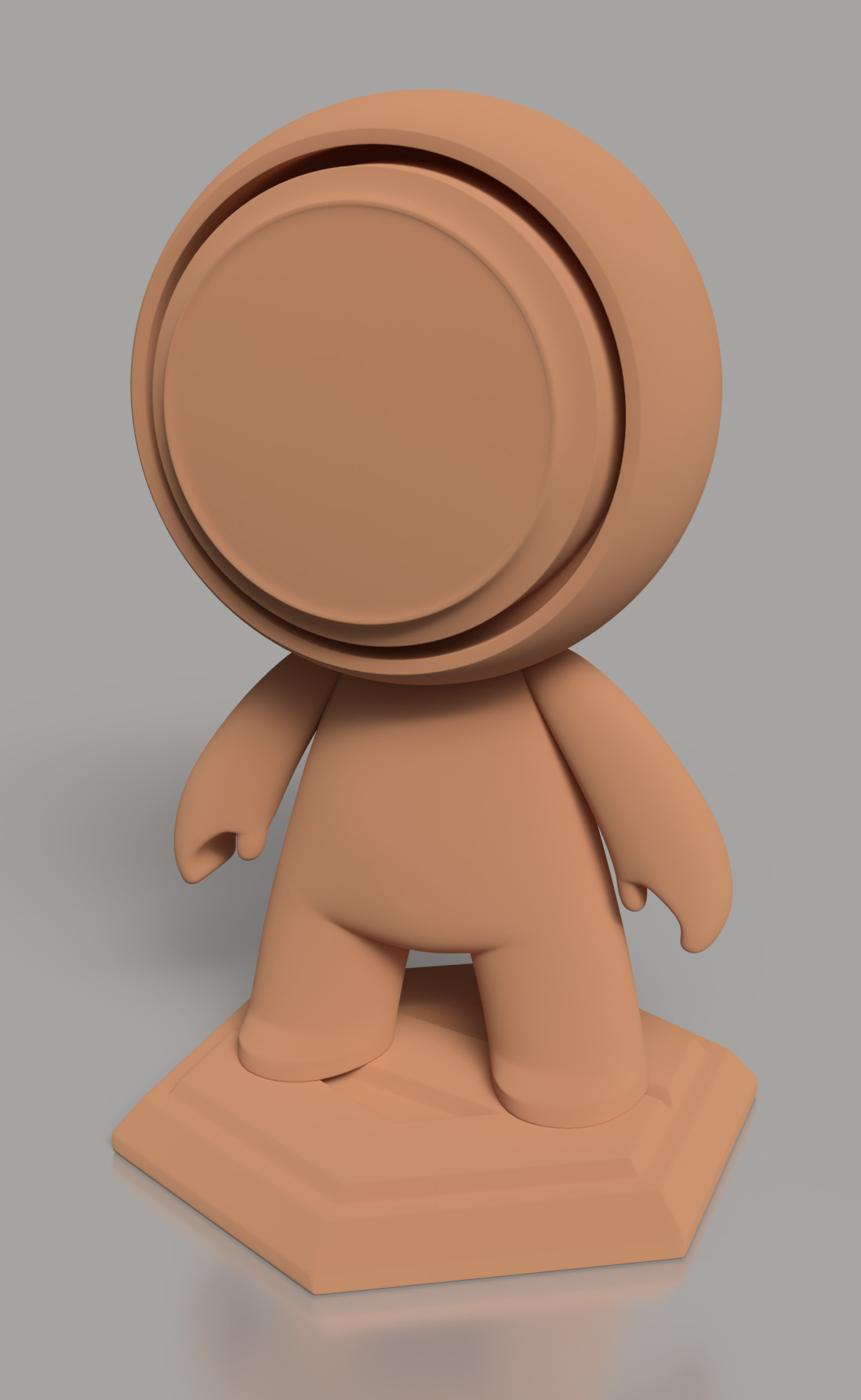}
    \subcaption{Fujii QON}
  \end{subfigure}%
  \begin{subfigure}[t]{0.25\linewidth}
    \includegraphics[width=\linewidth]{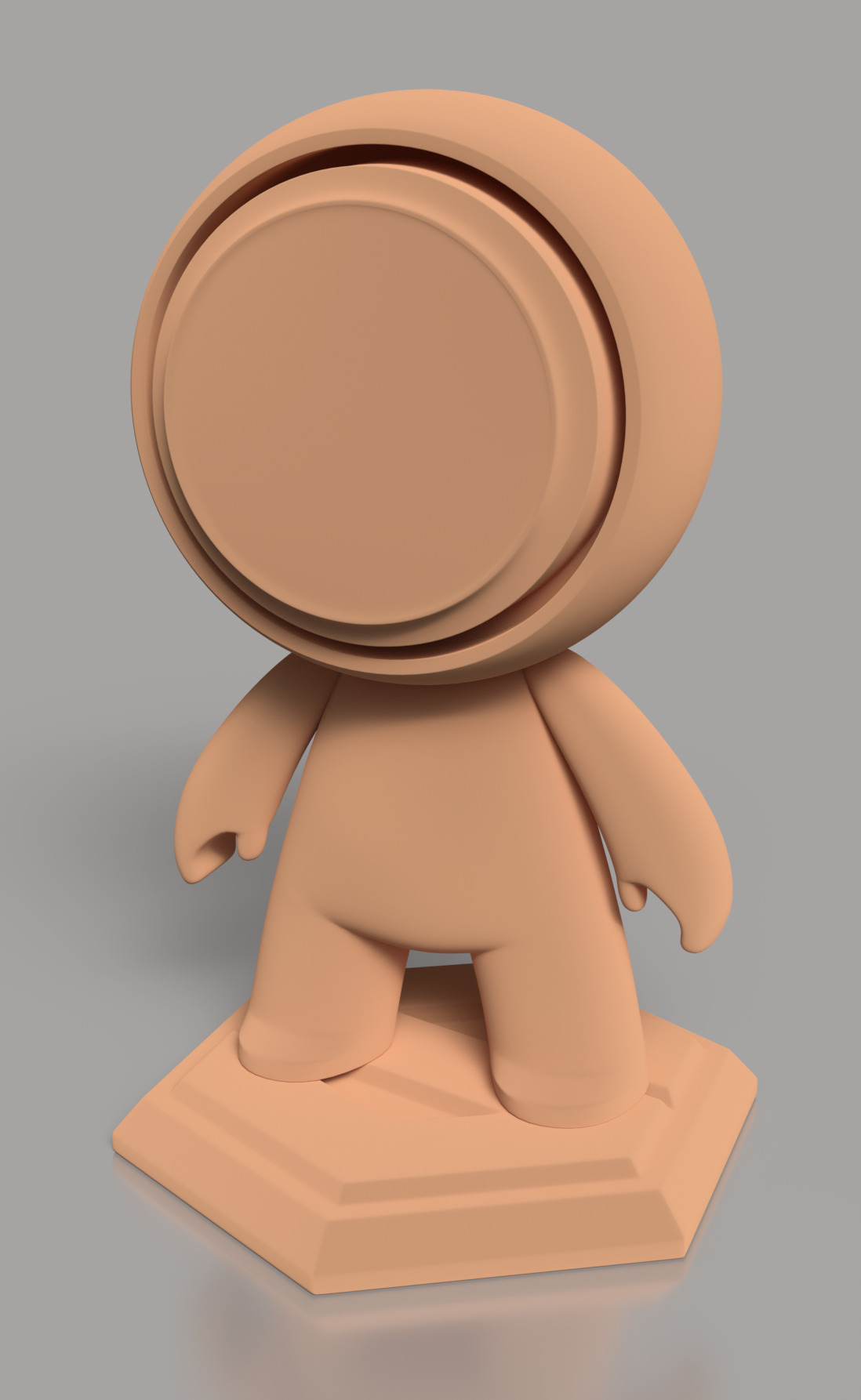}
    \subcaption{FON}
  \end{subfigure}%
  \begin{subfigure}[t]{0.25\linewidth}
    \includegraphics[width=\linewidth]{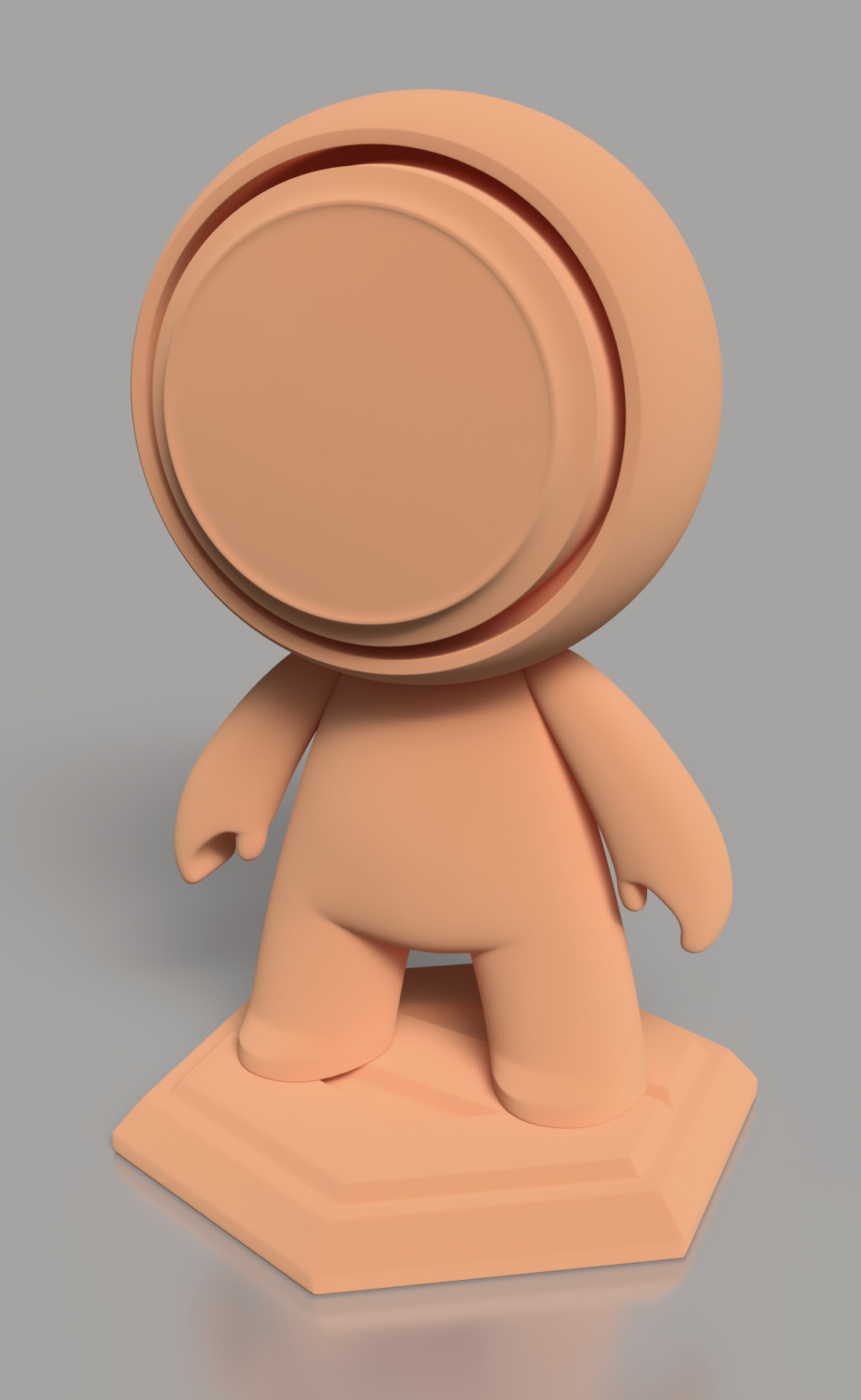}
    \subcaption{EON (exact)}
  \end{subfigure}%
  \begin{subfigure}[t]{0.25\linewidth}
    \includegraphics[width=\linewidth]{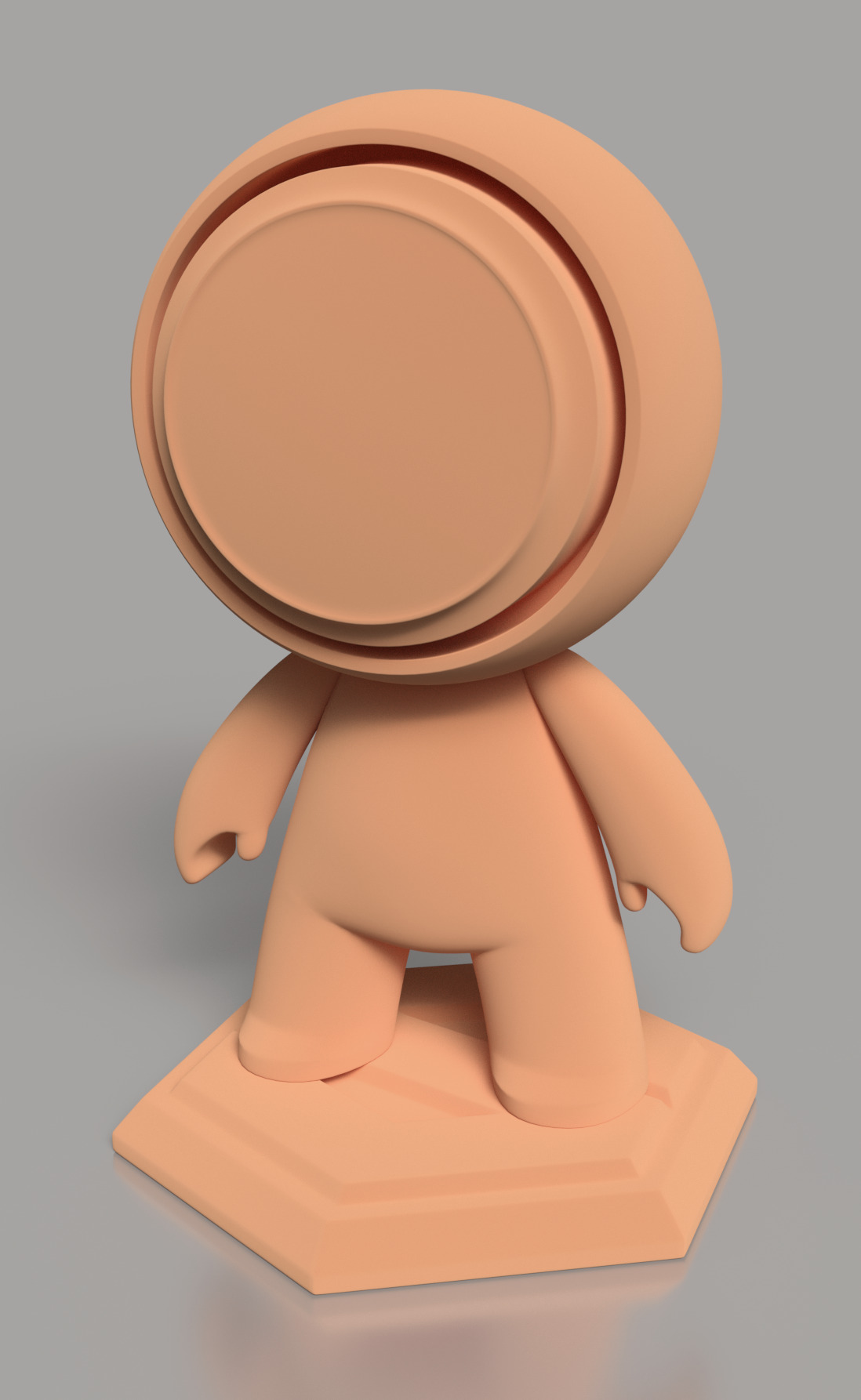}
    \subcaption{EON (approx)}
  \end{subfigure}
  \caption{Visual comparison of the Lambert model to the various Oren--Nayar models.}
  \label{fig:brdf_renders}
\end{figure}

\end{introduction}

\section{Existing Oren--Nayar Models}
\begin{existing-models}

\subsection{Qualitative Oren--Nayar (QON) \label{sec:QON}}

\citet{OrenNayar94} provided a simplified ``qualitative'' version of their model (abbreviated QON), which is cheaper to evaluate and thus more suitable for real-time use cases. We briefly review here the form and properties of this model, and derive a novel analytical formula for its albedo.

The QON model takes the form (for incident and outgoing rays $\omega_i$ and $\omega_o$, respectively)
\begin{equation}
  \mathbf{f}_\mathrm{q}(\omega_i, \omega_o) = \frac{\boldsymbol{\rho}}{\pi} \bigl( A_\mathrm{q} + B_\mathrm{q}\,g_\mathrm{q}(\omega_i, \omega_o) \bigr),
\end{equation}
where $\boldsymbol{\rho}$ is an RGB color parameter\footnote{Note that RGB color parameters such as $\boldsymbol{\rho}$ are denoted in bold throughout.} controlling the overall albedo and the coefficients $A_\mathrm{q}, B_\mathrm{q}$ are given in terms of a roughness parameter $\sigma$:
\begin{align*}
  A_\mathrm{q} &= 1 - 0.5 \frac{\sigma^2}{\sigma^2 + 0.33}  , \nonumber \\
  B_\mathrm{q} &= 0.45 \frac{\sigma^2}{\sigma^2 + 0.09}  .
\end{align*}
Thus in the limit $\sigma\rightarrow 0$, $\mathbf{f}_\mathrm{q} \rightarrow \boldsymbol{\rho}/\pi$, which is the Lambert BRDF. In terms of the polar angles $\theta_i$, $\theta_o$ of the incident and outgoing rays to the normal and the azimuthal angles $\phi_i$, $\phi_o$, the term $g_\mathrm{q}(\omega_i, \omega_o)$ is
\begin{equation}
  g_\mathrm{q}(\omega_i, \omega_o) = \cos^+(\phi_i-\phi_o) \sin{\alpha} \tan{\beta},
\end{equation}
where\footnote{A notation borrowed from \citet{RTR4}.} $x^+ \coloneqq \mathrm{max}(x, 0)$, $\alpha \coloneqq \max(\theta_i, \theta_o)$, and $\beta \coloneqq \min(\theta_i, \theta_o)$.
Note that the QON BRDF $\mathbf{f}_\mathrm{q}$ is symmetrical under the interchange $\omega_i \leftrightarrow \omega_o$, and thus reciprocal. It can equivalently be written more compactly in vector form with $g_\mathrm{q} = s/t_\mathrm{q}$:
\begin{equation} \label{QON_vector_form}
  \mathbf{f}_\mathrm{q}(\omega_i, \omega_o) = \frac{\boldsymbol{\rho}}{\pi} \left( A_\mathrm{q} + B_\mathrm{q} \frac{s}{t_\mathrm{q}} \right)
\end{equation}
with the term $s$ given by (with normal $N$)
\begin{eqnarray*}
  s = \omega_i \cdot \omega_o - (N \cdot \omega_i) (N \cdot \omega_o)
    = \cos(\phi_i-\phi_o) \,\sin\theta_i \,\sin\theta_o
\end{eqnarray*}
and term $1/t_\mathrm{q}$ given by
\begin{equation} \label{t_q}
  \frac{1}{t_\mathrm{q}} =
    \begin{cases}
      0                                                       & \text{if $s \le 0$}\\
      1 / \max\left(N \cdot \omega_i, N \cdot \omega_o\right) & \text{if $s > 0$.}
    \end{cases}
\end{equation}
According to \citet{OrenNayar94}, the $\sigma$ parameter is technically an angle giving the standard deviation of the distribution of microfacet angles to the horizontal. We adopt the range $\sigma \in [0, \pi/2]$ as suggested by \citet{FUJII}.

\begin{figure}[tb]
  \centering
    \includegraphics[width=0.48\linewidth]{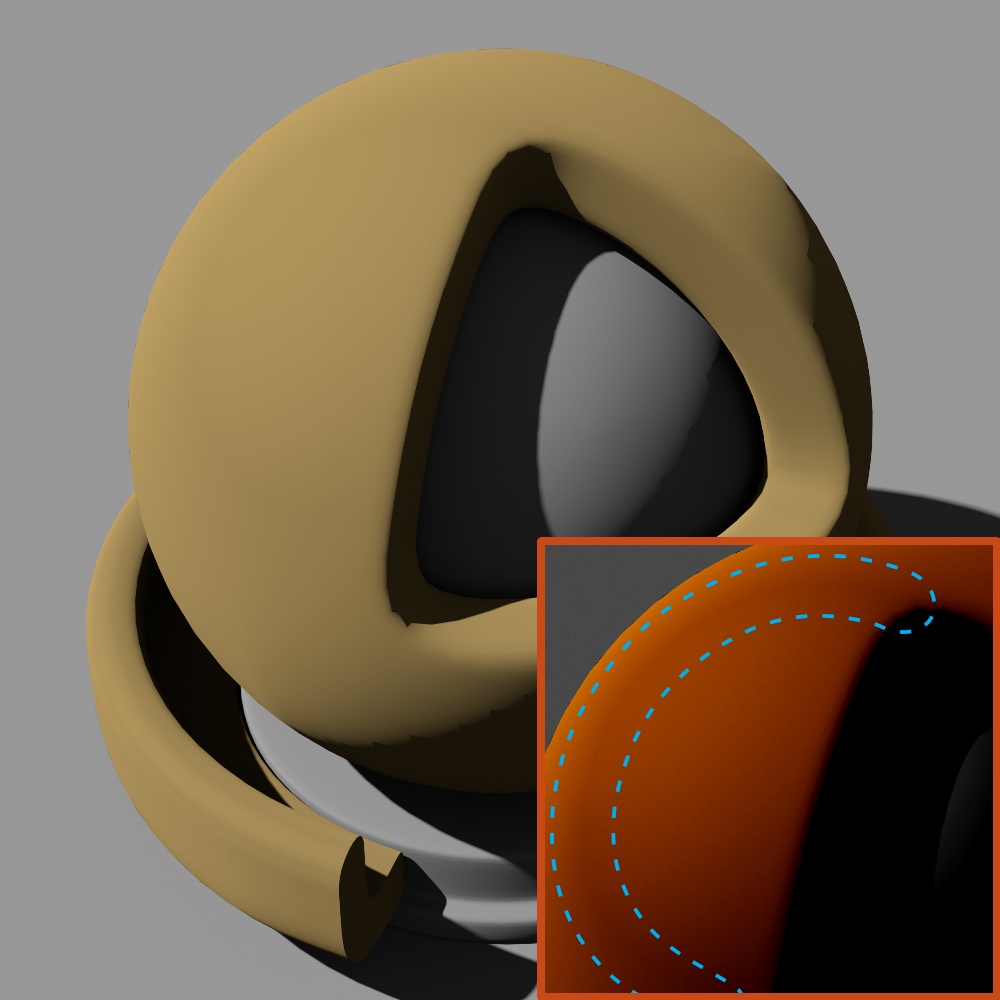}
    \includegraphics[width=0.48\linewidth]{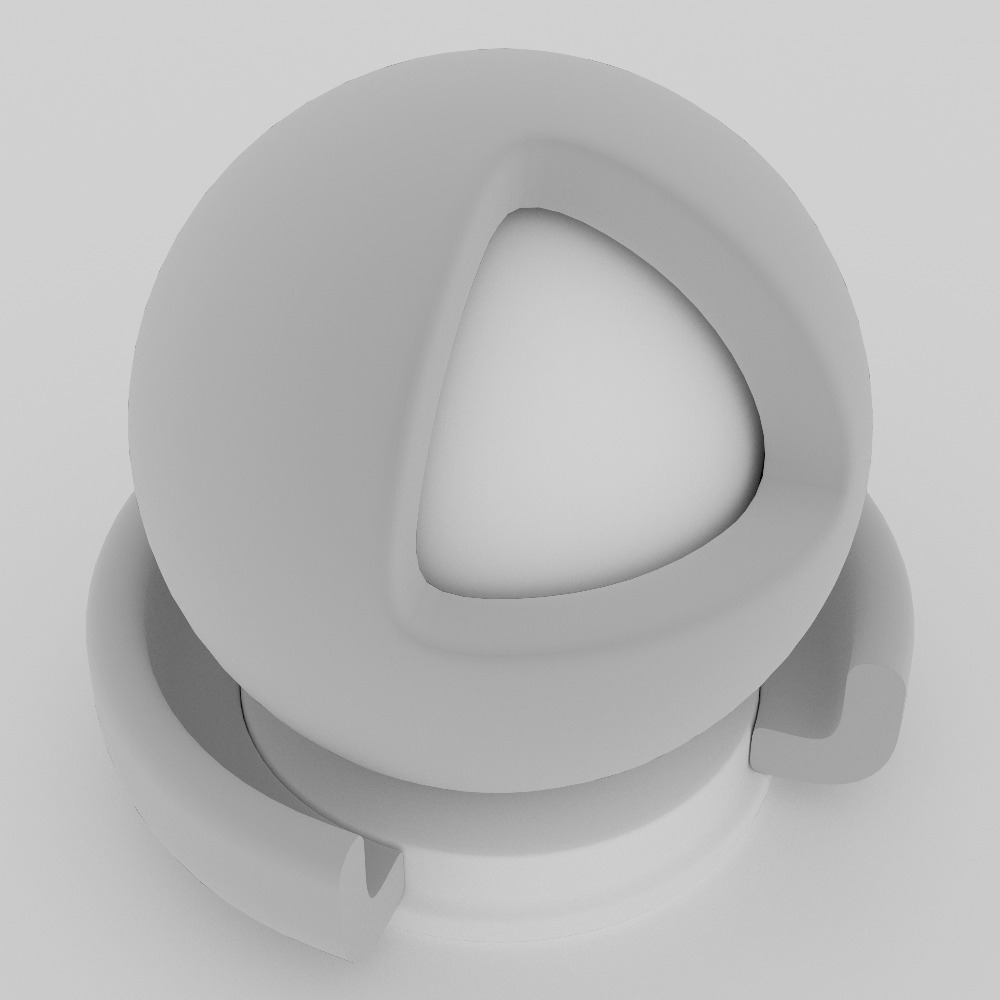}
  \caption{QON model single-bounce render and white-furnace test at $\sigma=\pi/2$. The dotted blue line in the enhanced contrast inset indicates where the subtle ``dark ring'' artifact occurs.}
  \label{fig:QON_renders}
\end{figure}
The left panel of Figure~\ref{fig:QON_renders} shows a render of an object (the brown shader ball) with the QON BRDF at $\sigma=\pi/2$. The right panel shows a QON white-furnace test (i.e., uniform illumination of the $\boldsymbol{\rho}=1$, $\sigma=\pi/2$ QON material, at 50 bounces) where, with an energy-preserving BRDF, the object would blend into the background and disappear at infinite bounce count. Clearly the QON model loses a significant amount of energy. The left panel of Figure~\ref{fig:QON_renders} shows the subtle ``dark ring'' artifact near the top left of the brown shader ball, which was pointed out by \citet{FUJII}. This is due to the lack of continuity of the BRDF as it passes through $s=0$.

We can verify this (numerically and analytically) by computing the directional albedo of the QON model $\mathbf{E}_\mathrm{q}(\omega_o)$, which is the energy reflected into direction $\omega_o$ under uniform illumination. This is given by the following integral of the BRDF w.r.t.\ to projected solid angle element $\mathrm{d}\omega^\perp_i$, over the positive hemisphere $\mathcal{H_+}$ of $\omega_i$:
\begin{equation} \label{QON_albedo}
  \mathbf{E}_\mathrm{q}(\omega_o) = \int_\mathcal{H_+} \mathbf{f}_\mathrm{q}(\omega_i, \omega_o) \,\mathrm{d}\omega^\perp_i  = \boldsymbol{\rho} \left( A_\mathrm{q} + \frac{B_\mathrm{q}}{\pi}G_\mathrm{q}(\omega_o)\right),
\end{equation}
where
\begin{equation}
  G_\mathrm{q}(\omega_o) = \int_\mathcal{H_+} g_\mathrm{q}(\omega_i, \omega_o) \, \mathrm{d}\omega^\perp_i  .
\end{equation}
The integral for $G_\mathrm{q}(\omega_o)$ factors into separate integrals over $\theta_i$ and $\phi_i$:
\begin{equation}
\int_0^{\pi/2} \sin\theta_i \cos\theta_i \,
                               \sin\alpha \tan\beta \; \mathrm{d}\theta_i
\int_0^{2\pi} \cos^+(\phi_i-\phi_o) \; \mathrm{d}\phi_i   .
\end{equation}
The $\theta_i$ integral breaks into two terms according to whether $\theta_i < \theta_o$ or $\theta_i \ge \theta_o$ (due to the definitions of $\alpha$ and $\beta$), which are easily evaluated. The $\phi_i$ integral reduces to $\int^{\pi/2}_{-\pi/2} \cos\phi_i \,\mathrm{d}\phi_i = 2$. The resulting formula for $G_\mathrm{q}$ is
\begin{equation} \label{QON_albedo_G}
  G_\mathrm{q}(\theta_o) = \sin\theta_o \left( \theta_o - \sin\theta_o \cos\theta_o \right) + \frac{2}{3}\tan\theta_o \left(1 - \sin^3\theta_o \right) .
\end{equation}
Note that this is undefined at exactly grazing angle $\theta_o=\pi/2$, but the limit is well-defined: $G_\mathrm{q}(\theta_o) \rightarrow \pi/2$ as $\theta_o \rightarrow \pi/2$. Thus, we have derived an analytical formula for the directional albedo of the QON model.
\begin{figure}[tb]
  \centering
  \includegraphics[width=0.49\linewidth]{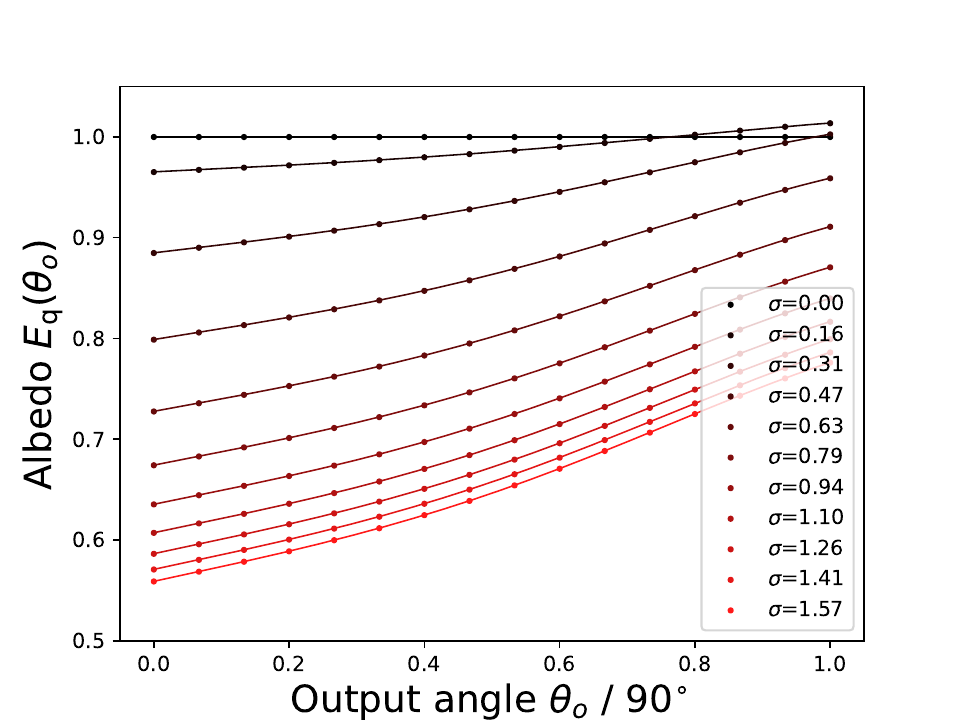}\hfill
  \includegraphics[width=0.49\linewidth]{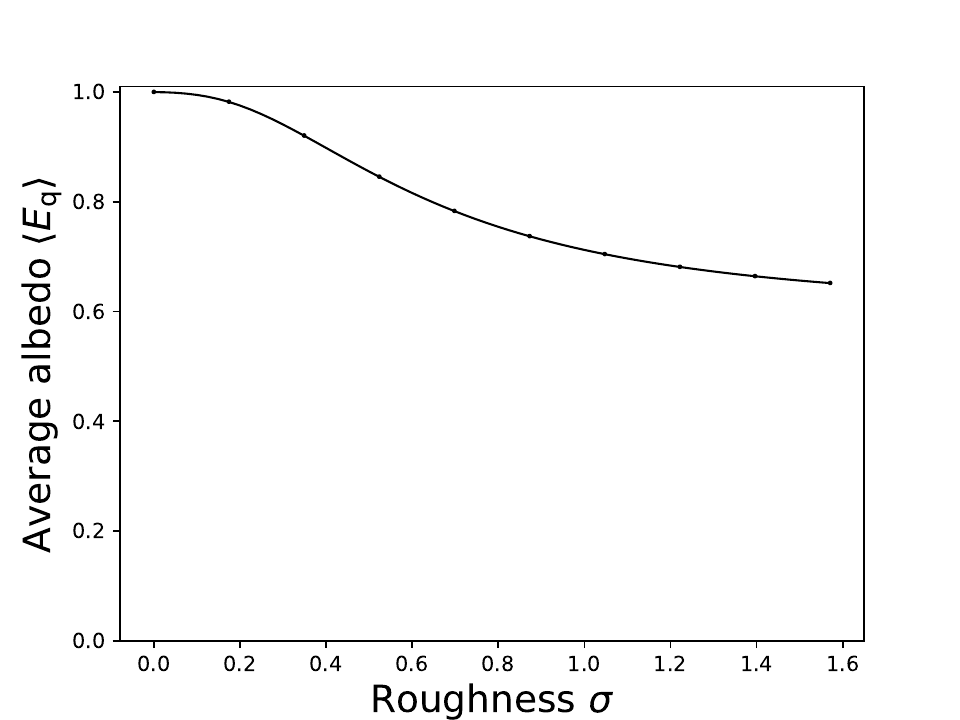}
\caption{QON model albedos. Directional albedo $\hat{E}_\mathrm{q}$ (left) and average albedo $\langle \hat{E}_\mathrm{q} \rangle$ (right), assuming $\boldsymbol{\rho}=1$ (the lines are the analytical formula given by Equations~\ref{QON_albedo} and~\ref{QON_albedo_G}, and the dots are the numerically evaluated integral).}
\label{fig:QON_albedos}
\end{figure}

Figure~\ref{fig:QON_albedos} shows the directional albedo $\hat{E}_\mathrm{q}(\omega_o)$ as a function of view angle $\theta_o$, for various $\sigma$ roughnesses (the hat on $\hat{E}_\mathrm{q}$ denotes that the albedo is evaluated at $\boldsymbol{\rho}=1$).

The term $\langle \hat{E}_\mathrm{q} \rangle = \int_\mathcal{H_+} \hat{E}_\mathrm{q}(\omega_o) \,\mathrm{d}\omega^\perp_o/\pi$ is the \emph{average albedo}, which can be computed analytically as
\begin{eqnarray}
  \langle \hat{E}_\mathrm{q} \rangle = A_\mathrm{q} + \left(\frac{2}{3} - \frac{64}{45\pi}\right) B_\mathrm{q}  .
\end{eqnarray}
Figure~\ref{fig:QON_albedos} shows the variation of the average albedo $\langle \hat{E}_\mathrm{q} \rangle$ as a function of $\sigma$. This corresponds to the total fraction of incident light energy that is reflected (into any direction) under uniform incident illumination. As was apparent from the white-furnace test, the QON model suffers from significant energy loss at high $\sigma$, i.e., does not preserve energy. It also violates energy conservation (i.e., the albedo exceeds 1) for grazing angles and low $\sigma$.\footnote{It can be shown that the fullON model also violates energy conservation near grazing angles, except at high roughness. It also does not preserve energy.}

\subsection{Fujii Oren--Nayar (FON) \label{sec:FON}}

A modification to the qualitative Oren--Nayar model that reduces the dark ring artifact and avoids violations of energy conservation was proposed by \citet{FUJII} (henceforth abbreviated FON). This has the same form as Equation~\ref{QON_vector_form}, i.e.,
\begin{equation} \label{FON_vector_form}
  \mathbf{f}_\mathrm{F}(\omega_i, \omega_o) = \frac{\boldsymbol{\rho}}{\pi} \left( A_\mathrm{F} + B_\mathrm{F} \frac{s}{t_\mathrm{F}} \right),
\end{equation}
but with modified $t$ term (compared to Equation~\ref{t_q}), changing the $s \le 0$ behavior,
\begin{equation}
  \frac{1}{t_\mathrm{F}} =
    \begin{cases}
      1                                                       & \text{if $s \le 0$}\\
      1 / \max\left(N \cdot \omega_i, N \cdot \omega_o\right) & \text{if $s > 0$}
    \end{cases}
\end{equation}
and with modified $A, B$ coefficients written in terms of a roughness parameter $r$:
\begin{align*}
  A_\mathrm{F} &= \frac{1}{1 + \left(\frac{1}{2} - \frac{2}{3\pi}\right)r} , \nonumber \\
  B_\mathrm{F} &= r A_\mathrm{F}  .
\end{align*}
In this model, the roughness parameter $r$ is no longer an angle (like the $\sigma$ of fullON or QON) but instead an interpolation weight in the range $r \in [0,1]$.

\begin{figure}[tb]
  \centering
    \includegraphics[width=0.3\linewidth]{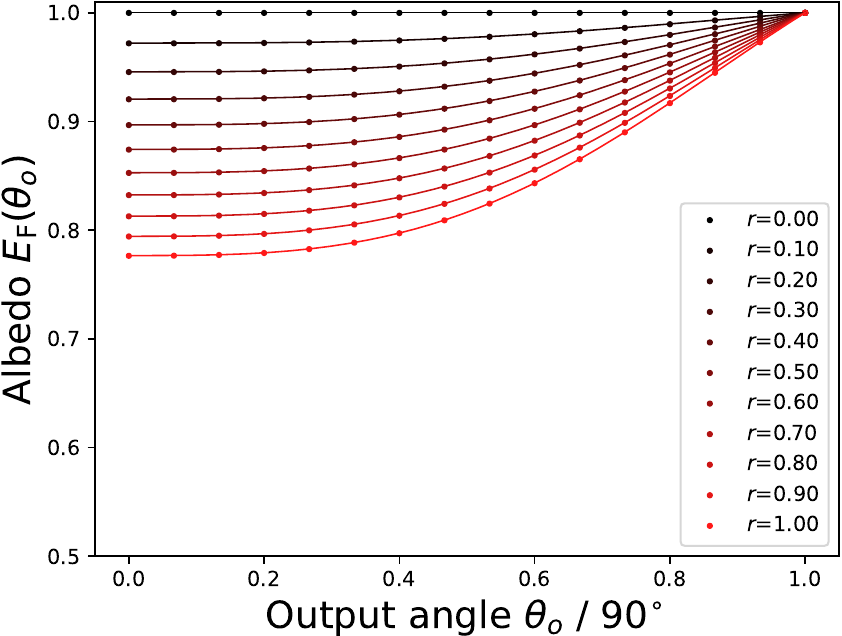}\hfill
    \includegraphics[width=0.302\linewidth]{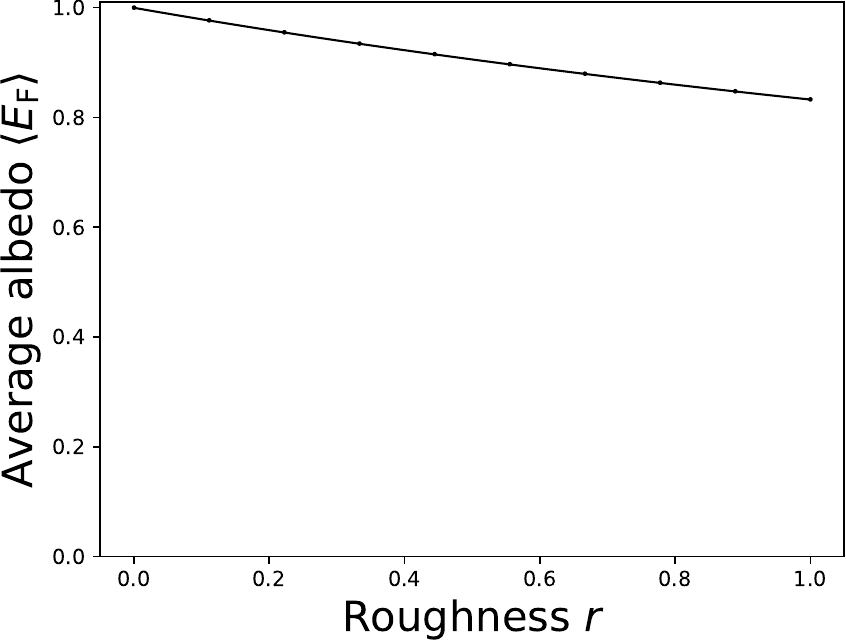}\hfill
    \includegraphics[width=0.34\linewidth]{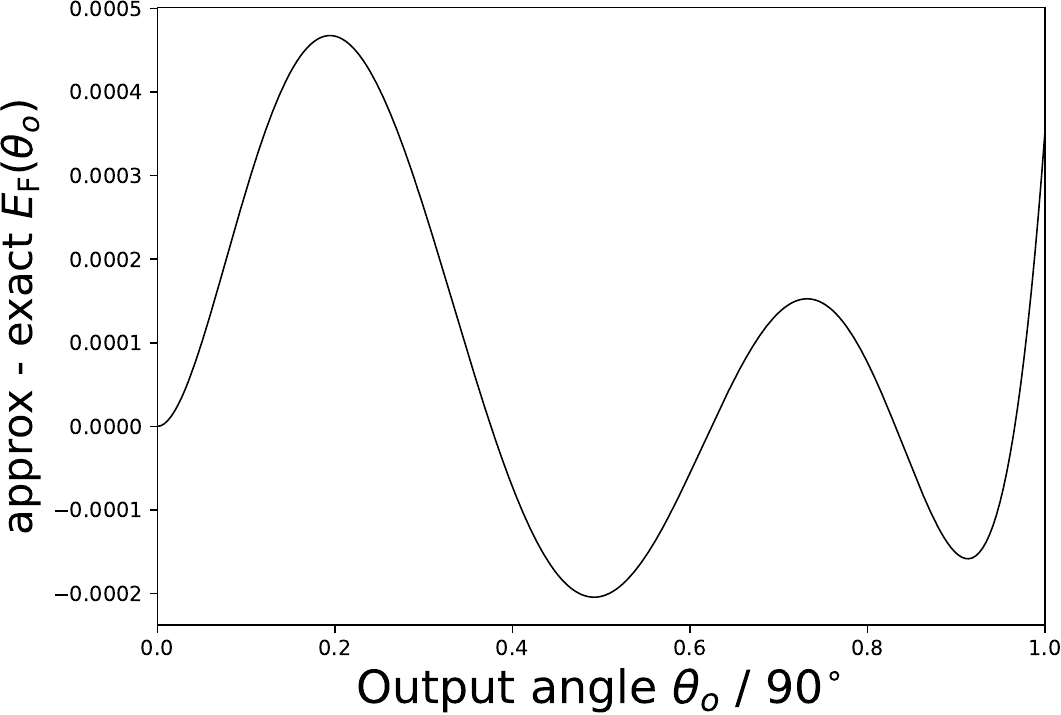}
  \caption{FON model albedos: Directional albedo $\hat{E}_\mathrm{F}$ (left) and average albedo $\langle \hat{E}_\mathrm{F} \rangle$ (center), assuming $\boldsymbol{\rho}=1$ (the lines are the analytical formula given by Equations~\ref{FON_albedo} and~\ref{FON_albedo_G}, and the dots are the numerically evaluated integral). The right panel shows the difference between the analytical and approximate forms of the albedo given by Equations~\ref{FON_albedo_G} and~\ref{FON_albedo_G_approx}, respectively.}
  \label{fig:FON_albedos}
\end{figure}

The directional albedo of the FON model, $\mathbf{E}_\mathrm{F}(\omega_o)$, is given by
\begin{equation} \label{FON_albedo}
  \mathbf{E}_\mathrm{F}(\omega_o) = \boldsymbol{\rho} \left( A_\mathrm{F} + \frac{B_\mathrm{F}}{\pi}G_\mathrm{F}(\omega_o)\right) = \boldsymbol{\rho} \, \hat{E}_\mathrm{F}(\omega_o),
\end{equation}
where (referring to Equation~\ref{QON_albedo_G})
\begin{equation} \label{FON_albedo_G}
  G_\mathrm{F}(\theta_o) = G_\mathrm{q}(\theta_o) - \frac{2}{3} \sin\theta_o  .
\end{equation}
This is plotted in the left panel of Figure~\ref{fig:FON_albedos}, for the $\boldsymbol{\rho}=1$ case. Note that $\hat{E}_\mathrm{F}(\omega_o) \rightarrow 1$ as $\theta_o \rightarrow \pi/2$ (for any roughness $r$), i.e., the model is energy preserving only at grazing view angles (and energy conserving at all other view angles). In fact, the coefficients $A_\mathrm{F}$, $B_\mathrm{F}$ were apparently determined by requiring that $\hat{E}_\mathrm{F}(\pi/2) = 1$ and $B_\mathrm{F} = r A_\mathrm{F}$.

Here we provide a good approximation to the function $G_\mathrm{F}(\theta_o)$ that is more efficient to evaluate than the expression in Equation~\ref{FON_albedo_G} (since the only trigonometric term is $\cos\theta_o$, which is trivial in local space), given by the sum
\begin{equation} \label{FON_albedo_G_approx}
  G_\mathrm{F}(\theta_o) \approx \pi \sum_{k=1}^4 g_k (1-\cos\theta_o)^k  ,
\end{equation}
with coefficients $g_k$ as provided in the following table:~\begin{table}[h!]
  \small
  \centering
  \begin{tabular}{|c c c c|}
   \hline
   $g_1$ & $g_2$ & $g_3$ & $g_4$ \\ [0.5ex]
   \hline
   $0.0571085289$ & $0.491881867$ & $-0.332181442$ & $0.0714429953$ \\
   \hline
  \end{tabular}
\end{table}

The difference between the exact albedo evaluated using Equation~\ref{FON_albedo_G} and the approximate albedo evaluated using Equation~\ref{FON_albedo_G_approx} is less than 0.1\% over the whole angular range, as shown in the right panel of Figure~\ref{fig:FON_albedos}.
\begin{figure}[tb]
  \centering
    \includegraphics[width=0.495\linewidth]{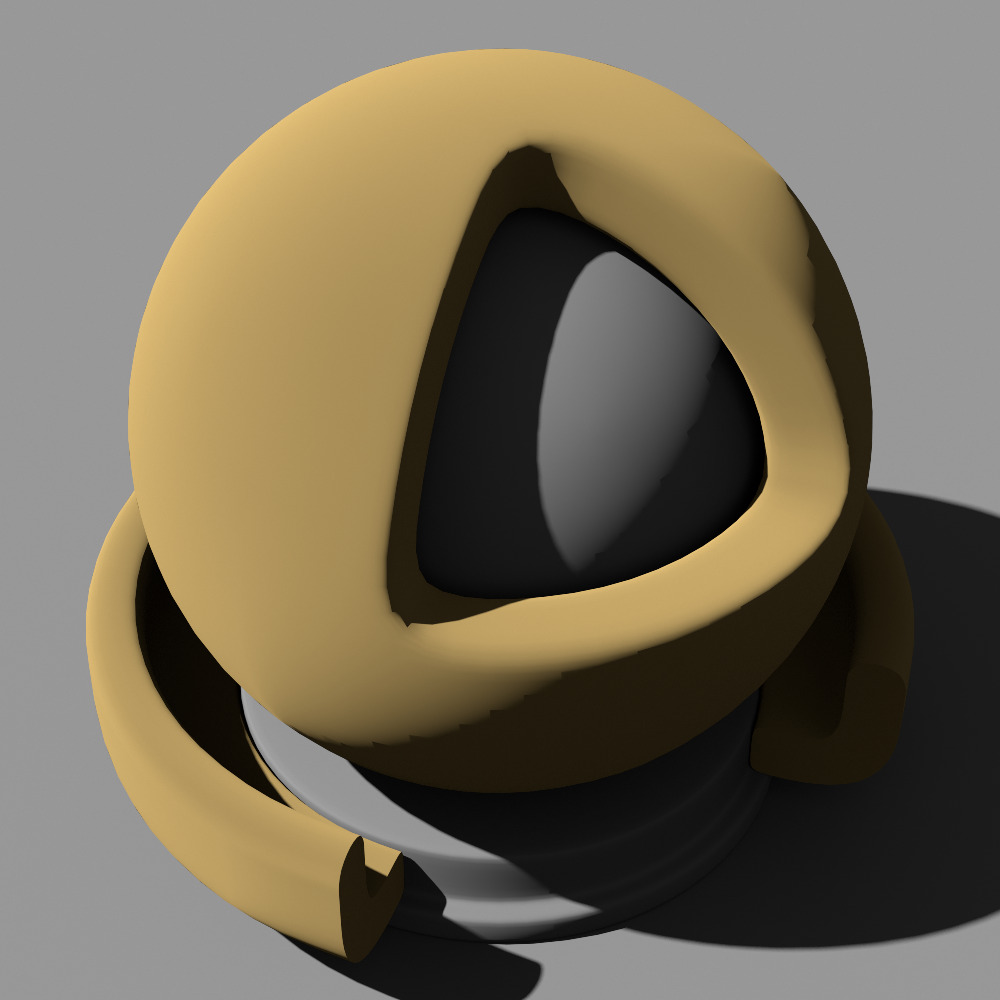}\hfill
    \includegraphics[width=0.495\linewidth]{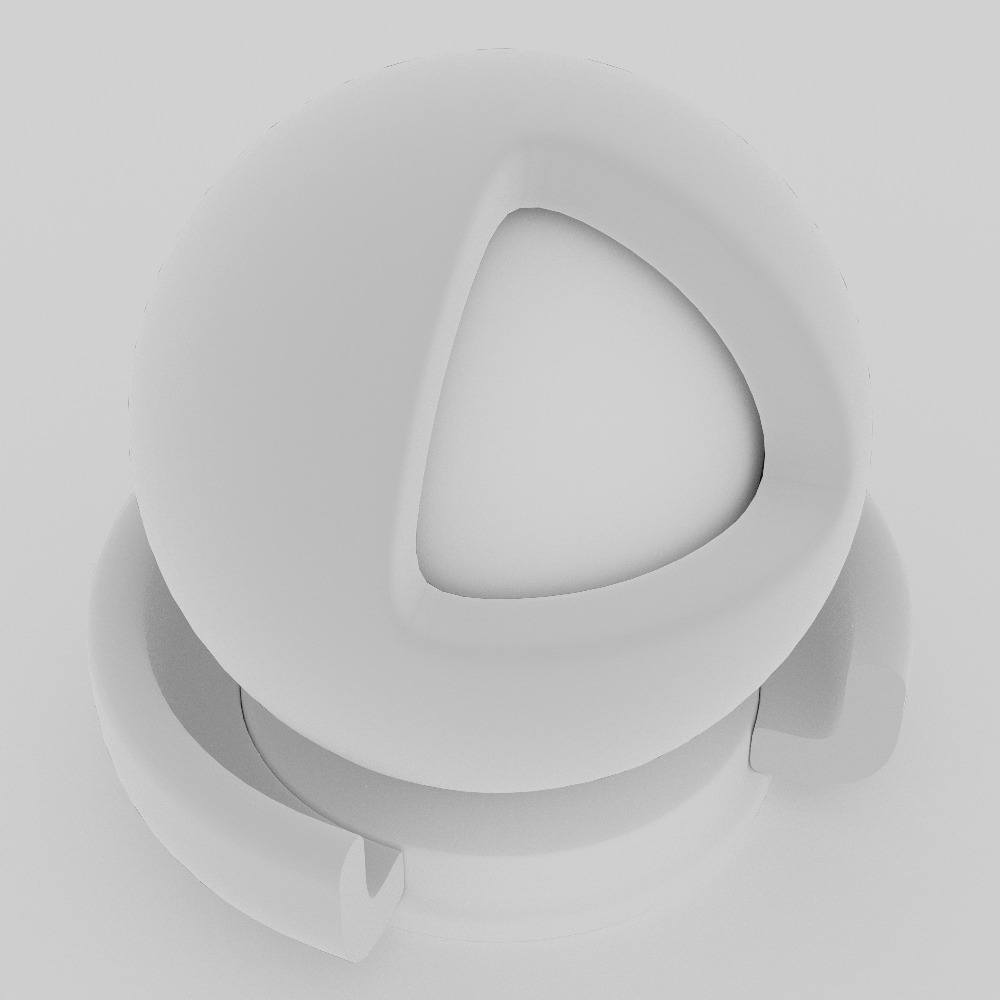}
  \caption{FON model single-bounce renders at $r=1$, with a directional light and colored albedo (left), and white-furnace test (right).}
  \label{fig:FON_renders}
\end{figure}

The average albedo of the FON model, $\langle \hat{E}_\mathrm{F} \rangle = \int_\mathcal{H_+} \hat{E}_\mathrm{F}(\omega_o) \,\mathrm{d}\omega^\perp_o/\pi$, can be evaluated as
\begin{eqnarray} \label{FON_average_albedo}
  \langle \hat{E}_\mathrm{F} \rangle = A_\mathrm{F} + \left(\frac{2}{3} - \frac{28}{15\pi}\right) B_\mathrm{F}  .
\end{eqnarray}
This is plotted in Figure~\ref{fig:FON_albedos}, which shows there is still approximately 20\% energy loss at high roughness values.
Figure~\ref{fig:FON_renders} shows the shaderball render with the FON model at $r=1$ and the corresponding white-furnace test. This model fixes the ring artifact and also performs better on the furnace test, but still loses energy at non-grazing angles.

\end{existing-models}

\section{Energy-Preserving Oren--Nayar Model (EON) \label{sec:EON}}
\begin{eon}

The FON model still suffers from significant energy loss due to modeling only a low number of scattering events between microfacets. We introduce the \emph{energy-preserving Oren--Nayar} model (or EON for short) that fixes this issue by following the energy compensation approach of \citet{Kulla17} for microfacet conductor models, where the missing energy due to multiple scattering is added back via a compensation term designed to be reciprocal.
In the EON model, we formulate the total BRDF as the sum of the FON model BRDF of Equation~\ref{FON_vector_form} and a term representing scattering orders not captured by the FON model:
\begin{equation} \label{EON_brdf}
  \mathbf{f}_\mathrm{EON}(\omega_i, \omega_o) = \mathbf{f}_\mathrm{F}(\omega_i, \omega_o) + \mathbf{f}^\mathrm{ms}_\mathrm{F}(\omega_i, \omega_o)  .
\end{equation}
The multiple-scattering lobe is a reciprocal function constructed to have a constant average albedo $\boldsymbol{\mathcal{F}}$, of the form
\begin{equation} \label{fms_form}
  \mathbf{f}^\mathrm{ms}_\mathrm{F}(\omega_i, \omega_o) =
  \frac{\boldsymbol{\mathcal{F}}}{\pi}
  \left(\frac{1 - \hat{E}_\mathrm{F}(\omega_i)}{1 - \langle\hat{E}_\mathrm{F}\rangle}\right)
  \left(\frac{1 - \hat{E}_\mathrm{F}(\omega_o)}{1 - \langle\hat{E}_\mathrm{F}\rangle}\right)  .
\end{equation}
The factor $\boldsymbol{\mathcal{F}}$ corresponds to the fraction of total energy reflected due to the multiple-scattering lobe, thus for $\boldsymbol{\rho}=1$ we require that $\boldsymbol{\mathcal{F}} = 1 - \langle\hat{E}_\mathrm{F}\rangle$ in order that the average albedo of the total BRDF equals 1.

\begin{figure}[t]
  \centering
    \includegraphics[width=0.495\linewidth]{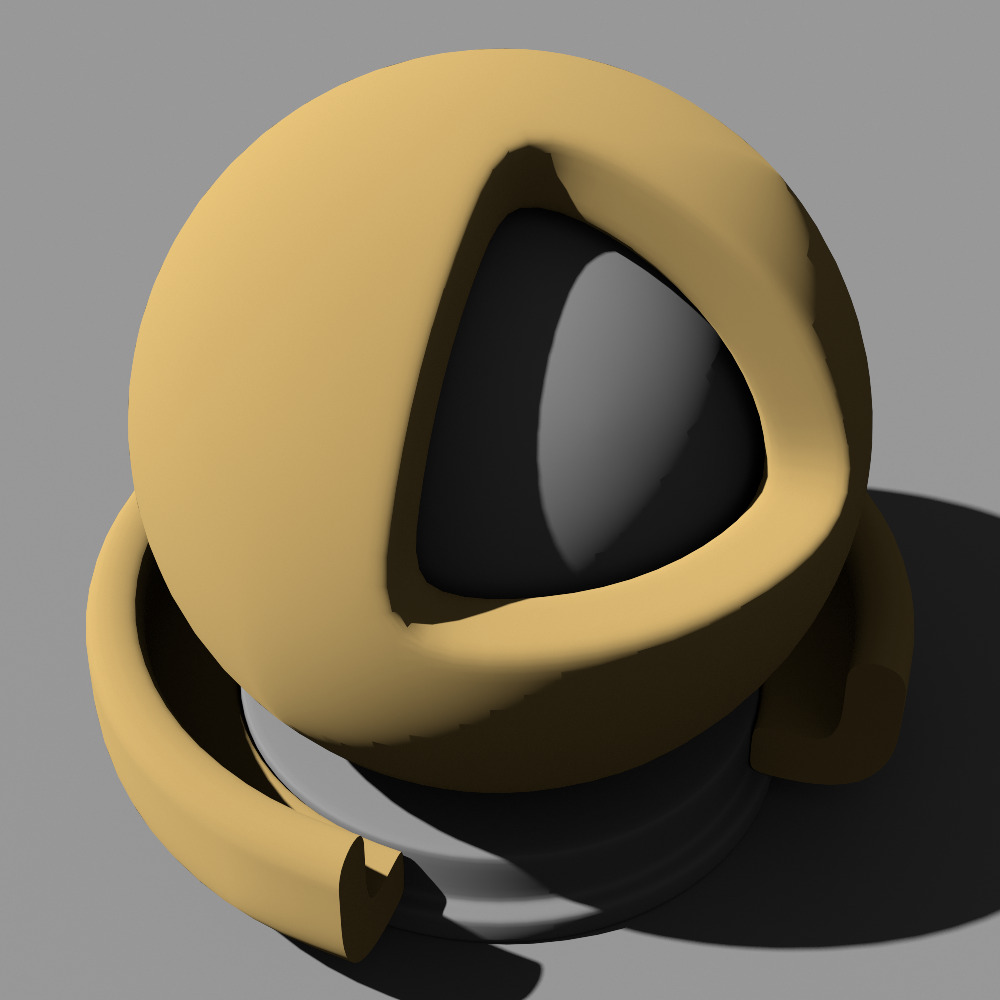}\hfill
    \includegraphics[width=0.495\linewidth]{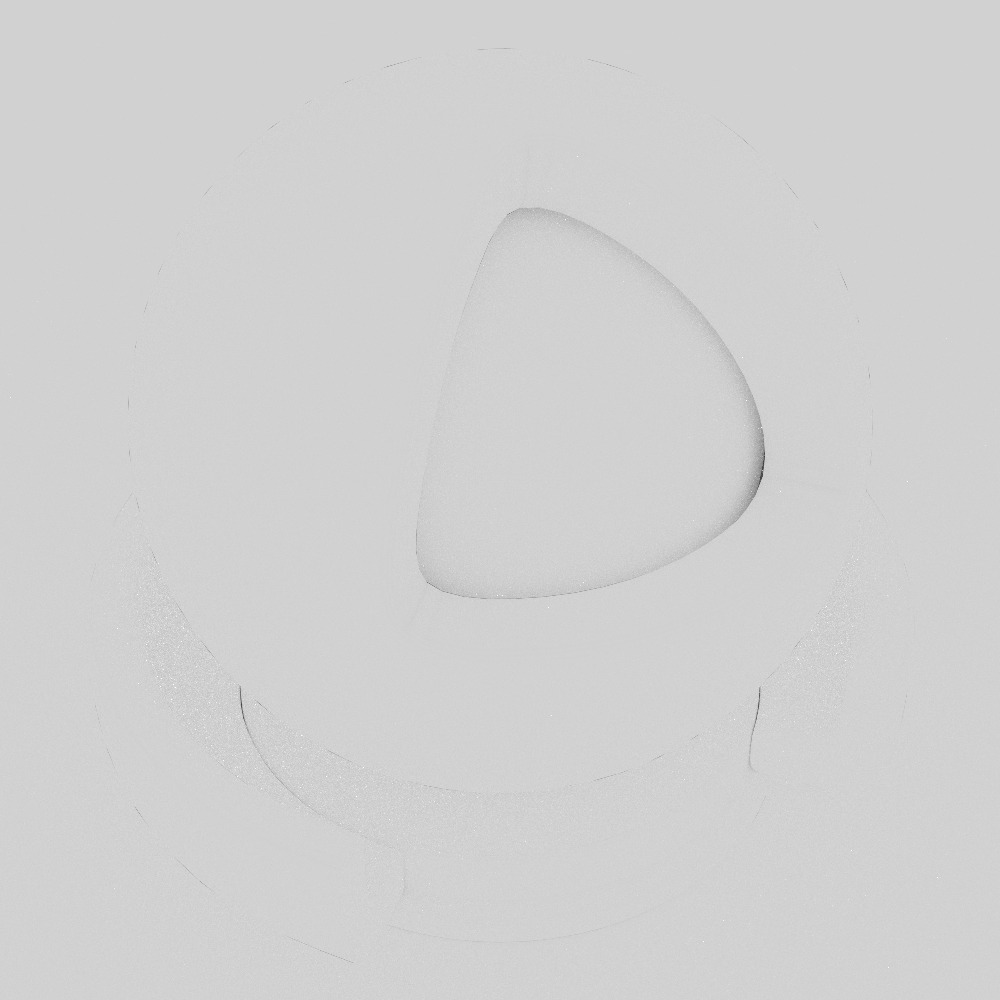}
  \caption{EON model single-bounce renders at $r=1$, with a directional light and colored albedo (left) and white-furnace test (right).}
  \label{fig:EON_renders}
\end{figure}

To derive a suitable approximate form for $\boldsymbol{\mathcal{F}}$, we follow the approach described by \citet{Jakob2014Comprehensive} and \citet{Kulla17}. We assume that the FON BRDF corresponds to single scattering only.
Then, considering the overall energy balance at each bounce on the microfacet surface (approximating the incident illumination in each case as uniform), at each vertex a fraction $\boldsymbol{\rho}$ of the energy is absorbed, a fraction $\langle \hat{E}_\mathrm{F} \rangle$ escapes the surface without further interaction (as for single scattering), and a fraction $1 - \langle \hat{E}_\mathrm{F} \rangle$ hits the surface again. Thus we can approximate $\boldsymbol{\mathcal{F}}$ as the following geometric series of terms, where the 1-bounce term is omitted since the single-scattering BRDF accounts for that:
\begin{eqnarray}
  \boldsymbol{\mathcal{F}} &=&
  \underbrace{\hbox{\sout{$\boldsymbol{\rho} \langle\hat{E}_\mathrm{F}\rangle$}}}_\text{1-bounce} +
  \underbrace{\boldsymbol{\rho}   \left(1 - \langle\hat{E}_\mathrm{F}\rangle\right)   \boldsymbol{\rho}\,\langle\hat{E}_\mathrm{F}\rangle}_\text{2-bounces} +
  \underbrace{\boldsymbol{\rho}^2 \left(1 - \langle\hat{E}_\mathrm{F}\rangle\right)^2 \boldsymbol{\rho}\,\langle\hat{E}_\mathrm{F}\rangle}_\text{3-bounces} + \cdots
  \nonumber \\
  &=& \frac{\boldsymbol{\rho}^2 \langle\hat{E}_\mathrm{F}\rangle \left(1 - \langle\hat{E}_\mathrm{F}\rangle\right)}{1 - \boldsymbol{\rho} \left(1 -\langle \hat{E}_\mathrm{F}\rangle\right)}   .
\end{eqnarray}
As $\boldsymbol{\rho} \rightarrow 1$, $\boldsymbol{\mathcal{F}} \rightarrow 1 - \langle\hat{E}_\mathrm{F}\rangle$ as required. A factor of $\boldsymbol{\rho}^2$ occurs in the numerator, since the multiple-scattering lobe corresponds only to double scattering and higher order \cite{Hill2018,Kulla17}.
The final form of the multiple-scattering lobe is thus:
\begin{equation} \label{fms_form2}
  \mathbf{f}^\mathrm{ms}_\mathrm{F}(\omega_i, \omega_o) =
  \frac{\boldsymbol{\rho}^2}{\pi}
  \frac{\langle\hat{E}_\mathrm{F}\rangle}{1 - \boldsymbol{\rho} \left(1 -\langle \hat{E}_\mathrm{F}\rangle\right)}
 \frac{\left(1 - \hat{E}_\mathrm{F}(\omega_i)\right) \left(1 - \hat{E}_\mathrm{F}(\omega_o)\right)}
{\left(1 - \langle\hat{E}_\mathrm{F}\rangle\right)}  .
\end{equation}
The corresponding directional albedo of this multiple-scattering lobe $f^\mathrm{ms}_\mathrm{F}$ is
\begin{equation} \label{E_ms}
  \mathbf{E}^\mathrm{ms}_\mathrm{F}(\omega_o) =
  \boldsymbol{\mathcal{F}} \frac{\left(1 - \hat{E}_\mathrm{F}(\omega_o)\right)}{1 - \langle\hat{E}_\mathrm{F}\rangle}  .
\end{equation}
Figure~\ref{fig:EON_renders} shows the shader ball render and white-furnace test with this model at $r=1$.
Since $\boldsymbol{\mathcal{F}} \rightarrow 1 - \langle\hat{E}_\mathrm{F}\rangle$ as $\boldsymbol{\rho} \rightarrow 1$, the total directional albedo $E_\mathrm{F}(\omega_o) + E^\mathrm{ms}_\mathrm{F}(\omega_o)\rightarrow 1$ also, so the EON model passes the white-furnace test. The EON model also inherits from the FON model the fix to the ring artifact of QON.


\begin{figure}[t]
  \captionsetup[subfigure]{font=small,labelfont={bf,sf}}
  \centering
  \begin{subfigure}[t]{0.49\linewidth}
    \includegraphics[width=\linewidth]{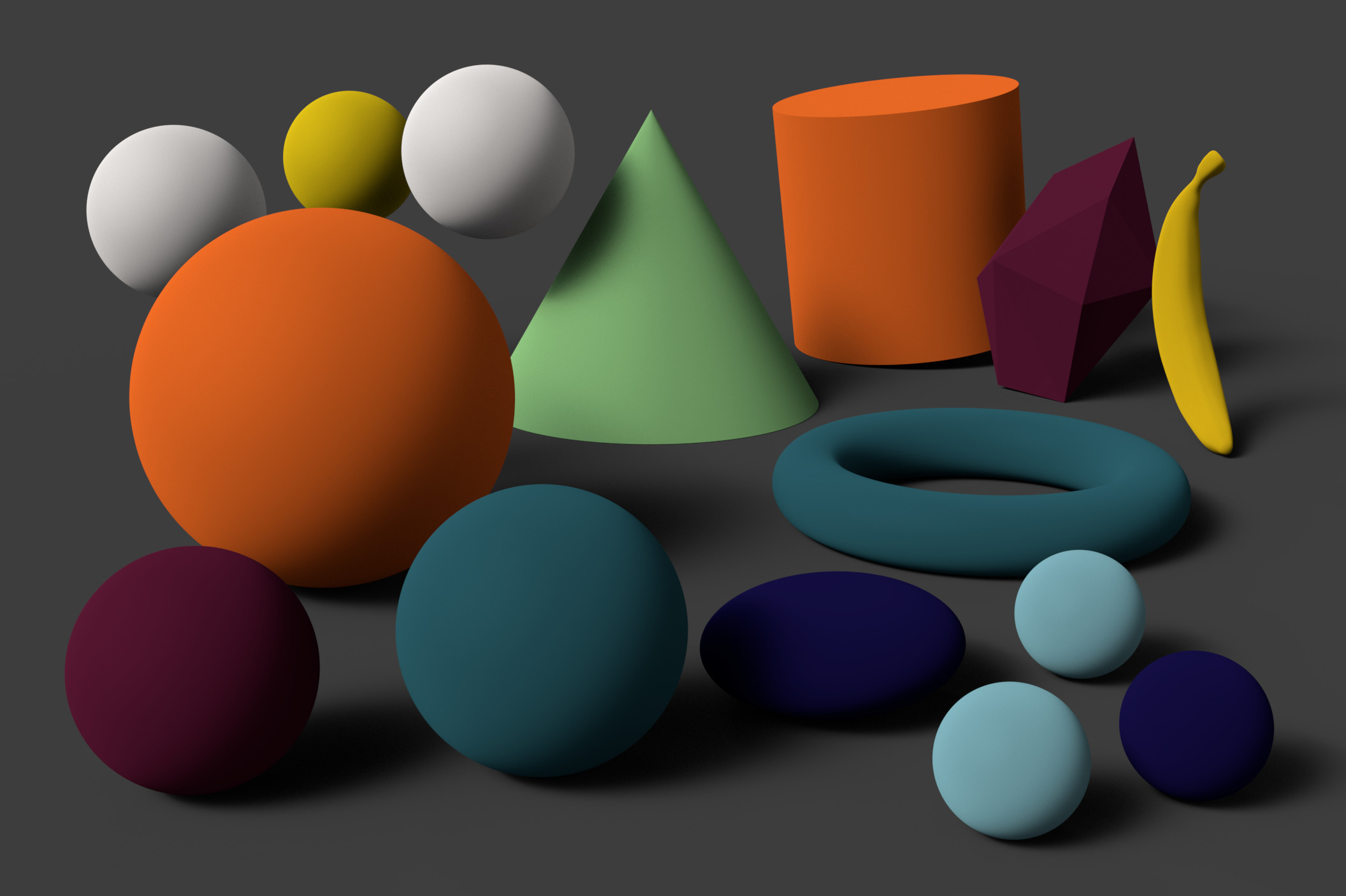}
    \subcaption{QON}
  \end{subfigure}%
  \hspace{0.1cm}
  \begin{subfigure}[t]{0.49\linewidth}
    \includegraphics[width=\linewidth]{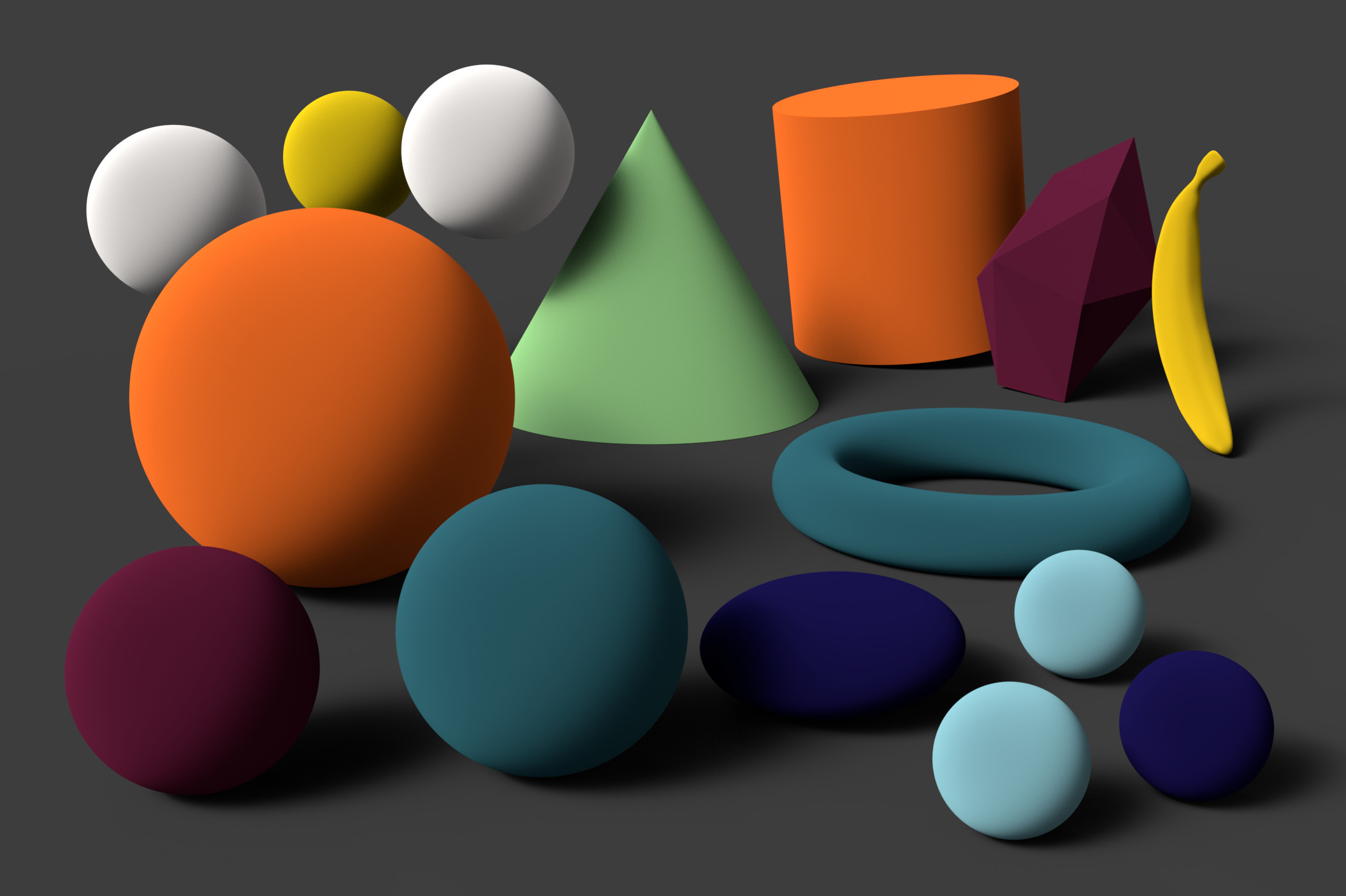}
    \subcaption{EON, single scattering only}
  \end{subfigure} \\ 
  \par\bigskip
  \begin{subfigure}[t]{0.49\linewidth}
    \includegraphics[width=\linewidth]{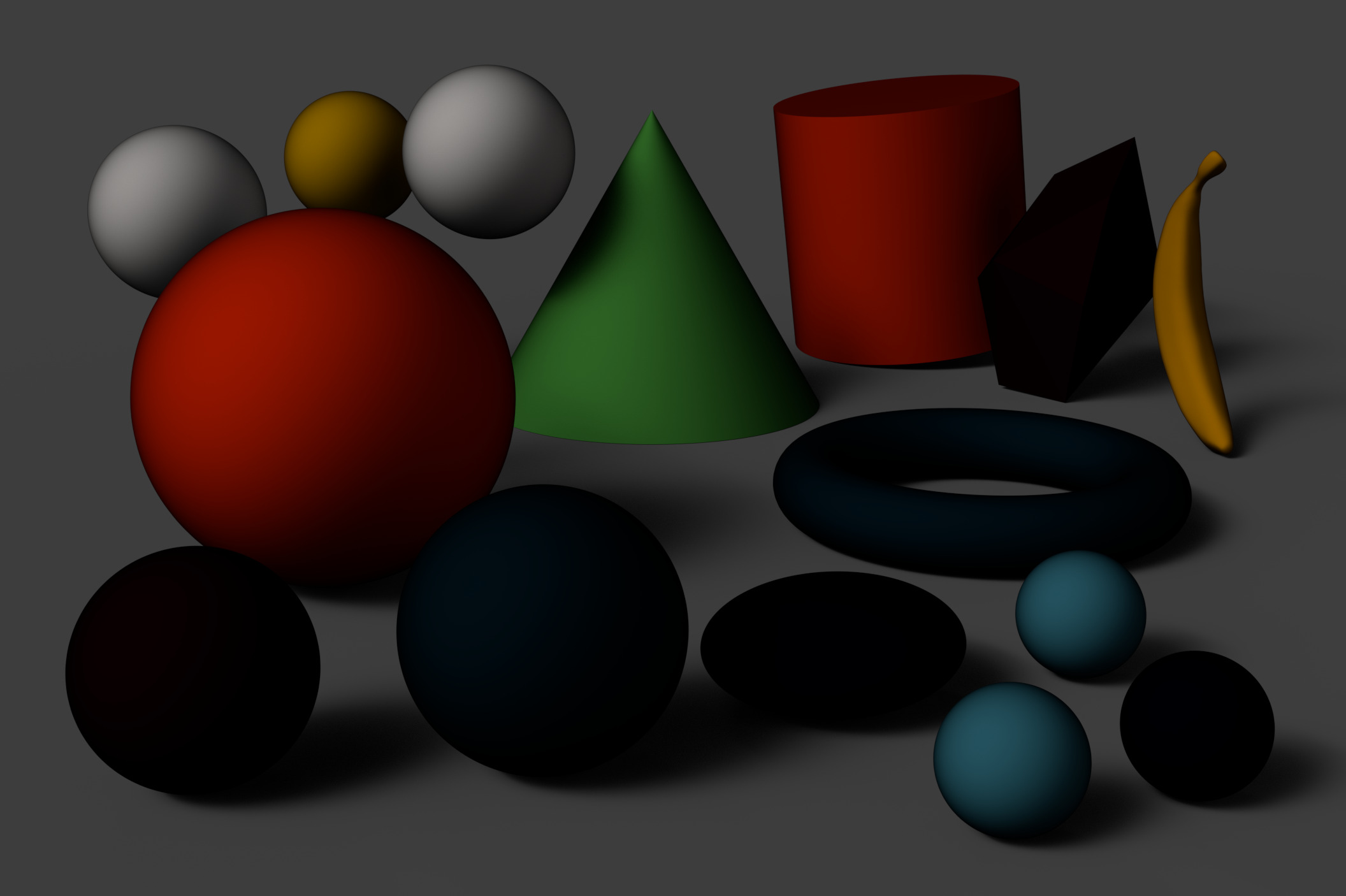}
    \subcaption{EON, multiple scattering only}
  \end{subfigure}%
  \hspace{0.1cm}
  \begin{subfigure}[t]{0.49\linewidth}
    \includegraphics[width=\linewidth]{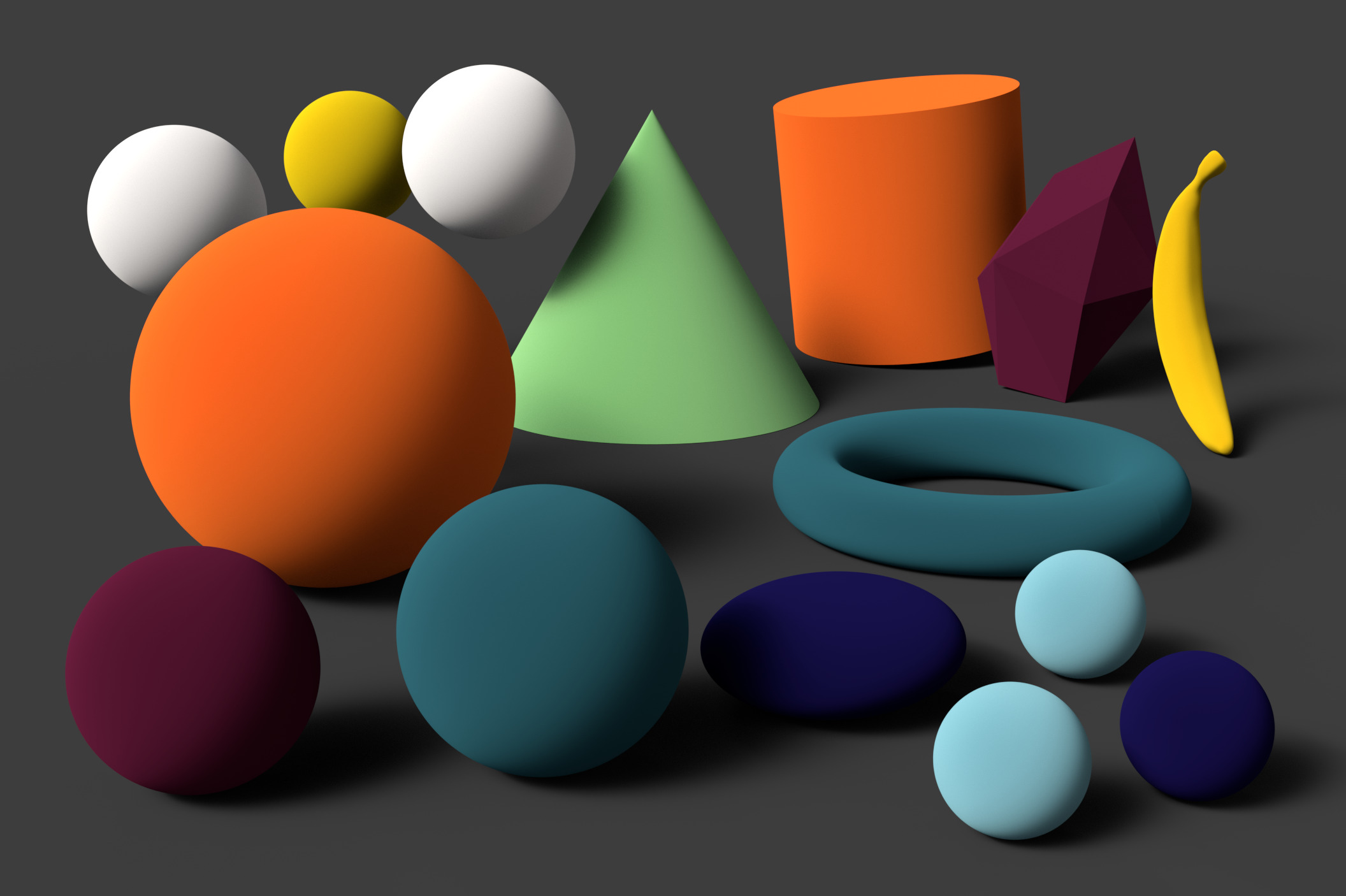}
    \subcaption{EON}
  \end{subfigure}
  \caption{Comparison of QON to the single- and multiple-scattering components of EON.}
  \label{fig:EON_multiple_scattering_color_shift}
\end{figure}

The $f^\mathrm{ms}_\mathrm{F}$ term will produce a color shift due to the extra saturation caused by multiple scattering (as shown on objects with different albedos in Figure~\ref{fig:EON_multiple_scattering_color_shift}). This is physically reasonable, and in practice the extra saturation is only particularly noticeable in cases with high roughness and high chromatic albedo. For use cases where such a color shift is not artistically desirable, we provide the formulas in Appendix~\ref{sec:albedo_inversion} for inverting the desired albedo to obtain the corresponding $\boldsymbol{\rho}$ parameter required to achieve it. Alternatively, the model can simply be evaluated with $\boldsymbol{\rho}=1$ (i.e., white) and the result multiplied by the artist-specified color.

Listing~\ref{listing:EON} gives a self-contained GLSL implementation of the EON BRDF (Equation~\ref{EON_brdf}), and we also provide a function to compute the EON albedo (Equation~\ref{EON_albedo}).

\begin{lstlisting}[
  basicstyle=\ttfamily\tinyplus,
  nolol=true,
  frame=trBL,
  mathescape=true,
  caption={GLSL code for evaluation of the EON model BRDF of Section~\ref{sec:EON}.},
  label={listing:EON},
  %float=p
]
  const float pi            = 3.14159265f;
  const float rcppi         = 1.0f / pi;
  const float constant1_FON = 0.5f - 2.0f / (3.0f * pi);
  const float constant2_FON = 2.0f / 3.0f - 28.0f / (15.0f * pi);

  float E_FON_exact(float mu, float r)
  {
      float AF = 1.0f / (1.0f + constant1_FON * r); // FON $A$ coefficient
      float BF = r * AF;                            // FON $B$ coefficient
      float Si = sqrt(1.0f - (mu * mu));
      float G = Si * (acos(mu) - Si * mu)
              + (2.0f / 3.0f) * ((Si / mu) * (1.0f - (Si * Si * Si)) - Si);
      return AF + (BF * rcppi) * G;
  }
  float E_FON_approx(float mu, float r)
  {
      float mucomp = 1.0f - mu;
      const float g1 = 0.0571085289f;
      const float g2 = 0.491881867f;
      const float g3 = -0.332181442f;
      const float g4 = 0.0714429953f;
      float GoverPi = mucomp * (g1 + mucomp * (g2 + mucomp * (g3 + mucomp * g4)));
      return (1.0f + r * GoverPi) / (1.0f + constant1_FON * r);
  }

  // Evaluates EON BRDF value, given inputs:
  //          rho = single-scattering albedo parameter
  //            r = roughness in [0, 1]
  //        exact = flag to select exact or fast approx. version
  // Note that this implementation assumes throughout that the directions are
  // specified in a local space where the $z$-direction aligns with the surface normal.
  vec3 f_EON(vec3 rho, float r, vec3 wi_local, vec3 wo_local, bool exact)
  {
      float mu_i = wi_local.z;                               // Input angle cos
      float mu_o = wo_local.z;                               // Output angle cos
      float s = dot(wi_local, wo_local) - mu_i * mu_o;       // QON $s$ term
      float sovertF = s > 0.0f ? s / max(mu_i, mu_o) : s;    // FON $s/t$
      float AF = 1.0f / (1.0f + constant1_FON * r);          // FON $A$ coefficient
      vec3 f_ss = (rho * rcppi) * AF * (1.0f + r * sovertF); // Single-scatter lobe
      float EFo = exact ? E_FON_exact(mu_o, r):              // FON $w_o$ albedo (exact)
                          E_FON_approx(mu_o, r);             // FON $w_o$ albedo (approx)
      float EFi = exact ? E_FON_exact(mu_i, r):              // FON $w_i$ albedo (exact)
                          E_FON_approx(mu_i, r);             // FON $w_i$ albedo (approx)
      float avgEF = AF * (1.0f + constant2_FON * r);         // Average albedo
      vec3 rho_ms = (rho * rho) * avgEF / (vec3(1.0f) - rho * (1.0f - avgEF));
      const float eps = 1.0e-7f;
      vec3 f_ms = (rho_ms * rcppi) * max(eps, 1.0f - EFo)    // Multi-scatter lobe
                                   * max(eps, 1.0f - EFi)
                                   / max(eps, 1.0f - avgEF);
      return f_ss + f_ms;
  }

  // Computes EON directional albedo:
  vec3 E_EON(vec3 rho, float r, vec3 wi_local, bool exact)
  {
      float mu_i = wi_local.z;                       // Input angle cos
      float AF = 1.0f / (1.0f + constant1_FON * r);  // FON $A$ coefficient
      float EF = exact ? E_FON_exact(mu_i, r):       // FON $w_i$ albedo (exact)
                         E_FON_approx(mu_i, r);      // FON $w_i$ albedo (approx)
      float avgEF = AF * (1.0f + constant2_FON * r); // Average albedo
      vec3 rho_ms = (rho * rho) * avgEF / (vec3(1.0f) - rho * (1.0f - avgEF));
      return rho * EF + rho_ms * (1.0f - EF);
  }
\end{lstlisting}

\end{eon}

\section{Importance Sampling the EON Model}
\label{sec:EON_sampling}
\begin{eon-sampling}

The original Oren--Nayar model is typically importance-sampled using standard cosine-weighted hemispherical sampling. While this perfectly importance-samples the model in the zero-roughness (Lambertian) case, it is far from optimal in the high-roughness case near grazing angles. This is shown by the red curve in Figure~\ref{fig:importance_sampling_variance}, which is the variance of the throughput weight $w_j = \left(\omega_i\cdot N\right)^+\mathbf{f}_\mathrm{EON}/p_j(\omega_i)$ (setting $\boldsymbol{\rho}=1$) for the EON BRDF when sampled with cosine-weighted hemispherical sampling. The variance increases by a factor of over 100 at grazing angles compared to normal incidence. This is due to the cosine-weighted sampling greatly underestimating the PDF of grazing directions generated by the EON BRDF.
Uniformly sampling the hemisphere produces the dotted black variance curve, which has lower variance than cosine-weighted sampling at grazing angles, but has higher variance at non-grazing angles.

Cosine-weighted hemispherical sampling amounts to sampling from the hemispherical distribution function
\begin{equation} \label{clamped_cosine_lobe}
  D_H(\omega_H) = \frac{\left(\omega_H\cdot N\right)^+}{\pi}  .
\end{equation}
An improved importance sampling scheme can be obtained via a technique known as \emph{Linearly Transformed Cosines} (LTC) \cite{Heitz16LTC}, whereby a given direction vector $\omega_H$ sampled from this clamped cosine lobe is linearly transformed using a matrix $\mathbf{M}$,
yielding $\omega_i$:\footnote{Note that we assume the usual convention for unidirectional path tracing, where the direction of the outgoing ray $\omega_o$ is known, and the incident ray direction $\omega_i$ is the one being sampled. Both the incident ray direction $\omega_i$ and the outgoing ray direction $\omega_o$ are oriented to point away from the surface, i.e., the incident ray is in the opposite direction to incident photons, while the outgoing ray is parallel to outgoing photons. Also note that the sampling would work equally well for the reverse case, when tracing paths in the direction of light flow.}
\begin{equation} \label{LTC_direction_transform}
  \omega_i = \frac{\mathbf{M}\,\omega_H} {\norm{\mathbf{M}\,\omega_H}}  .
\end{equation}
The resulting lobe can be adjusted to best fit a cosine-weighted target BRDF by varying the matrix coefficients and can be efficiently sampled from via $D_H$. The PDF of the sampled direction $\omega_i$ is then given by \cite{Heitz16LTC}
\begin{equation} \label{LTC_pdf}
  D(\omega_i) = D_H \left( \frac{\mathbf{M}^{-1}\,\omega_i} {\norm{\mathbf{M}^{-1}\,\omega_i}} \right) \frac{\partial\omega_H}{\partial\omega_i},
\end{equation}
where the factor $\frac{\partial\omega_H}{\partial\omega_i}$ is the Jacobian of the transformation from $\omega_H$ to $\omega_i$:
\begin{equation}
  \frac{\partial\omega_H}{\partial\omega_i} = \frac{|\mathbf{M}^{-1}|} {\norm{\mathbf{M}^{-1}\,\omega_i}^3}  .
\end{equation}

\enlargethispage{4pt}

Rotating azimuthally into a space where the $\omega_o$ direction (of the outgoing ray) has the form $\omega_o = (\sin\theta_o, 0, \cos\theta_o)$, we take the LTC transformation matrix to have the most general form for an isotropic BRDF \cite{Heitz16LTC}:
\begin{equation} \label{LTC_matrix}
  \mathbf{M} =
\begin{bmatrix}
  a  & 0  & b      \\
  0       & c  & 0 \\
  d  & 0  & 1
\end{bmatrix},
\end{equation}
where the $a, c$ coefficients stretch the lobe radially in the $x$-$y$ plane, while the $b, d$ coefficients shear the lobe in the $x$-$z$ plane.
%
The coefficients $a, b, c, d$ are functions of the outgoing angle $\theta_o$ and roughness $r$.
These functions can be adjusted via a fitting process to make the LTC lobe best match the shape of the cosine-weighted EON BRDF, to better importance-sample the backscattering peak.
\begin{figure}[tb]
  \centering
    \includegraphics[width=0.327\linewidth]{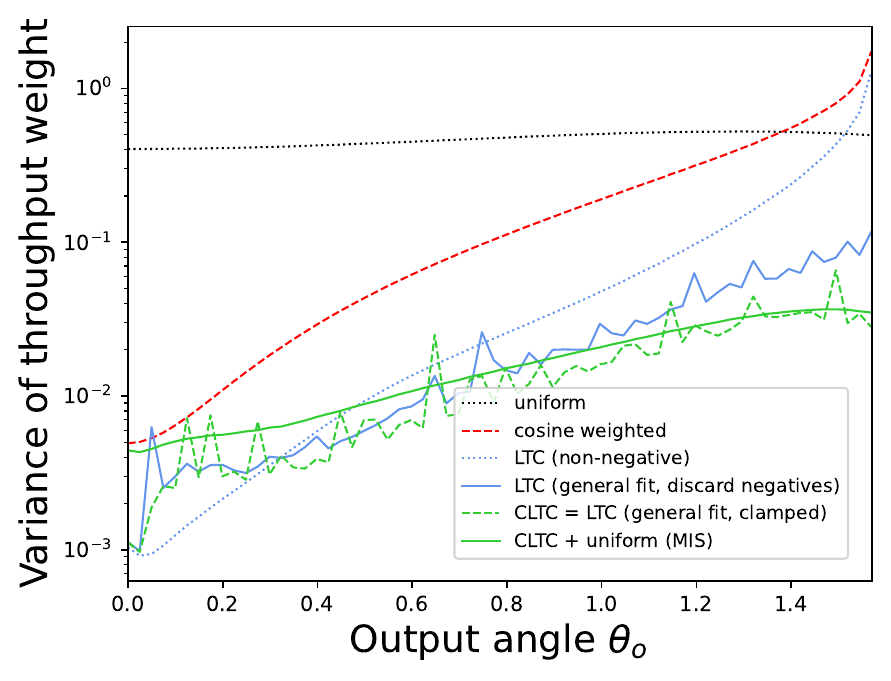}
    \includegraphics[width=0.323\linewidth]{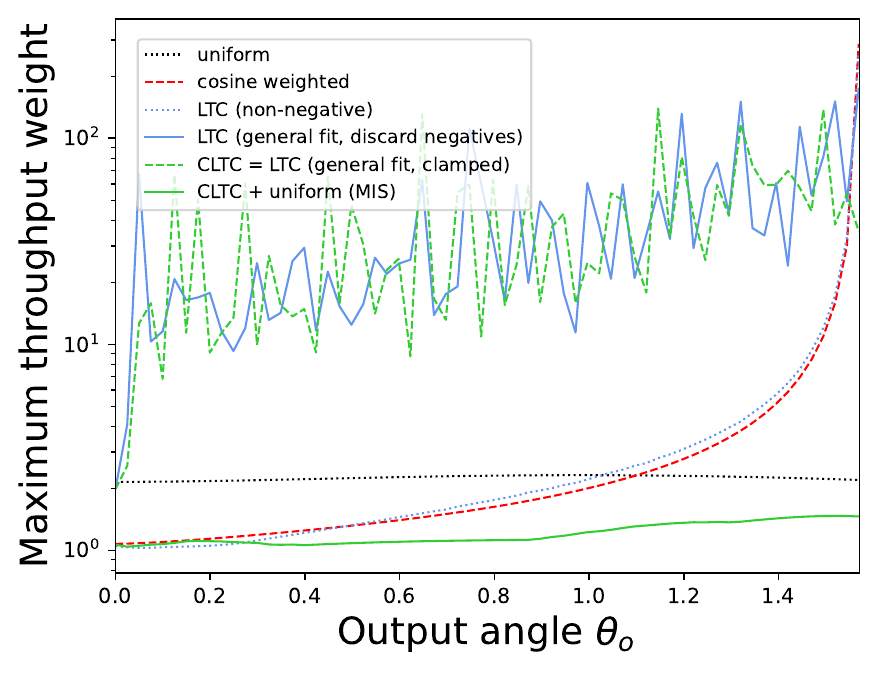}
    \includegraphics[width=0.332\linewidth]{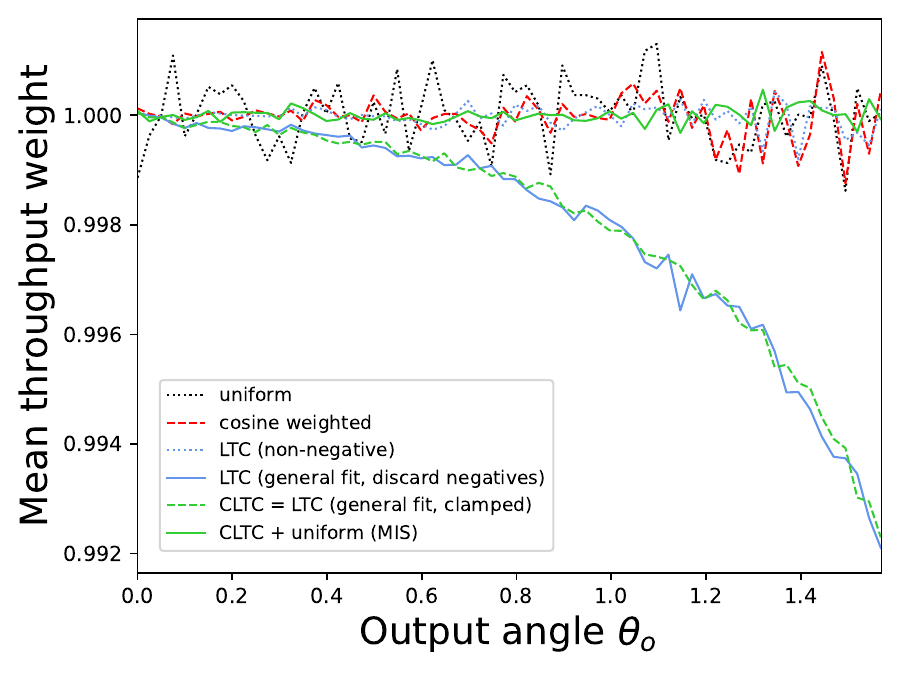}
  \caption{Statistics of the throughput weight $w_j$ over $10^6$ samples of the EON BRDF as a function of angle, using the various techniques: variance of the weight (left), the maximum weight (center), and the mean weight (right).}
  \label{fig:importance_sampling_variance}
\end{figure}

For our fitting we employed Mathematica's \lstinline{NonlinearModelFit}\footnote{With the default \lstinline{Automatic} method and error function.} in two stages: first generating a table of best-fit LTC matrices across $\mu = \cos\theta_o$ and $r$, and then fitting polynomial or rational functions to this data for each of the coefficients. The goal of this second step was to offer a self-contained and portable form for practitioners by avoiding the need for a table lookup. The end result is provided by the GLSL function \lstinline{ltc_coeffs} in Listing~\ref{listing:EON_LTC_fit}.
  \begin{lstlisting}[
      basicstyle=\ttfamily\tiny,
      nolol=true,
      float=b,
      frame=trBL,
      mathescape=true,
      caption={LTC lobe coefficients $a, b, c, d$ as a function of $\mu = \cos\theta_o$ and $r$, for importance sampling of the EON model.},
      label={listing:EON_LTC_fit}
    ]
  void ltc_coeffs(float mu, float r,
                  out float a, out float b, out float c, out float d)
  {
      a = 1.0f + r*(0.303392f + (-0.518982f + 0.111709f*mu)*mu + (-0.276266f + 0.335918f*mu)*r);
      b = r*(-1.16407f + 1.15859f*mu + (0.150815f - 0.150105f*mu)*r)/(mu*mu*mu - 1.43545f);
      c = 1.0f + r*(0.20013f + (-0.506373f + 0.261777f*mu)*mu);
      d = r*(0.540852f + (-1.01625f + 0.475392f*mu)*mu)/(-1.0743f + (0.0725628f + mu)*mu);
  }
\end{lstlisting}

\enlargethispage{20pt}

One possible simplification is to set the coefficient $d=0$, which guarantees that the sampled directions of the LTC lobe lie in the positive hemisphere \cite{Zeltner2022}.
This produces the ``LTC (non-negative)'' variance curve in dotted blue in Figure~\ref{fig:importance_sampling_variance}.
Unfortunately, while this is an improvement on cosine-weighted and uniform sampling, it still suffers from high variance at grazing angles, since the PDF is still very low in this region.

\begin{figure}[!tb]
  \centering
  \includegraphics[width=\linewidth]{figures/cltc_sampling_with_labelling_aligned.pdf}
  \caption{\footnotesize Geometry of CLTC importance sampling:
  \textbf{(a)} The LTC lobe (directions scaled by $D(\omega_i)$) for a near grazing angle $\omega_o$ and $r = 1$. We seek to restrict sample directions to the positive half-space of the $x$-$y$ plane: $\omega_i \in \mathcal{H}_+$.
  \textbf{(b)} In the untransformed space of $D_H$, these directions correspond to $\omega_H \in \mathcal{H}^{-1}_+$, the positive half-space of plane $\mathcal{P}$ with normal $\hat{n}_\mathcal{P}$.
  It follows that we should perform cosine-weighted sampling of the upper hemisphere clipped by $\mathcal{P}$ (shown in green).
  We achieve this by uniformly sampling the 2D projection of this subregion on the $x$-$y$ plane and then
  \textbf{(c)} reprojecting back onto the upper hemisphere. \textbf{(d)} Finally, we proceed as usual by applying the linear transform to obtain the sampled output directions: $\omega_i = \mathbf{M} \, \omega_H$.
  }
  \label{fig:CLTC_sampling_geometry}
  \vspace*{-8pt}
\end{figure}
To significantly reduce variance, the more general LTC matrix with a nonzero $d$ coefficient is required, which produces an LTC lobe that extends \emph{below} the hemisphere. The variance of this is shown in the ``LTC (general fit)'' curve in solid blue in Figure~\ref{fig:importance_sampling_variance}. This greatly reduces the variance peak at grazing angles, but generates a rather spiky variance curve.
Also, on sampling from this lobe, directions that are generated below the hemisphere are effectively wasted as they have a throughput weight of zero. This is not acceptable for renders at low sample count.

To prevent the generation of samples below the hemisphere, we modify the LTC sampling to appropriately restrict the sampling domain of the clamped cosine lobe. We refer to this as \emph{Clipped Linearly Transformed Cosine} (CLTC) sampling. Figure~\ref{fig:CLTC_sampling_geometry} gives an overview of the method. This produces the CLTC green dashed variance curves in Figure~\ref{fig:importance_sampling_variance}.
The idea is that directions $\omega_i \in \mathcal{H}_+$ that lie in the positive half-space of the $x$-$y$ plane (Figure~\ref{fig:CLTC_sampling_geometry}(a)) correspond to untransformed directions $\omega_H \in \mathcal{H}^{-1}_+$ that are in the positive half-space of a plane $\mathcal{P}$ (Figure~\ref{fig:CLTC_sampling_geometry}(b)).
The (unnormalized) normal to $\mathcal{P}$ is\footnote{Using the identity given in Equation~17 of the original LTC paper of \citet{Heitz16LTC}.}
\begin{equation} \label{normal_p}
\hat{n}_\mathcal{P} = \mathbf{M}^{-1}\,\hat{x} \times \mathbf{M}^{-1}\,\hat{y} = \mathrm{det}(\mathbf{M}^{-1})\mathbf{M}^T\,\hat{z}.
\end{equation}
In the case of our fit, $\mathrm{\det}(\mathbf{M}^{-1}) > 0$ over all $\theta_o$ and $r$, so given Equation~\ref{normal_p} and the definition of $\mathbf{M}$ from Equation~\ref{LTC_matrix}, $\hat{n}_\mathcal{P} \propto \mathbf{M}^T\,\hat{z} = (d, 0, 1)$.
As proof that directions $\omega_i$ remain in the positive half-space under transformation by $\mathbf{M}^{-1}$, it is sufficient to show that $\mathbf{M}^{-1}\,\hat{z} \cdot \hat{n}_\mathcal{P} > 0$. This is the case since $\mathbf{M}^{-1}\,\hat{z} \cdot \hat{n}_\mathcal{P} = \mathrm{det}(\mathbf{M}^{-1})$ and, as previously stated,  $\mathrm{\det}(\mathbf{M}^{-1}) > 0$.

It follows that the $\omega_H$ region we must sample from is given by the positive hemisphere \emph{clipped} to the positive half-space of $\mathcal{P}$. In other words, the full hemisphere with a \emph{spherical lune} omitted, as shown in Figure~\ref{fig:CLTC_sampling_geometry}(b). The opening angle of this removed lune is angle $\theta_\mathcal{P}$ between the plane $\mathcal{P}$ and the $x$-$y$ plane, given according to the previous description by $\tan\theta_\mathcal{P} = d$.

\enlargethispage{-30pt}

To sample from the remaining hemispherical subregion with the cosine-weighted PDF $D_H(\omega_H)$ applied, we use the Nusselt analog, in which the correct PDF is obtained by sampling uniformly from the projection of this subregion onto the $x$-$y$ plane, then reprojecting the sampled 2D point back onto the upper hemisphere (Figure~\ref{fig:CLTC_sampling_geometry}(c)).

Recall that the solid angle element $\mathrm{d}\omega_H$ projects to an area element on the $x$-$y$ plane given by $\mathrm{d}A= \left(\omega_H\cdot N\right)^+\mathrm{d}\omega_H$. The area-measure PDF $p(\mathbf{x})$ of the projected points $\mathbf{x}$ on the $x$-$y$ plane must satisfy $p(\mathbf{x}) \mathrm{d}A = p(\omega_H) \mathrm{d}\omega_H$, so $p(\mathbf{x}) = p(\omega_H) / \left(\omega_H\cdot N\right)^+$. Thus, since $p(\omega_H) \propto D_H(\omega_H)$, the area-measure PDF is constant and equal to the inverse area of the projected region.

As shown in Figure~\ref{fig:CLTC_sampling_geometry}(b), the projected shape to sample from takes the form of a half-circle (outlined in green) and a half-ellipse (outlined in red) with semi-major axis $1$ and semi-minor axis $\cos\theta_\mathcal{P} = 1/\sqrt{1+d^2}$. The total area of the projected shape is therefore $A = \frac{\pi}{2} \left(1 + \cos\theta_\mathcal{P}\right)$. On reprojection, the sampling PDF for the direction $\omega_H$ is thus given by
\begin{equation}
p(\omega_H) = \frac{\left(\omega_H\cdot N\right)^+}{A} = \frac{2\,D_H(\omega_H)}{1 + 1/\sqrt{1+d^2}}  .
\end{equation}
(Note that as $r \rightarrow 0$, $d \rightarrow 0$, so the PDF reduces to that of cosine-weighted sampling.)
The final sampled direction $\omega_i$ (Figure~\ref{fig:CLTC_sampling_geometry}(d)) is obtained by applying the linear transform of Equation~\ref{LTC_direction_transform}, and $p(\omega_i) = D(\omega_i)$ is computed according to Equation~\ref{LTC_pdf}.

  \begin{lstlisting}[
      basicstyle=\ttfamily\tiny,
      nolol=true,
      frame=trBL,
      mathescape=true,
      caption={GLSL code for CLTC importance sampling.},
      label={listing:CLTC_importance_sampling},
      float=t
    ]
  mat3 orthonormal_basis_ltc(vec3 w)
  {
      float lenSqr = dot(w.xy, w.xy);
      vec3 X = lenSqr > 0.0f ? vec3(w.x, w.y, 0.0f) * inversesqrt(lenSqr) : vec3(1,0,0);
      vec3 Y = vec3(-X.y, X.x, 0.0f); // cross(Z, X)
      return mat3(X, Y, vec3(0, 0, 1));
  }

  vec4 cltc_sample(vec3 wo_local, float r, float u1, float u2)
  {
      float a, b, c, d; ltc_coeffs(wo_local.z, r, a, b, c, d);   // Coeffs. of LTC $M$
      float R = sqrt(u1); float phi = 2.0f * pi * u2;            // CLTC sampling
      float x = R * cos(phi); float y = R * sin(phi);            // CLTC sampling
      float vz = 1.0f / sqrt(d*d + 1.0f);                        // CLTC sampling factors
      float s = 0.5f * (1.0f + vz);                              // CLTC sampling factors
      x = -mix(sqrt(1.0f - y*y), x, s);                          // CLTC sampling
      vec3 wh = vec3(x, y, sqrt(max(1.0f - (x*x + y*y), 0.0f))); // $\omega_H$ sample via CLTC
      float pdf_wh = wh.z / (pi * s);                            // PDF of $\omega_H$ sample
      vec3 wi = vec3(a*wh.x + b*wh.z, c*wh.y, d*wh.x + wh.z);    // $M \, \omega_H$ (unnormalized)
      float len = length(wi);                                    // $\norm{M \, \omega_H} = 1/\norm{M^{-1} \, \omega_H}$
      float detM = c*(a - b*d);                                  // $|M|$
      float pdf_wi = pdf_wh * len*len*len / detM;                // $\omega_i$ sample PDF
      mat3 fromLTC = orthonormal_basis_ltc(wo_local);            // $\omega_i$ -> local space
      wi = normalize(fromLTC * wi);                              // $\omega_i$ -> local space
      return vec4(wi, pdf_wi);
  }

  float cltc_pdf(vec3 wo_local, vec3 wi_local, float r)
  {
      mat3 toLTC = transpose(orthonormal_basis_ltc(wo_local));                 // $\omega_i$ -> LTC space
      vec3 wi = toLTC * wi_local;                                              // $\omega_i$ -> LTC space
      float a, b, c, d; ltc_coeffs(wo_local.z, r, a, b, c, d);                 // Coeffs. of LTC $M$
      float detM = c*(a - b*d);                                                // $|M|$
      vec3 wh = vec3(c*(wi.x - b*wi.z), (a - b*d)*wi.y, -c*(d*wi.x - a*wi.z)); // $\mathrm{adj}(M) \, \omega_i$
      float lenSqr = dot(wh, wh);                                              // $|M| \norm{M^{-1} \, \omega_i}$
      float vz = 1.0f / sqrt(d*d + 1.0f);                                      // CLTC sampling factors
      float s = 0.5f * (1.0f + vz);                                            // CLTC sampling factors
      float pdf = detM*detM/(lenSqr*lenSqr) * max(wh.z, 0.0f) / (pi * s);      // $w_i$ sample PDF
      return pdf;
  }
  \end{lstlisting}

A simple algorithm for uniform-sampling this union of a half-circle and half-ellipse is given by \citet{Heitz2018GGX}, which we provide as GLSL code in Listing~\ref{listing:CLTC_importance_sampling}.

The green dashed CLTC curves of variance and maximum throughput weight in Figure~\ref{fig:importance_sampling_variance} show that this method generates high outliers compared to uniform sampling, which will correspond to fireflies in the render. There is also a slight bias in the estimator as exhibited in the deviation of the mean throughput weight from 1, since there are directions where the LTC lobe is zero but the EON lobe is nonzero.


To prevent the variance spikes and bias, we combine the CLTC lobe sampling with a uniform hemispherical lobe\footnote{This is a form of \emph{defensive importance sampling}~\cite{Owen2000}.} via \emph{one-sample} multiple importance sampling (MIS), producing the ``CLTC + uniform (MIS)'' variance curve in solid green in Figure~\ref{fig:importance_sampling_variance}. With MIS, the maximum throughput weight (center) is smoothed and reduced, and the throughput weight itself (right) is debiased.

The probability of selecting the uniform lobe as a function of $\theta_o$ and roughness was found via optimization.
In practice, we minimized the maximum throughput weight across $\mu = \cos\theta_o$ for $r = 1$, and then fit a polynomial function to the result.\footnote{Again using Mathematica's \lstinline{NonlinearModelFit}.} Finally, we augmented this with an additional $r^{0.1}$ term to smoothly transition to the Lambertian case as $r \rightarrow 0$. Listing~\ref{listing:EON_importance_sampling} gives the complete GLSL implementation.


  \begin{lstlisting}[
      basicstyle=\ttfamily\tiny,
      nolol=true,
      frame=trBL,
      float=tp,
      mathescape=true,
      caption={GLSL code for importance sampling of the EON model, as in Section~\ref{sec:EON_sampling}.},
      label={listing:EON_importance_sampling}
    ]

    vec3 uniform_lobe_sample(float u1, float u2)
    {
        float sinTheta = sqrt(1.0f - u1*u1); float phi = 2.0f * pi * u2;
        return vec3(sinTheta * cos(phi), sinTheta * sin(phi), u1);
    }

    // Samples (via CLTC) from EON BRDF, given inputs:
    //       rho = single-scattering albedo parameter
    //  wo_local = direction of outgoing ray (directed away from vertex)
    //         r = roughness in [0, 1]
    //    u1, u2 = IID uniform random numbers in [0,1]
    // Returns vec4(vec3(wi_local), pdf)
    vec4 sample_EON(vec3 wo_local, float r, float u1, float u2)
    {
        float mu = wo_local.z;
        float P_u = pow(r, 0.1f) * (0.162925f + (-0.372058f + (0.538233f - 0.290822f*mu)*mu)*mu);
        float P_c = 1.0f - P_u;                    // Probability of CLTC sample
        vec4 wi; float pdf_c;
        if (u1 <= P_u) {
          u1 = u1 / P_u;
          wi = uniform_lobe_sample(u1, u2);        // Sample wi from uniform lobe
          pdf_c = cltc_pdf(wo_local, wi.xyz, r); } // Evaluate CLTC PDF at wi
        else {
          u1 = (u1 - P_u) / P_c;
          wi = cltc_sample(wo_local, r, u1, u2);   // Sample wi from CLTC lobe
          pdf_c = wi.w; }
        const float pdf_u = 1.0f / (2.0f * pi);
        wi.w = P_u*pdf_u + P_c*pdf_c;              // MIS PDF of wi
        return wi;
    }

    // PDF corresponding to the above sampling routine
    float pdf_EON(vec3 wo_local, vec3 wi_local, float r)
    {
        float mu = wo_local.z;
        float P_u = pow(r, 0.1f) * (0.162925f + (-0.372058f + (0.538233f - 0.290822f*mu)*mu)*mu);
        float P_c = 1.0f - P_u;
        float pdf_c = cltc_pdf(wo_local, wi_local, r);
        const float pdf_u = 1.0f / (2.0f * pi);
        return P_u*pdf_u + P_c*pdf_c;
    }
  \end{lstlisting}

Figure~\ref{fig:importance_sampling_variance_versus_roughness} shows how the variance and maximum throughput weight of the algorithm of Listing~\ref{listing:EON_importance_sampling} vary as a function of the roughness parameter $r$, compared to regular cosine-weighted sampling (note that the throughput weight is $1$ at $r=0$ for both cosine-weighted and CLTC sampling).

As $r \rightarrow 0$, both methods produce zero variance and throughput weight (since our algorithm reduces to cosine-weighted sampling in the zero roughness limit). At nonzero roughness, our method produces both variance and maximum throughput whose peaks are much lower than the grazing peaks for cosine-weighted sampling.
\begin{figure}[!htb]
  \centering
    \includegraphics[width=0.495\linewidth]{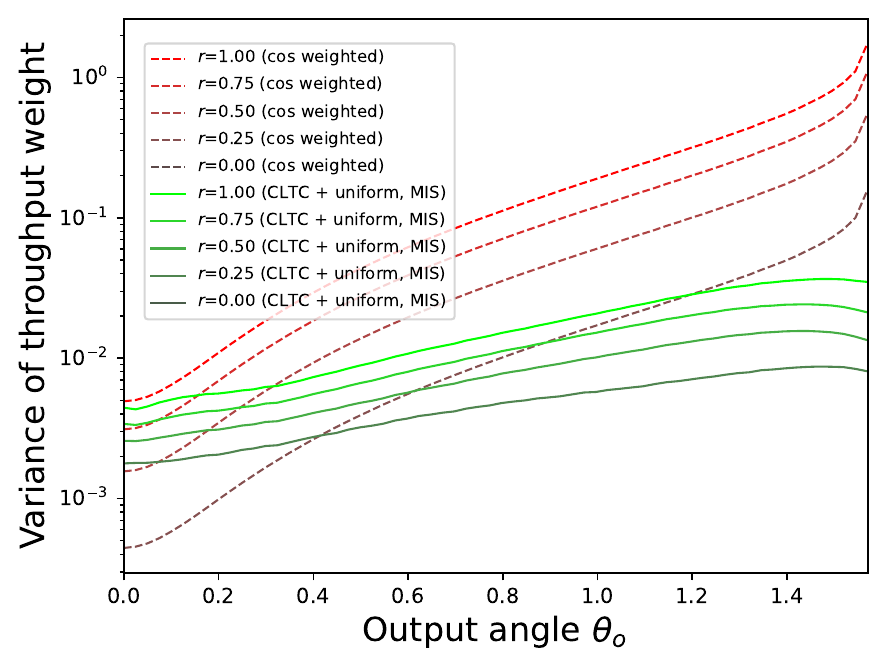}
    \includegraphics[width=0.495\linewidth]{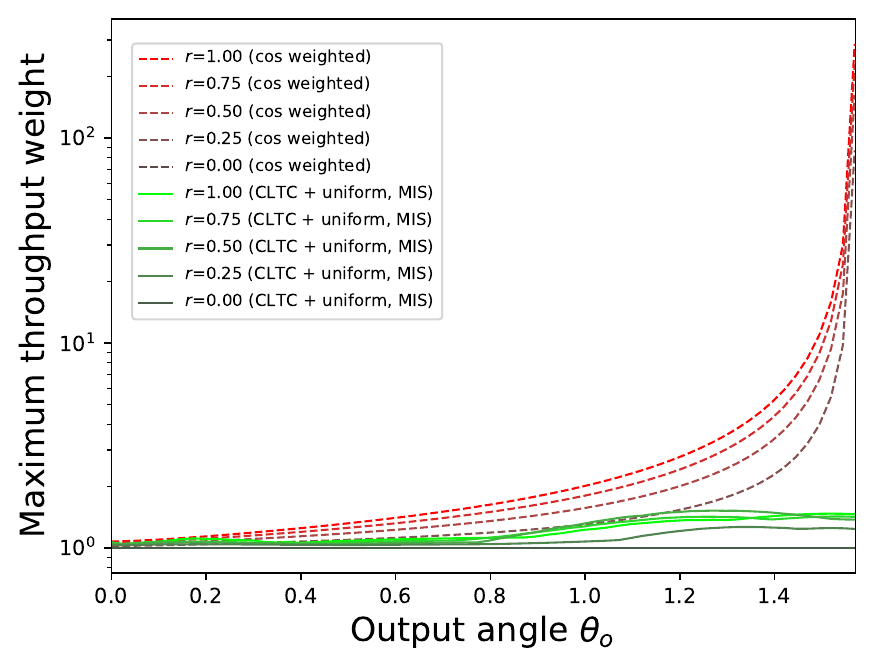}
  \caption{The effect of the roughness $r$ on the variance (left) and maximum throughput weight (right) (of $10^6$ samples) of our CLTC (plus MIS with uniform) sampling method, compared to traditional cosine-weighted sampling. \vspace{10pt}}
  \label{fig:importance_sampling_variance_versus_roughness}
\end{figure}

Figure~\ref{fig:CLTC_sampling_renders} shows the variance reduction of CLTC sampling versus cosine sampling in renders at high bounce count and low sample-per-pixel count.

Table~\ref{table:timings} shows timings for the various BRDFs discussed, as nanoseconds per BRDF evaluation or sampling invocation (where the sampling time includes the time to evaluate the BRDF in the sampled direction also). The profiles were computed by porting the GLSL implementations of the BRDFs to C++ using the GLM library \cite{GLM_library} and compiling with maximum optimizations enabled. Timings were performed on a single thread of an AMD 3970X, averaged over several billion samples (using random roughness values, albedo values, and input directions).

Unsurprisingly, the Lambert model BRDF is unbeatably fast to evaluate, while the fullON method is slowest. The exact EON BRDF is about $1.3\times$ faster to evaluate than fullON, and approximate EON is about $7\times$ faster.

The QON and FON variants are faster to evaluate than EON, but, as discussed, they suffer from energy loss and artifacts that render them impractical. The CLTC sampling of the approximate EON model takes roughly twice as long as cosine sampling of the same model (and roughly the same time as cosine sampling of the fullON model).

However, as discussed above (see Figure~\ref{fig:importance_sampling_variance_versus_roughness}), the variance reduction of the CLTC sampling method exceeds a factor of 100 at grazing angles, thus the time to reach a given noise level will be greatly reduced.

\begin{table}[!b]
  \normalsize
  \centering
    \begin{tabular}{|l c c c|}
      \hline
      BRDF  &  Eval (ns) & Cosine Sample (ns) & CLTC Sample (ns) \\ [0.5ex]
      \hline\hline
      Lambert                &   1.4      &   67.2   &  --- \\
      \hline
      FON                    &   6.3      &   67.2   &  --- \\
      \hline
      QON                    &   6.3      &   67.4   &  --- \\
      \hline
      fullON        &   141.7   &   211.218 &  --- \\
      \hline
      EON (approx)           &  20.8      &   105.8  &  217.0 \\
      \hline
      EON (exact)            &   108.7    &   193.5  &  297.0 \\
      \hline
    \end{tabular}
      \caption{BRDF evaluation and sampling timings.}
  \label{table:timings}
\end{table}


\end{eon-sampling}


\begin{figure}[!tb]
  \centering
    \includegraphics[width=0.495\linewidth]{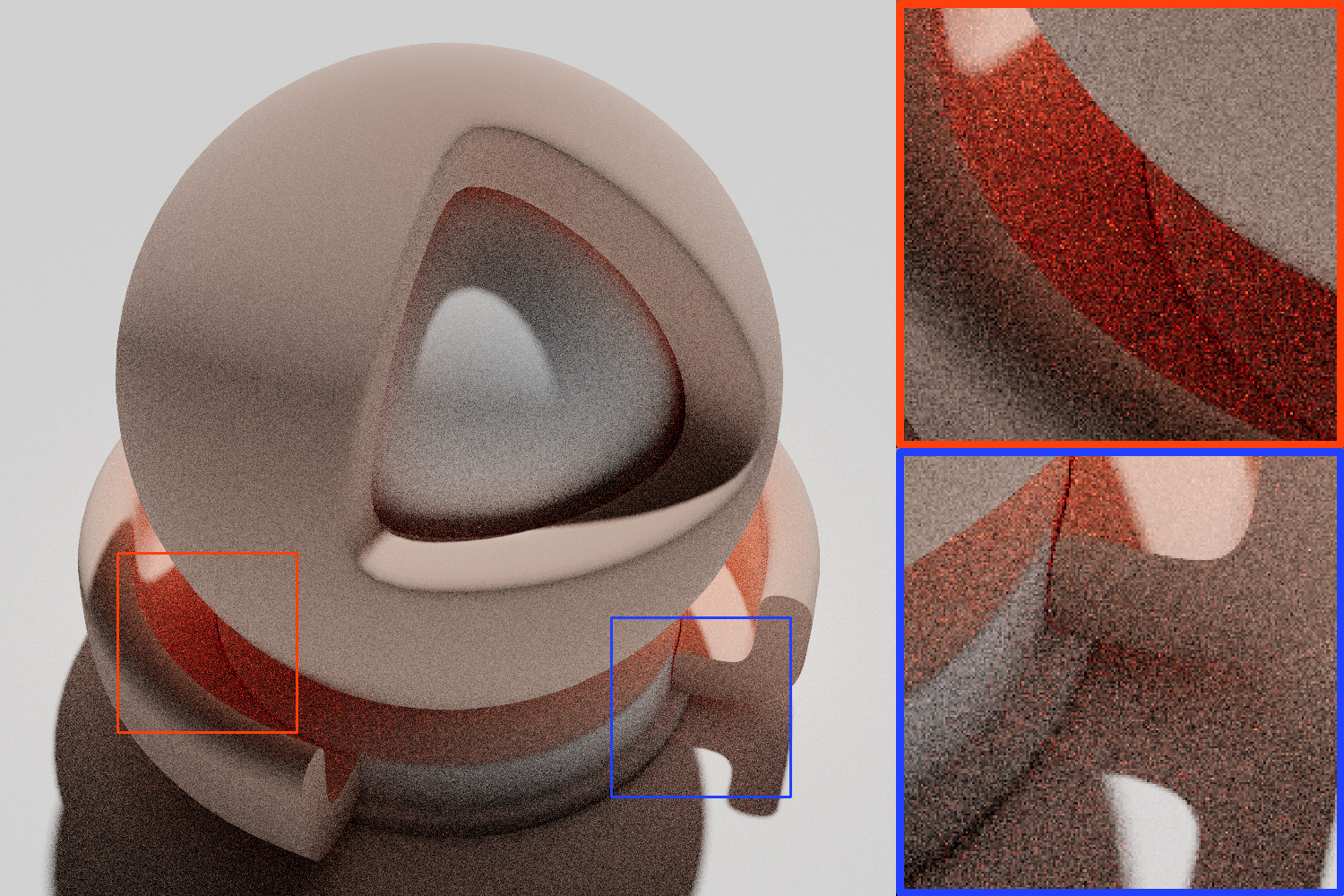}
    \includegraphics[width=0.495\linewidth]{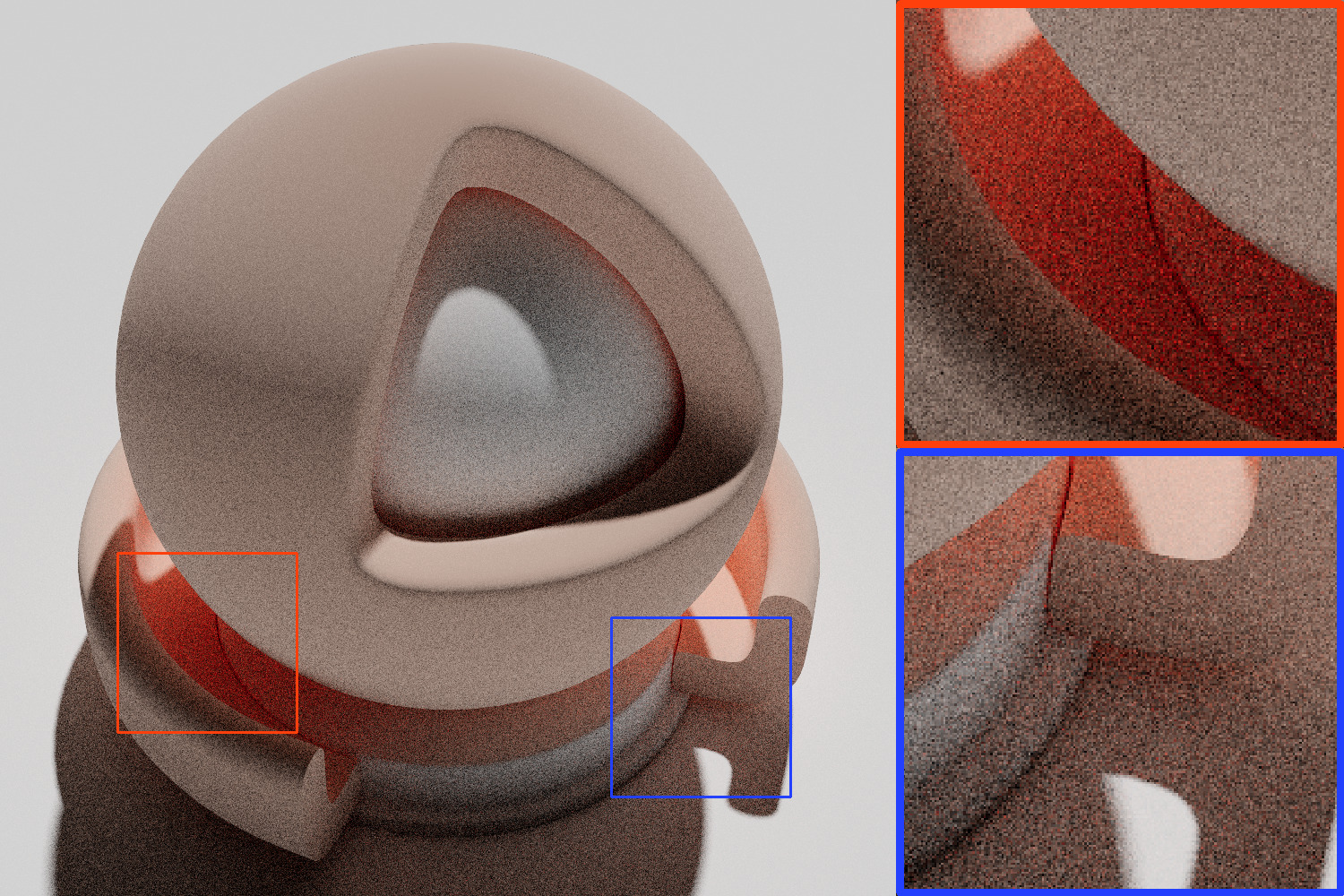} \\
    \vspace{0.06cm}
    \includegraphics[width=0.495\linewidth]{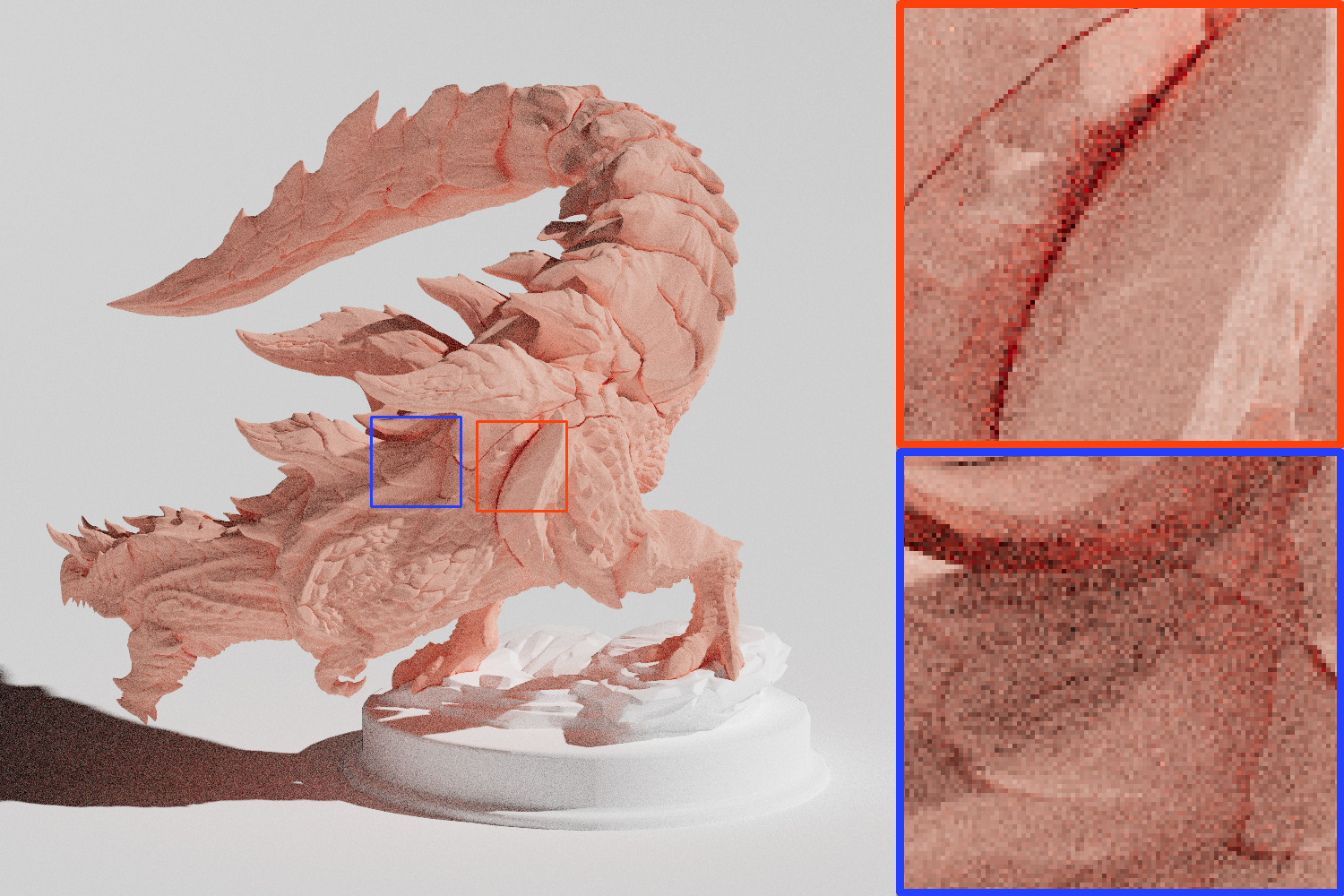}
    \includegraphics[width=0.495\linewidth]{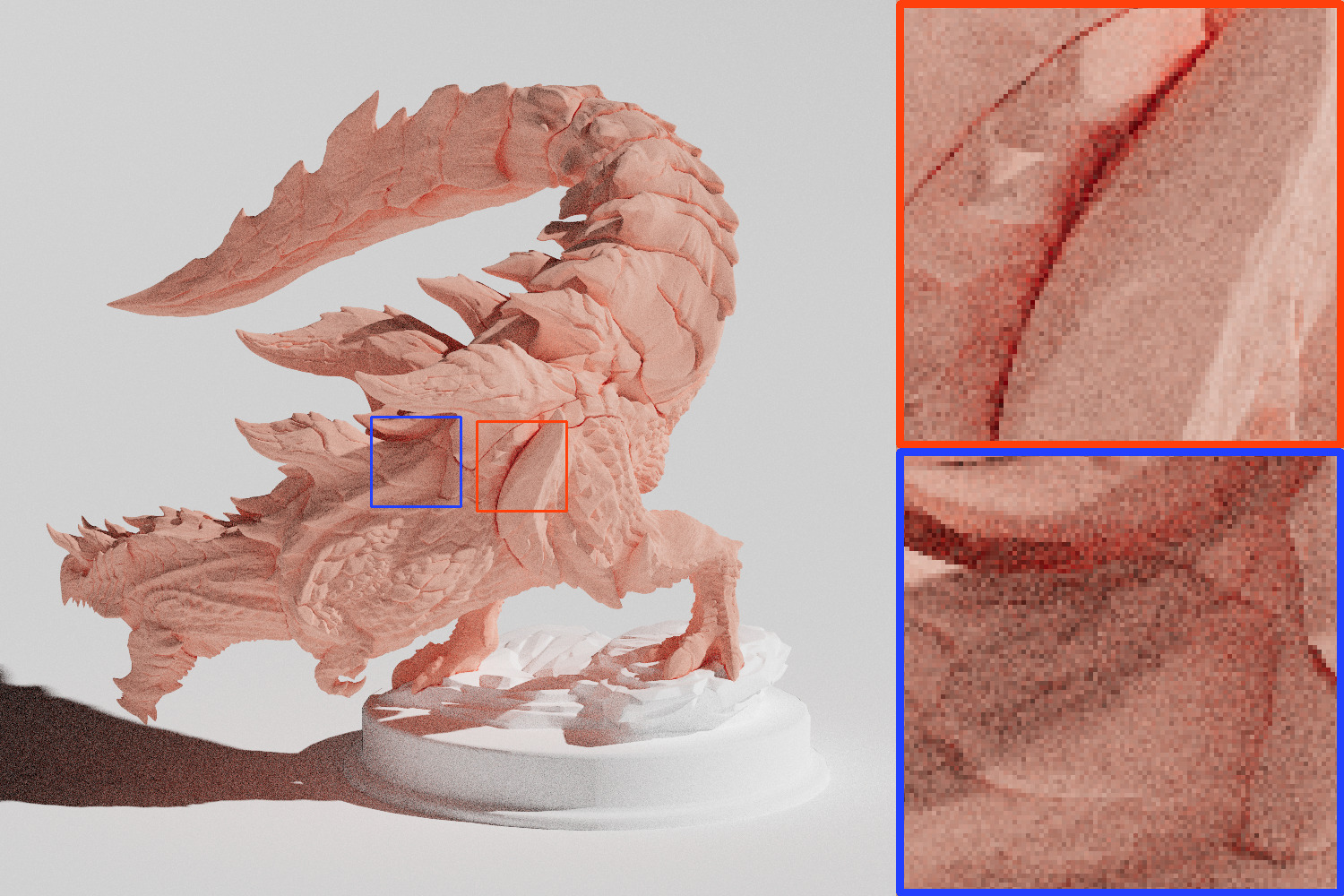}
  \caption{EON model at maximal roughness ($r=1$), rendered with traditional cosine-weighted sampling (left) and our CLTC sampling (right), with 50 samples per pixel and 50 bounces. The variance reduction of CLTC sampling produces a significantly smaller amount of noise and fireflies than cosine-weighted sampling.}
  \label{fig:CLTC_sampling_renders}
\end{figure}

\vspace*{2pt}

\section{Conclusion}
\label{sec:conclusion}

\enlargethispage{-10pt}

In this paper we have introduced the \emph{energy-preserving Oren--Nayar} model (or EON), a new variant of the classic model of \citet{OrenNayar94}, that addresses many of the practical limitations of the original. We consider the EON model to be more practical than other existing variants of the Oren--Nayar model because it is energy preserving, reciprocal, artifact-free, easy to implement, and efficient to evaluate, making it a strong candidate for use in production rendering applications.

We also provided a practical importance sampling scheme for this model, based on a Linearly Transformed Cosine (LTC) lobe. We introduced \emph{Clipped Linearly Transformed Cosine} (CLTC) sampling to prevent negative hemisphere sampling, and combined this with uniform hemispherical sampling via MIS to achieve a much lower variance and maximum throughput weight than standard cosine-weighted sampling (as illustrated in Figure~\ref{fig:CLTC_sampling_renders}).

For these reasons it has been chosen as the rough diffuse model used in the proposed industry standard OpenPBR \"uber-shader
specification \cite{OpenPBR}.

The full source code presented in the listings can be downloaded from GitHub \cite{portsmouth-EON-diffuse}.

\textit{Future Work.}
The overall shape of our new BRDF is influenced by the underlying assumptions of the Oren--Nayar model. It would be interesting to compare to alternative physically based models, especially the recent Lambertian sphere \cite{LambertianSpheresBRDF} and VMF diffuse \cite{VMFDiffuse2024} models, which are based on the scattering from a collection of spherical Lambertian particles. We note here though that, as far as we are aware, those models are not explicitly energy preserving, are considerably more computationally intensive to evaluate than EON, and do not come with an importance sampling scheme more sophisticated than cosine sampling. Indeed, the CLTC sampling scheme developed here could possibly be repurposed to provide improved importance sampling for these models.

Future refinements to the EON model itself could include better aligning our new model with an underlying microfacet model. This could allow more physically accurate color shifting by separating the single, double, and higher-order scattering components, with a more physically accurate shape for the multiple-scattering lobes.

While this paper focuses on the Oren--Nayar family of diffuse BRDFs, the broader question of how best to model the appearances of the wide variety of real-world diffuse materials remains open, requiring a balance between fidelity and practical considerations such as usability and computational cost. We encourage additional research to refine these tradeoffs and further advance physically based diffuse reflectance modeling.


\section*{Acknowledgements}

We thank Chris Kulla, Eugene d'Eon, Luke Emrose, and Ivo Kondapaneni for helpful discussions.
In several figures we used the Standard Shader Ball asset of \citet{Mazzone2023}.
The dragon of Figure~\ref{fig:teaser} is the Glavenus model by \citet{Glavenus}, from myminifactory.com.
The model in Figure~\ref{fig:brdf_renders} is the ``Mat'' model from Adobe \cite{MATmesh}.

\newpage
\small
\bibliographystyle{jcgt}
\bibliography{paper}

\appendix

\section{Albedo inversion}

\label{sec:albedo_inversion}

For color matching purposes, it can be helpful to have a map from some desired observed color $\mathbf{C}$ (i.e., albedo), in some simple configuration like normal incidence view with a white skydome, to the $\boldsymbol{\rho}$ parameter which achieves that. We provide formulas for this here.

According to Eqn.~\ref{E_ms}, the total directional albedo of the EON model is
\begin{equation} \label{EON_albedo}
\mathbf{E}_\mathrm{EON}(\omega) =  \boldsymbol{\rho} \frac{ (1-\boldsymbol{\rho})\hat{E}_\mathrm{F}(\omega) +  \boldsymbol{\rho} \langle \hat{E}_\mathrm{F} \rangle} { (1-\boldsymbol{\rho}) +  \boldsymbol{\rho} \langle \hat{E}_\mathrm{F} \rangle} \ ,
\end{equation}
where the normalized directional and average albedos of the FON model, $\hat{E}_\mathrm{F}$ and $\langle \hat{E}_\mathrm{F} \rangle$ are defined in Eqns.~\ref{FON_albedo}~and~\ref{FON_average_albedo}. At normal incidence we have (with $\alpha = \frac{1}{2} - \frac{2}{3\pi}$, $\beta = \frac{2}{3} - \frac{28}{15\pi}$)
\begin{eqnarray}
  \hat{E}_\mathrm{F}(N)              &=& 1/(1 + \alpha r)               \ , \nonumber \\
  \langle \hat{E}_\mathrm{F} \rangle &=& (1 + \beta r)/(1 + \alpha r)   \ .
\end{eqnarray}
Solving $\mathbf{E}_\mathrm{EON}(N) = \mathbf{C}$ for the $\boldsymbol{\rho}$ that generates the desired color viewed at normal incidence using the quadratic formula gives
\begin{equation}
  \boldsymbol{\rho} = \frac{1}{2A} \left( -\mathbf{B} + \sqrt{\mathbf{B}^2 + 4A \mathbf{C}} \right) \ ,
\end{equation}
where
\begin{eqnarray}
  A &=& \langle \hat{E}_\mathrm{F} \rangle - \hat{E}_\mathrm{F}(N) \ge 0 \nonumber \\
  \mathbf{B} &=& \hat{E}_\mathrm{F}(N) + \mathbf{C} \left( 1 - \langle \hat{E}_\mathrm{F} \rangle \right)  \ge 0 \ .
\end{eqnarray}
At low roughness parameter $r$, both $\hat{E}_\mathrm{F}(N), \langle \hat{E}_\mathrm{F} \rangle \rightarrow 1$, thus $A \rightarrow 0$ producing a divergence. This can be avoided by switching at low-$r$ to the Taylor approximation $\boldsymbol{\rho} = \mathbf{C}(\mathbf{B}^2 - A\mathbf{C})/\mathbf{B}^3$ (or simply $\boldsymbol{\rho} = \mathbf{C}$, the low-$r$ limit).
A useful approximation for this normal-incidence case is (with $< 0.3\%$ error)
\begin{equation}
\boldsymbol{\rho} \approx \mathbf{C} + 0.258831 \left(0.98995 - \mathbf{C}\right) \mathbf{C} r \ .
\end{equation}

\begin{figure}[!hb]
  \centering
    \includegraphics[width=0.495\linewidth]{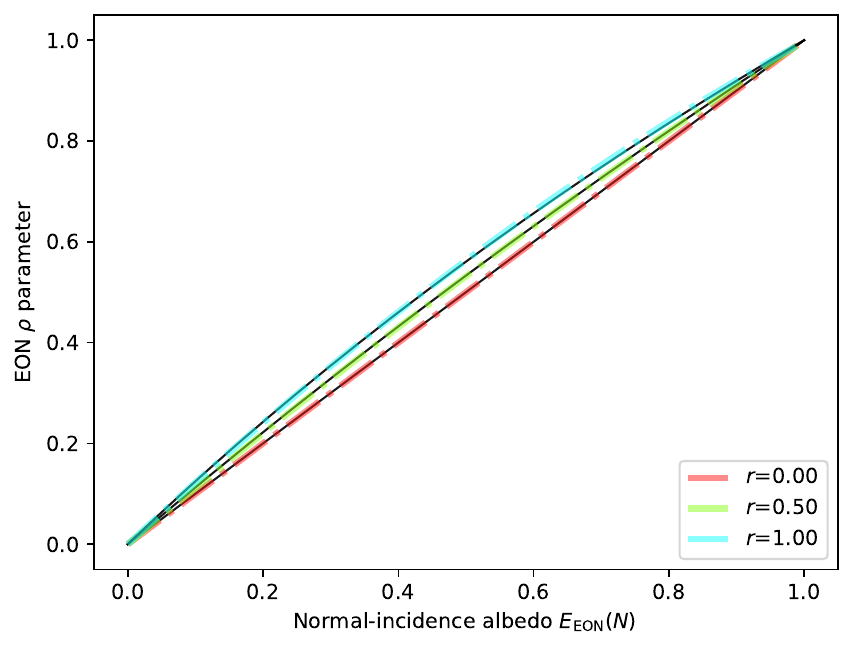}
    \includegraphics[width=0.495\linewidth]{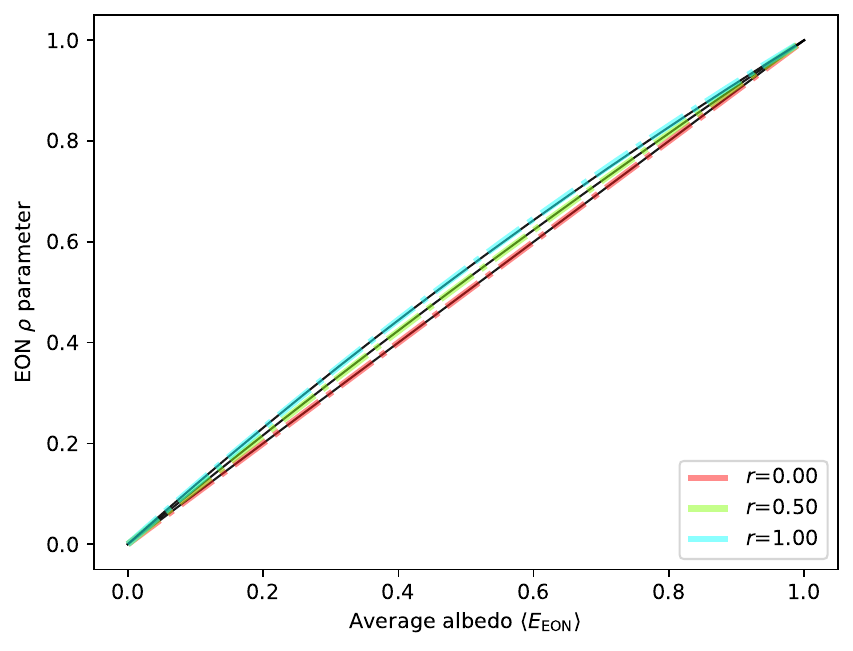}
  \caption{The $\boldsymbol{\rho}$ parameter required to achieve a desired color viewed at normal incidence (left) or as an average albedo (right), as a function of roughness $r$. The black lines are the exact solutions, while the dashed colored lines show the approximations given in the text.
  \label{fig:albedo_inversion}}
\end{figure}

If instead we constrain the average albedo, i.e. set $\langle \mathbf{E}_\mathrm{EON} \rangle = \mathbf{C}$, then we obtain the solution
\begin{equation}
\boldsymbol{\rho} = \frac{\mathbf{C}} {\mathbf{C} + \langle \hat{E}_\mathrm{F} \rangle \left(1 - \mathbf{C}\right) } \ .
\end{equation}
A useful approximation for this average albedo case is (again with $< 0.3\%$ error)
\begin{equation}
\boldsymbol{\rho} \approx \mathbf{C} + 0.189468 \left(1 - \mathbf{C}\right) \mathbf{C} r \ .
\end{equation}
These formulas and their accuracy are illustrated in Figure~\ref{fig:albedo_inversion}.

\clearpage
\section*{Author Contact Information}

\hspace{-2mm}\begin{tabular}{p{0.32\textwidth}p{0.32\textwidth}p{0.32\textwidth}}
Jamie Portsmouth \newline
\href{mailto:jamports@mac.com}{jamports@mac.com}
&
Peter Kutz \newline
\href{mailto:peter.kutz@gmail.com}{peter.kutz@gmail.com} \newline
&
Stephen Hill \newline
\href{mailto:steve@selfshadow.com}{steve@selfshadow.com} \newline
\end{tabular}

\afterdoc

\end{document}